\documentclass[12pt,preprint]{aastex}

\slugcomment{}

\shorttitle{Evolution of {\em Swift}/BAT blazars}
\shortauthors{Ajello et al.}

\begin{document}


\title{The Evolution of {\em Swift}/BAT blazars and the origin of the
MeV background.} 


\author{M. Ajello\altaffilmark{1}, L. Costamante\altaffilmark{2}, 
R.~M. Sambruna\altaffilmark{3}, N. Gehrels\altaffilmark{3},
J. Chiang\altaffilmark{1}, A. Rau\altaffilmark{4}, 
A. Escala\altaffilmark{1,5}, J. Greiner\altaffilmark{6}, 
J. Tueller\altaffilmark{3}, J.~V. Wall\altaffilmark{7}, and
R.~F. Mushotzky\altaffilmark{3}
}

\email{majello@slac.stanford.edu}
\altaffiltext{1}{SLAC National Laboratory and Kavli Institute
for Particle Astrophysics and Cosmology, 2575 Sand Hill Road, Menlo Park,
CA 94025, USA}

\altaffiltext{2}{W.W. Hansen Experimental Physics Laboratory \& Kavli Institute
for Particle Astrophysics and Cosmology, Stanford University, USA}

\altaffiltext{3}{Astrophysics Science Division, Mail Code 661, NASA Goddard Space Flight Center, Greenbelt, MD 20771, USA }

\altaffiltext{4}{Caltech Optical Observatories, MS 105-24, California, 
Institute of  Technology, Pasadena, CA 91125, USA}

\altaffiltext{5}{Departamento de Astronom\'{\i}a, Universidad de Chile, Casilla 36-D, Santiago, Chile.}

\altaffiltext{6}{Max Planck Institut f\"{u}r Extraterrestrische Physik, P.O. Box 1603, 85740, Garching, Germany}

\altaffiltext{7}{Physics and Astronomy Department,
University of British Columbia, 6224 Agricultural Road,
Vancouver V6T 1Z1, Canada}


\begin{abstract}
We use 3\,years of data from the {\em Swift}/BAT survey to 
select a complete sample of X-ray  blazars above 15\,keV. 
This sample comprises  26 Flat-Spectrum Radio Quasars (FSRQs) and 12 BL Lac 
objects detected over a redshift range of 0.03$<$z$<$4.0.
We use this sample to determine, for the first time in the  15--55\,keV band, 
the evolution of
blazars. We find that, contrary to the Seyfert-like AGNs detected by BAT,
the population of blazars shows strong positive evolution. This evolution 
is comparable to the evolution of luminous optical QSOs and 
luminous X-ray selected AGNs.
We also find evidence for an epoch-dependence of the evolution 
as determined previously for radio-quiet AGNs.
We interpret both these findings as a strong link between accretion and jet activity.
In our sample, the FSRQs evolve strongly, while our best-fit shows
that BL Lacs might not evolve at all. The blazar population accounts
for 10--20\,\% (depending on the evolution of the BL Lacs) of the Cosmic
X--ray background (CXB) in the 15--55\,keV band. We find that FSRQs can
explain the entire CXB emission for energies above 500\,keV solving the
mystery of the generation of the MeV background.  The evolution of luminous
FSRQs shows a peak in redshift ($z_c$=4.3$\pm0.5$) which is larger
than the one observed in QSOs and X--ray selected AGNs.
We argue that FSRQs can be used
as tracers of massive elliptical galaxies in the early Universe.
\end{abstract}


\section{Introduction}

Blazars constitute the most extreme class of active galactic nuclei (AGNs).
Their broad-band and highly variable emission is due to a relativistic
jet pointing close to our line of sight \citep[e.g.][]{blandford78}.
In the framework of the AGN unified model,
 which ascribes the observed features of AGNs to  orientation effects
\citep{antonucci93,urry95}, the properties of misaligned blazars 
are consistent with those of radio galaxies. Indeed, 
the two blazar sub-populations,
BL Lacertae (BL Lac)  objects and flat spectrum radio quasars (FSRQs),
are thought to be the beamed counterparts of low- and high-luminosity
radio galaxies, respectively \citep{wall97,willott01}. 
Both classes of objects are normally found only
in the nuclei of giant elliptical galaxies
\citep[e.g.][]{falomo00,odowd02}.

Blazars have been extensively studied at radio \citep{dunlop90,wall05},
soft X-ray \citep{giommi94,rector00,wolter01,caccianiga02,beckmann03,padovani07} 
and GeV energies  \citep{hartman99}.  It seems consolidated that FSRQs evolve
positively \citep[i.e. there were more blazars in the past,][]{dunlop90} 
up to a redshift cut-off which depends on luminosity 
\citep[e.g.][]{padovani07,wall08b}.  In this respect FSRQs evolve
similarly to the population of  X--ray selected, radio-quiet, AGNs 
\citep{ueda03,hasinger05,lafranca05}.
On the other hand, the evolution of BL Lac objects remains a matter 
of debate, since  they were found  to evolve
negatively in a few cases \citep[e.g.][]{rector00,beckmann03} and not
evolving at all in other ones \citep{caccianiga02,padovani07}.

Deriving the luminosity function of a class of objects 
allows to understand the properties of the parent
population and to estimate the diffuse (unresolved) background produced
by the entire class.
Despite all the previous studies, the lack of a sensitive all-sky
hard X-ray survey has prevented, so far,  to gather a sizable sample
of blazars and to study their evolution in the $>$10\,keV band. 
For these reasons, the contribution of blazars to the X--ray background
(above $>10$\,keV) has never been quantified. The aim of this study
is to address all these questions using data from the {\it Swift}/BAT instrument.

\subsection{The high-energy Background}

Radio-quiet AGNs are more abundant than blazars and have been shown to be
the major constituent of the Cosmic X--ray Background 
\citep[CXB,][]{ueda03,treister05,gilli07}. This consolidates the
idea that the CXB emission is the result of accretion onto
super-massive black holes. More precisely, the X--ray emission of AGNs
is due to Compton up-scatter of UV photons (generated in the inner part
of the accretion disk) by high-energy electrons which populate
a region above the disk commonly referred to as corona.
This process, known as Comptonization, was first proposed by \cite{zdziarski86}. 
The bulk of the electron population present in the
corona is expected to be thermal and this naturally produces a 
cut-off in the spectrum of AGNs. The detection of the AGN cut-offs,
detected at energies between 50\,keV and 400\,keV
by OSSE  \citep[][and references therein]{madejski94,zdziarski00}
and the non-detection of Seyfert-like AGNs by EGRET \citep[]{lin93,dermer95a}
confirms this interpretation. 

Population synthesis models normally assume that all emission-line
AGNs have a cut-off in the 200--500\,keV energy range \citep[e.g.][]{ueda03,
gilli07}. The effect of the cut-off combined with the cosmic  evolution of AGNs
implies that the the contribution of radio-quiet AGNs to the CXB emission 
above $\sim$200\,keV is negligible.
Thus, the high-energy CXB emission in the 200--10000\,keV energy range
remains currently unexplained. A few candidates have been proposed to
explain this background. One is the $\gamma$-ray emission
 originated from nuclear decays from Type Ia supernovae 
\citep[SNe Ia;][]{clayton75,zdziarski96,watanabe99}.
However, on the basis of  measurements of the cosmic SN Ia rates,
recent studies showed that the background flux expected from 
SNs Ia is about an order of magnitude lower than the observed 
CXB emission \citep{ahn05a,strigari05}.
Annihilation of dark matter particles has also been discussed, 
but no viable light (with ``MeV'' mass)  dark
matter particle candidate has been found 
\citep[][and references therein]{ahn05b}.

Very recently \cite{inoue08} discussed the possibility that the 
hot corona may contain a small fraction (relative to the whole population)
of non-thermal electrons. These electrons might be powered by magnetic
reconnections in a similar way as it happens during solar flares 
\citep[e.g.][]{shibata95}. In this framework a faint non-thermal
component present in millions of AGNs might explain the observed 
background. However, due to the lack of sensitive instruments
surveying the MeV sky this hypothetical 
non-thermal emission of AGNs has never been detected.

Blazars, whose emission extend from the Radio to the TeV band,
are certainly contributing, despite their relative low space 
density, to the high-energy 
background (both in the X-ray and $\gamma$-ray energy band).
In particular, an important role is certainly played by the 
so called ``MeV blazars'' whose Inverse Compton (IC)
 peak is located in the MeV band
\citep[][]{bloemen95,sikora02,sambruna06}. 
An attempt to quantify the contribution of blazars to the
high-energy background has been performed by \cite{giommi06}
using a multi-frequency selected sample of blazars. Despite
the uncertainties related to the extrapolation from the microwave
to the hard X--ray energy band, they conclude that blazars can explain
$\sim$10\,\% of the CXB emission in the 2--10\,keV energy band and
possibly 100\,\% of the background above 500\,keV.

In this work, we use a complete sample
of blazars detected in the {\em Swift}/BAT survey
to derive, for the first time at these energies,
the X-ray luminosity function (XLF) and the cosmic evolution
of blazars and to assess their contribution to the high-energy background.
The paper is organized as follows. In $\S$~\ref{sec:sample} we 
describe how the blazar sample was selected among the BAT extragalactic
sources and discuss its incompleteness.  In $\S$~\ref{sec:vvm} we
introduce the Maximum Likelihood method which is used to determine
the blazar evolution. The luminosity function of the BAT blazars is derived in
$\S$~\ref{sec:results} and in $\S$~\ref{sec:cxb} it is used to quantify the 
contribution of blazars to the diffuse background.
We discuss the results of our analysis in 
$\S$~\ref{sec:discussion}.
Throughout this paper we assume a standard concordance cosmology 
(H$_0$=70\,km s$^{-1}$ Mpc$^{-1}$, $\Omega_M$=1-$\Omega_{\Lambda}$=0.3).

%
%
\section{The {\em Swift}/BAT sample}
\label{sec:sample}

The Burst Alert Telescope \citep[BAT;][]{barthelmy05}
onboard the {\em Swift} satellite \citep{gehrels04}, represents 
a major improvement in sensitivity for imaging of the hard X-ray sky. 
BAT is a coded mask telescope with a wide field of view 
(FOV, 120$^\circ\times$90$^{\circ}$ partially coded) aperture sensitive in
the 15--200\,keV domain. The main goal of BAT is to locate
Gamma-Ray Bursts (GRBs). While chasing new GRBs, BAT surveys
the hard X-ray sky with an unprecedented sensitivity. Thanks 
to its wide FOV and its pointing strategy, BAT monitors continuously
up to 80\% of the sky every day. Thanks to this quasi-random pointing
strategy and the large FOV, BAT exposure is uniform on the whole sky.

Results of the BAT survey \citep{markwardt05,ajello08a,tueller09}
show that BAT reaches a sensitivity of $\sim$1\,mCrab\footnotemark{}
\footnotetext{1\,mCrab in the 15--55\, keV band corresponds to 
1.27$\times10^{-11}$\,erg cm$^{-2}$ s$^{-1}$}
in 1\,Ms of exposure.
Given its sensitivity and the large exposure already accumulated in 
the whole sky, BAT poses itself as an excellent instrument for studying
populations whose emission is faint in  hard X-rays.

For the analysis presented here we used {\em Swift}/BAT survey observations
performed between March 2005 and March 2008. Data screening and
processing was performed according to the recipes presented in \cite{ajello08a}.
The chosen energy interval is 15-55\,keV. The lower
limit is dictated by the energy threshold of the detectors. The upper limit
was chosen as to avoid the presence of strong background lines which
could worsen the overall sensitivity \citep[see][for details about the 
BAT background]{ajello08c}. 
The all-sky image is obtained as  
the weighted average of all the shorter observations.
The average exposure time in our image is 4.3\,Ms, being 2.0\,Ms and 
6.8\,Ms the minimum and maximum exposure times respectively. 
The final image shows
a Gaussian normal noise and we identified source candidates
as excesses above  the 5\,$\sigma$ level.
All the candidates are then fit with the BAT point spread function 
(using the standard BAT
tool {\it batcelldetect}) to derive the best source position.
Moreover, in order to avoid problems related to source confusion and 
sample incompleteness, we considered only sources at high ($|$b$|$$>$15$^{\circ}$)
Galactic latitude.
This analysis is based on mean source fluxes determined
over the 3\,year period spanned by the survey.
Our high-latitude sample comprises 305 sources.
Of these 40 are Galactic sources (mainly X-ray binaries) and 6
are galaxy clusters (already comprised in the sample of \cite{ajello09}).
The remainder are 247 extragalactic sources and 12 unidentified objects.
The exact composition of the sample is reported in Tab.~\ref{tab:sampleb15}.

\begin{deluxetable}{lc}
\tablewidth{0pt}
\tablecaption{Composition of the BAT high-latitude  sample
 ($|$b$|\geq$15$^{\circ}$ and S/N$\geq$5). 
\label{tab:sampleb15}}
\tablehead{
\colhead{CLASS} &  \colhead{\# objects }}
\startdata
Total                         & 305 \\
Seyferts                      & 199 \\
Blazars                       & 38 \\
Galaxies                      & 4\tablenotemark{a}   \\
Galaxy Clusters               & 6   \\
Radio Galaxies                & 6   \\
Galactic Sources              & 40\tablenotemark{b}  \\
Unidentified                  & 12 \\

\enddata
\tablenotetext{a}{These objects are candidate radio-quiet AGNs (Seyferts), identified by means of a 2--10\,keV follow-up observation, for which an optical
spectrum, and thus redshift, is not yet available.}
\tablenotetext{b}{It includes all objects of Galactic nature (i.e. pulsars,
X-ray binaries, etc.)}.
\end{deluxetable}

\begin{figure}[ht!]
  \begin{center}
  	 \includegraphics[scale=0.7]{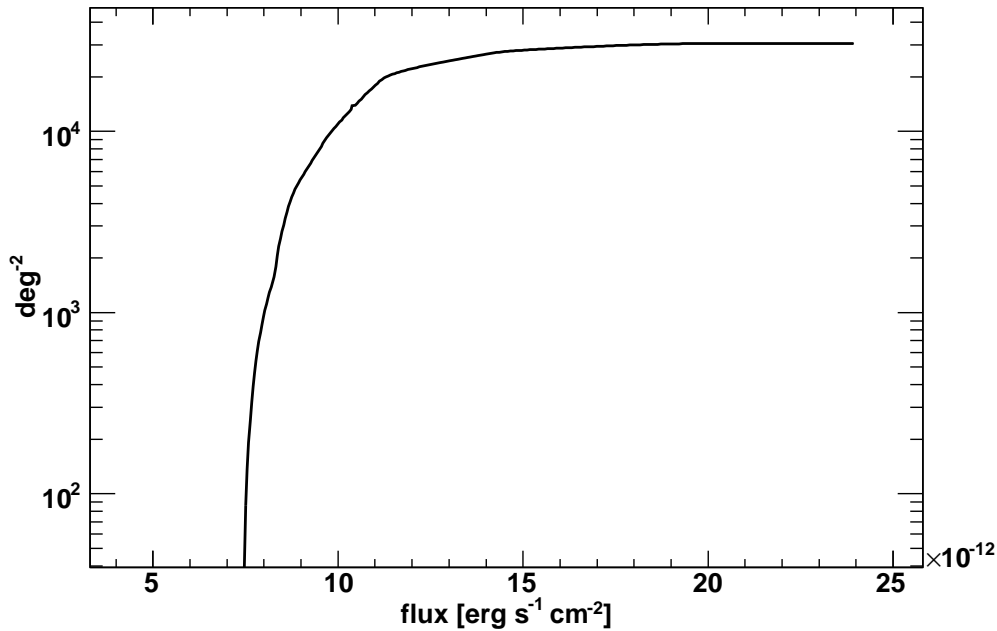}
  \end{center}
  \caption{Sky coverage of the BAT survey as a function of minimum detectable
flux for S/N$\geq 5$\,$\sigma$ and $|$b$|\geq$15$^{\circ}$ 
in the 15--55\,keV band.
\label{fig:skycov}}
\end{figure}

Being the BAT survey not a flux-limited survey, 
but rather a significance-limited  one, it is important to address
how the survey flux limit changes over the sky area. 
This is often referred to as sky coverage, that is
the distribution of the survey's area as a function of limiting flux.
Its knowledge  is very important when performing population studies
as the ones described in the  next sections.
The reader is referred to \cite{ajello08a} for how to derive the sky coverage
which as a function of the minimum detectable flux $F_{min}$
is defined as the sum of the area covered to fluxes $f_i<F_{min}$:
\begin{equation}
\Omega(<F_{min})=\sum_{i}^{N}A_i \ , \ \ \ f_i<F_{min}
\end{equation}
where N is the number of image pixels and $A_i$ 
is the area associated to each of them. 
A visual representation of the sky coverage is reported in 
Fig.~\ref{fig:skycov} which shows clearly the good sensitivity of BAT
which reaches, in our analysis (15--55\,keV), 
a limiting sensitivity of $\sim0.6$\,mCrab
(7.3$\times10^{-12}$\,erg cm$^{-2}$ s$^{-1}$).


\subsection{The Blazar sample}
Selection of blazars in the BAT energy band is not problematic.
Thanks to the deep, although sparse,
coverage of the sky in soft (0.1--2.5\,keV)
and medium (2--10\,keV)  X--rays, most of the BAT sources are already
well studied objects. The location accuracy of BAT was recently
characterized by \cite{tueller09}
using the all-sky sample of sources detected in 22 months of observations.
They find that the 96\,\% error radius for a
5\,$\sigma$, 10\,$\sigma$,  and 20\,$\sigma$ source is respectively
7.5\arcmin\ , 3.34\arcmin\ and 1.59\arcmin\ .
 For the identification of most of the BAT sources,
we  used the 
ROSAT All-Sky Survey Bright Source Catalogue \citep{voges99}, 
the Third IBIS Catalog \citep{bird07}, the BAT catalogs of
\cite{markwardt05} and \cite{tueller08,tueller09} as well as SIMBAD and NED.
We remark that all the identifications reported in the
aforementioned catalogs are based on a thorough study, both in the optical
and X--rays, of the properties of the BAT counterparts 
\citep[e.g.][]{ajello08b,winter08}.
Some 50 BAT sources, for which the above catalogs did not provide
an identification, have a publicly available {\em Swift}/XRT observation.
These observations provided a secure identification of the BAT objects
and the results of all these follow-ups will be discussed elsewhere
(see Tab.~\ref{tab:sampleb15} for a breakdown of the total sample) .
As shown in \cite{ajello08b} and 
\cite{winter08}, most of the BAT AGNs  show complex X--ray
spectra (with the presence of iron line, Compton reflection, and soft excesses) 
and often very large absorption 
(N$_{\rm H}\sim$10$^{23}$\,atoms cm$^{-2}$)  which securely associates
the BAT object with  a non-blazar AGN (e.g. Seyfert).

\begin{figure*}[ht!]
  \begin{center}
  \begin{tabular}{cc}
    \includegraphics[scale=0.43]{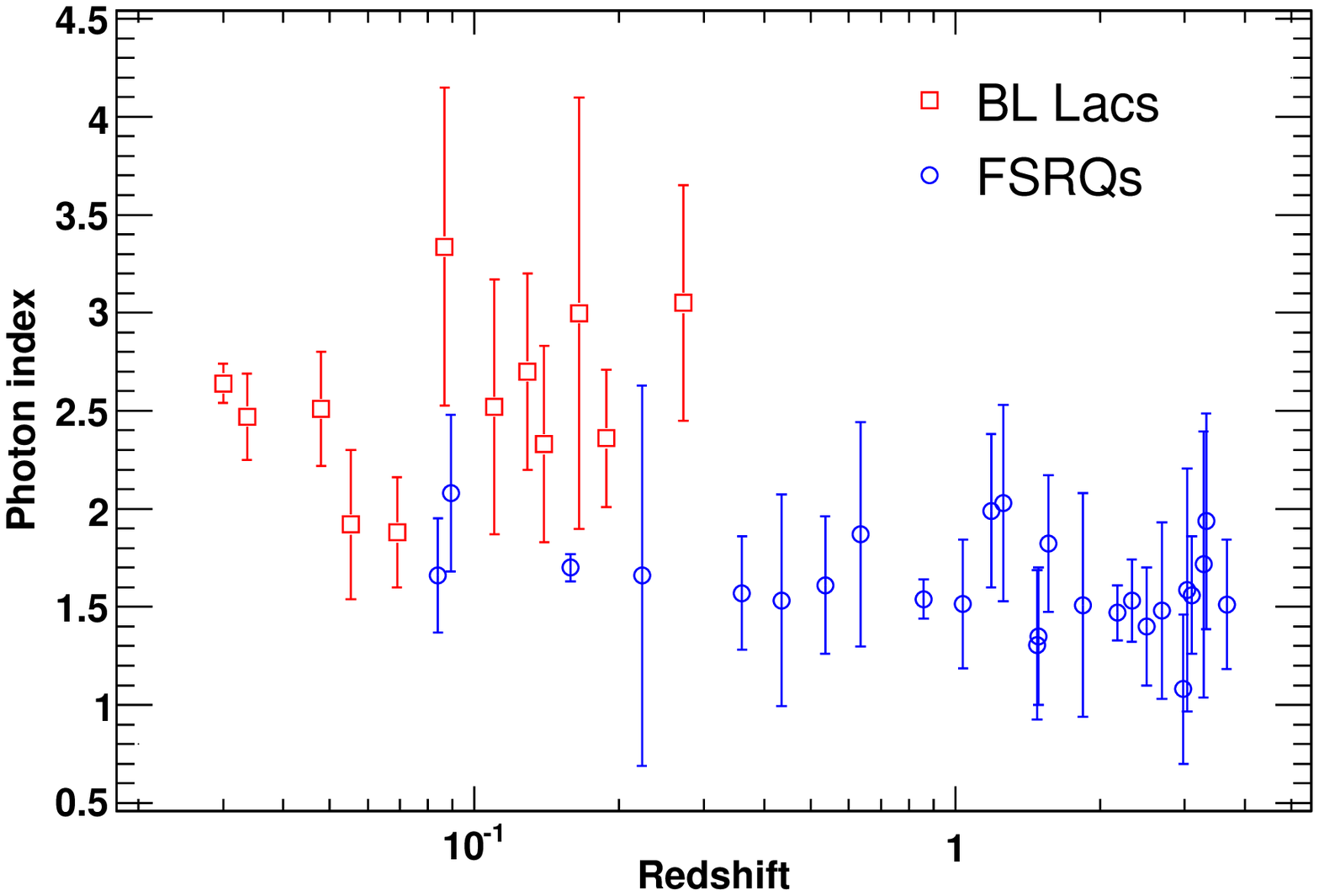} 
  	 \includegraphics[scale=0.43]{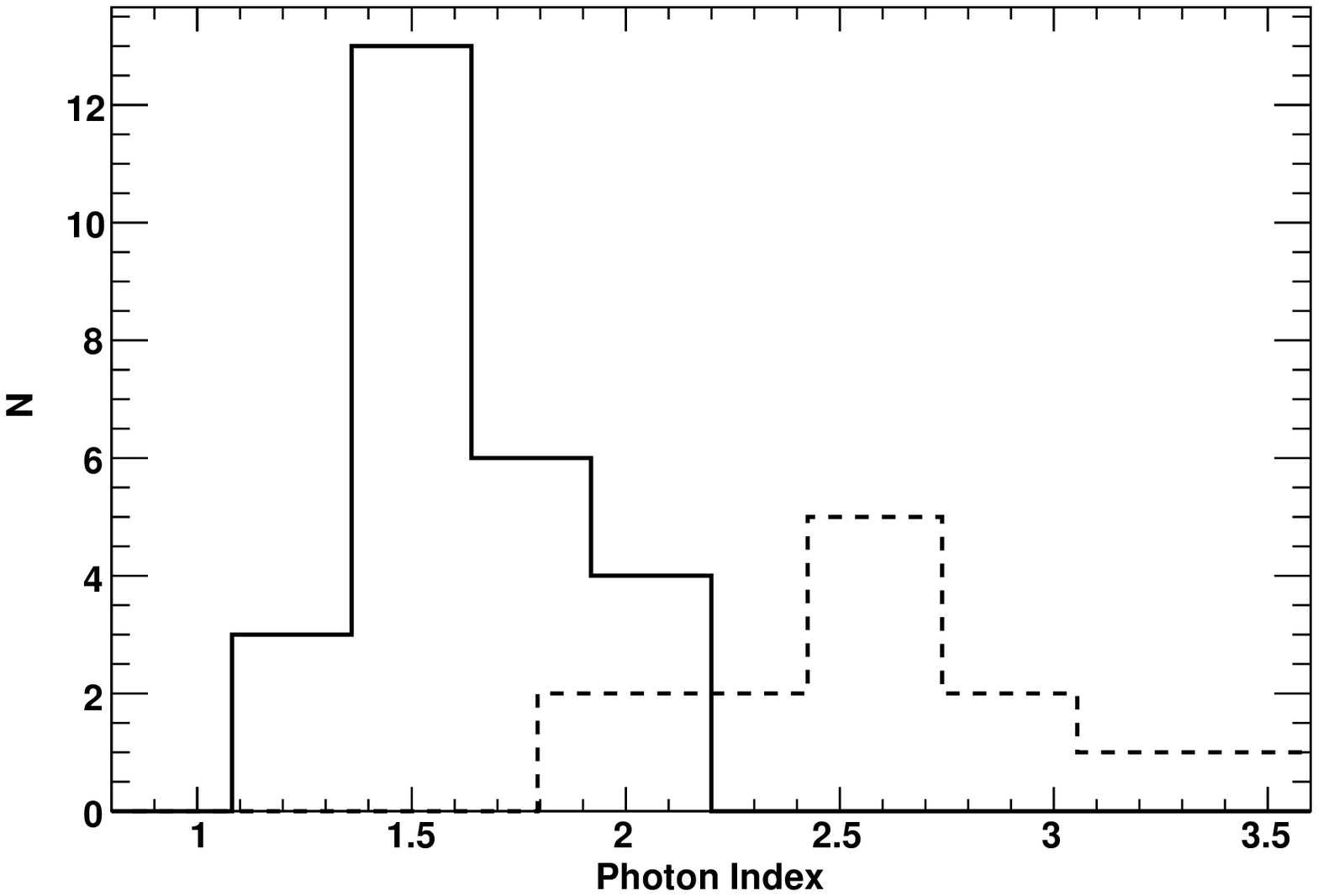}\\
\end{tabular}
  \end{center}
  \caption{
{\bf Left Panel:} Photon index versus redshift for the blazars
in the BAT sample. The low redshift sources, mainly BL Lacs 
(red squares), have
a soft spectrum (photon index of $\sim$2.5, while the high
redshift ones (mainly FSRQs, blue circles) have a hard spectrum. 
{\bf Right Panel:} Photon index distribution for FSRQs (solid line)
and BL Lacs (dashed line).
 \label{fig:phidx}}
\end{figure*}

\begin{figure}[ht!]
  \begin{center}
  	 \includegraphics[scale=0.9]{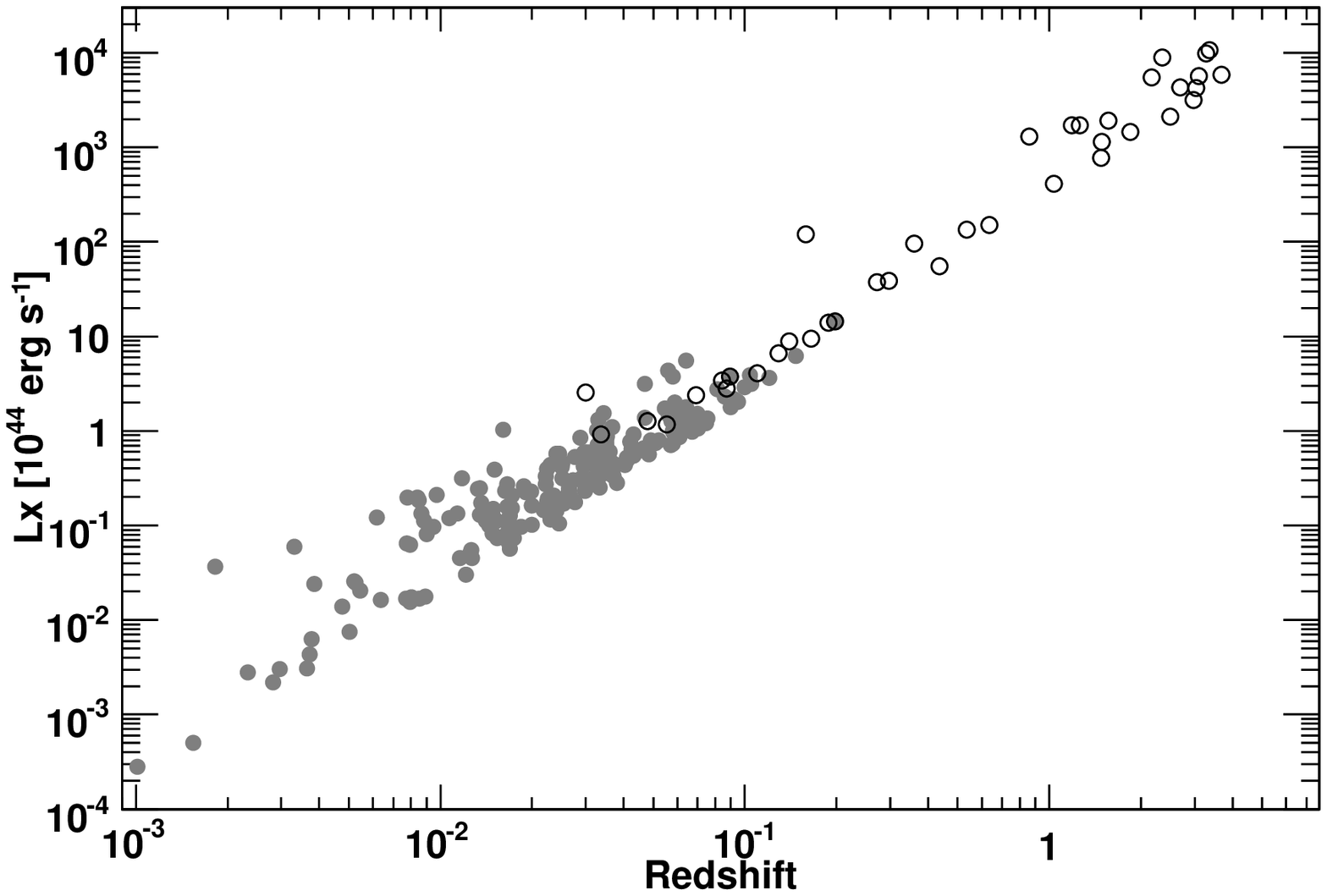}
  \end{center}
  \caption{Luminosity-redshift plane for the BAT blazars (empty circles) and
BAT Seyfert-like AGNs (filled gray circles).
\label{fig:lzplane}}
\end{figure}

In order to securely identify and classify BAT sources as blazars, 
we relied mainly  on the blazar catalog (BZCAT)
of \cite{massaro09} which contains only bona-fide blazars.
These are sources which
show a dominant, broad-band, non-thermal component associated with a 
(relativistic) jet,
a  variable nuclear activity and are detected at least in radio, optical
and X-rays. Given the fact that the BZCAT does not cover the entire sky,
we also used the CRATES catalog of
FSRQs \citep[see also below;][]{healey07} and when necessary individual
source publications.
The sample of BAT blazars is reported in Tab.~\ref{tab:cat}
along with the 
main properties of the sources (e.g. fluxes, signal-to-noise  ratios, etc.),
while the sample
of Seyferts and radio galaxies detected by BAT is reported for reference
in the Appendix~\ref{sec:app}.
For this analysis, BL Lacs are identified as objects
in which the equivalent width of the strongest emission line is
less than 5\,\AA\ and the optical spectrum shows a Ca II H/K break $<$0.4
\citep[e.g.][]{urry95,marcha96}.
The blazar sample comprises 26 FSRQs and 12 BL Lac objects and the full
details of the classification (including the references for it) are 
given in Tab.~\ref{tab:cat}.
Among the BL Lac objects, 9 are of the high-frequency-peaked (HBL) type 
while the rest are of the low-frequency peaked (LBL) type.
Figure~\ref{fig:phidx} (both panels) shows that the BL Lacs detected
by BAT have a softer spectrum with respect to the FSRQ population.
The average photon index of the BL Lac objects, in the BAT band
 is 2.5$\pm0.5$ while the one for FSRQs is 1.6$\pm0.3$. 
This is however not surprising, indeed 
it confirms the expectation that BAT samples the synchrotron component
in BL Lacs and the IC component in FSRQs \cite[e.g. see the BeppoSAX
results of][]{donato05}. 
Figure~\ref{fig:lzplane}
shows the luminosity-redshift plane for the BAT blazars in comparison
with the Seyfert-like AGNs detected by BAT. There are a few things
that can be noted. First, the whole BAT sample of AGNs (Seyferts and blazars)
spans almost 4 decades in redshift and 8 in luminosity. 
There are 16 objects detected at redshift larger than one and they are 
all FSRQs. On the other hand, no blazars are detected 
at low luminosities and low redshift.
This will be discussed in the next sections.

We cross-correlated the BAT blazar sample with the Third EGRET catalog
\citep{hartman99} using a two degree search radius around the BAT
positions to cope with the large error radius of the EGRET positions. 
We further made sure, by searching in the literature, that the EGRET
source, if any, is securely associated with the BAT counterpart
\citep[e.g.][]{sowards04}.
We found that only 12 BAT sources are included in the EGRET catalog being
these 3 BL Lac and 9 FSRQ objects (see Tab.~\ref{tab:cat}).
Moreover, we also checked  how many BAT blazars were also
detected by {\it Fermi}-LAT in 
the first three months of operations \citep{lat_3m}. Again
we found that only 12 sources (9 FSRQs and 3 BL Lacs) are in common
between the two samples (see Tab.~\ref{tab:cat}).
This implies that EGRET/{\it Fermi} and BAT do not sample  exactly the same 
blazar population. In particular, the blazars
BAT detects at the highest redshifts' are not (yet) present in the 
EGRET/{\it Fermi} catalogs.
This would mean that the high-redshift BAT blazars are faint in the
GeV band. This hypothesis will be discussed in the next sections.

We believe the incompleteness of our blazar catalog is negligible.
The fraction of unidentified sources in the total extragalactic
sample is $<$5\,\%.
 The most simple assumption is that these
sources are distributed in source classes exactly like the identified
sources. However, this does not take into consideration other
properties like radio emission. Indeed, no blazars are known
to be radio-quiet sources \citep{wolter01}.
On this basis we found that  none of the unidentified
sources has a radio-loud object within 8\arcmin\ 
or  is present in the blazar catalogs used above. 
In addition we also looked for spatial coincidences, within 8\arcmin\ ,
between the BAT unidentified objects and the CRATES catalog \citep{healey07}.
The CRATES catalog contains all known FSRQs object above a 4.8-GHz flux
of 65\,mJy and for $|$b$|$$>$10\,degrees. Again we found that none of the 
BAT unidentified objects is associated with a CRATES radio source.
We, thus, confidently believe that the incompleteness of the BAT
blazar sample is negligible (i.e. no new blazars are hiding among
the unidentified BAT sources).

\begin{deluxetable}{lcccclcccccc}
\tablewidth{0pt}
\tabletypesize{\tiny}
\rotate
\tablecaption{Blazar sample\label{tab:cat}}
\tablehead{
\colhead{SWIFT NAME}       & \colhead{R.A.}                &
\colhead{Decl.}            & \colhead{Flux}                &
\colhead{S/N}              & \colhead{ID}                  &
\colhead{Type}             & \colhead{Redshift}            &
\colhead{Photon index}     & \colhead{3EG\tablenotemark{a} ?} & \colhead{LAT\tablenotemark{b}?}
& \colhead{Ref.\tablenotemark{c}}\\
%
%
\colhead{}                        & \colhead{\scriptsize (J2000)}         &
\colhead{\scriptsize (J2000)}     & \colhead{\scriptsize(10$^{-11}$\,erg cm$^{-2}$ s$^{-1}$)} &
\colhead{}                        & \colhead{}                            &
\colhead{}                        & \colhead{} & \colhead{}& \colhead{}& \colhead{}
}
\startdata
J0010.4+1056 & 2.617 & 10.935 & 1.88$\pm0.21$ & 9.2 & QSO B0007+107 & FSRQ & 0.09 &  2.08$\pm0.40$ & & & B \\ 
J0018.8+8137 & 4.713 & 81.624 & 1.18$\pm0.19$ & 6.2 & S5 0014+81 & FSRQ & 3.36 &  1.93$\pm0.55$  & & & B\\ 
J0123.0+3421 & 20.752 & 34.351 & 1.27$\pm0.21$ & 5.9 & 1ES  0120+340 & BLLAC (HBL) & 0.27 &  3.05$\pm0.60$ & & & B\\ 
J0207.3+2929 & 31.848 & 29.500 & 1.25$\pm0.22$ & 5.6 & 4C 29.06 & BLLAC (LBL) & 0.11 &  2.52$\pm0.65$ & & & V\\ 
J0225.2+1849 & 36.302 & 18.823 & 1.42$\pm0.22$ & 6.5 & RBS 0315 & FSRQ & 2.69 &  1.48$\pm0.45$ & & & R\\ 
J0233.1+2017 & 38.292 & 20.295 & 1.65$\pm0.22$ & 7.4 & 1ES 0229+200 & BLLAC (HBL) & 0.14 &  2.33$\pm0.50$ & & & B\\ 
J0313.0-7645 & 48.250 &   -76.750   & 0.97$\pm0.12$ & 5.1 & PKS 0312-77  & FSRQ &  0.22 & 1.66$\pm0.97$ & & &C \\ 
J0336.5+3219 & 54.136 & 32.325 & 1.83$\pm0.24$ & 7.6 & 4C 32.14 & FSRQ & 1.26 &  2.03$\pm0.50$ & & & B\\ 
J0349.7-1157 & 57.449 & -11.951 & 1.32$\pm0.21$ & 6.3 & 1ES 0347-121 & BLLAC (HBL) & 0.19 &  2.36$\pm0.35$ & & &S\\ 
J0353.1-6829 & 58.287 & -68.490 & 1.34$\pm0.18$ & 7.5 & IGR J03532-6829 & BLLAC (HBL) & 0.09 &  3.33$\pm0.81$ & & & I\\ 
J0523.0-3626 & 80.755 & -36.447 & 1.61$\pm0.18$ & 9.0 & PKS 0521-365 & BLLAC (LBL)  & 0.06 &  1.92$\pm0.38$ & y & & F\\ 
J0525.5-4557 & 81.398 & -45.951 & 1.06$\pm0.17$ & 6.4 & PKS 0524-460 & FSRQ & 1.48 &  1.31$\pm0.38$& & & K\\ 
J0539.9-2838 & 84.999 & -28.650 & 1.27$\pm0.20$ & 6.2 & PKS 0537-286 & FSRQ & 3.10 &  1.56$\pm0.30$ & & & C\\ 
J0550.8-3217 & 87.716 & -32.289 & 2.08$\pm0.20$ & 10.5 & PKS 0548-322 & BLLAC (HBL) & 0.07 &  1.88$\pm0.28$ & & & S\\ 
J0635.9-7515 & 99.000 & -75.250 & 0.94$\pm0.18$ & 5.1 & PKS 0637-752 & FSRQ & 0.64 &  1.87$\pm0.57$& & & C\\ 
J0746.5+2550 & 116.648 & 25.848 & 1.49$\pm0.25$ & 5.9 & SDSS J074625.87+254902.2 & FSRQ & 2.98 &  1.08$\pm0.38$& & & B\\ 
J0805.3+6148 & 121.349 & 61.800 & 0.96$\pm0.19$ & 5.1 & GB6 J0805+6144 & FSRQ & 3.03 &  1.58$\pm0.62$ & & & B\\ 
J0841.4+7054 & 130.363 & 70.915 & 2.85$\pm0.18$ & 15.9 & 4C +71.07 & FSRQ & 2.17 &  1.47$\pm0.14$ &y & & B\\ 
J1104.5+3812 & 166.126 & 38.210 & 12.16$\pm0.17$ & 80.5 & Mrk 421 & BLLAC (HBL) & 0.03 &  2.64$\pm0.10$ & y & y & B\\ 
J1130.0-1448 & 172.512 & -14.815 & 2.17$\pm0.26$ & 8.3 & PKS 1127-145 & FSRQ & 1.19 &  1.99$\pm0.39$ &y & y & T\\ 
J1213.2+3238 & 183.300 & 32.648 & 0.90$\pm0.17$ & 5.4 & B2 1210+33 & FSRQ & 2.50 &  1.40$\pm0.30$ & & & B\\ 
J1224.9+2118 & 186.249 & 21.300 & 0.95$\pm0.18$ & 5.3 & QSO B1222+216 & FSRQ & 0.43 &  1.53$\pm0.54$ &y & & B\\ 
J1229.1+0202 & 187.283 & 2.047 & 18.31$\pm0.21$ & 87.0 & 3C 273 & FSRQ & 0.16 &  1.70$\pm0.07$ &y &y & B\\ 
J1256.1-0547 & 194.048 & -5.800 & 1.40$\pm0.23$ & 6.1 & 3C 279 & FSRQ & 0.54 &  1.61$\pm0.35$&y & y & C\\ 
J1428.8+4240 & 217.208 & 42.667 & 1.40$\pm0.16$ & 8.5 & H 1426+428 & BLLAC (HBL) & 0.13 &  2.70$\pm0.50$ & & & B\\ 
J1513.1-0903 & 228.275 & -9.061 & 2.49$\pm0.34$ & 7.3 & PKS 1510-089 & FSRQ & 0.36 &  1.57$\pm0.29$&y&y &B\\ 
J1654.1+3945 & 253.542 & 39.763 & 3.46$\pm0.21$ & 16.7 & Mrk 501 & BLLAC (HBL) & 0.03 &  2.47$\pm0.22$ & & y & B\\ 
J1959.7+6509 & 299.940 & 65.159 & 2.29$\pm0.20$ & 11.5 & 1ES 1959+650 & BLLAC (HBL) & 0.05 &  2.51$\pm0.29$&y&y &B\\ 
J2055.6-4710 & 313.918 & -47.182 & 1.48$\pm0.27$ & 5.5 & QSO B2052-47 & FSRQ & 1.49 &  1.35$\pm0.35$&y&y&C\\ 
J2114.1+8205 & 318.540 & 82.095 & 2.00$\pm0.19$ & 10.4 & S5 2116+81 & FSRQ & 0.08 &  1.66$\pm0.29$& & & M\\ 
J2129.3-1536 & 322.350 & -15.600 & 1.55$\pm0.27$ & 5.8 & PKS 2126-158 & FRSQ & 3.28 &  1.72$\pm0.68$& & & C\\ 
J2151.9-3027 & 327.999 & -30.457 & 3.72$\pm0.26$ & 14.3 & PKS 2149-306 & FSRQ & 2.35 &  1.52$\pm0.21$& & & C \\ 
J2229.6-0831 & 337.403 & -8.524 & 1.44$\pm0.23$ & 6.4 & PKS 2227-08 & FSRQ & 1.56 &  1.82$\pm0.35$& &y & C\\ 
J2232.3+1141 & 338.100 & 11.700 & 1.02$\pm0.20$ & 5.1 & CTA 102 & FSRQ & 1.04 &  1.51$\pm0.33$&y&y &B \\ 
J2252.0+2218 & 343.000 & 22.300 & 1.00$\pm0.19$ & 5.2 & MG3 J225155+2217 & FSRQ & 3.67 &  1.51$\pm0.33$& & &B\\ 
J2253.9+1608 & 343.487 & 16.135 & 4.81$\pm0.19$ & 24.7 & 3C 454.3 & FSRQ & 0.86 &  1.54$\pm0.10$&y&y&B\\ 
J2327.5+0935 & 351.900 & 9.600 & 1.03$\pm0.20$ & 5.2 & PKS 2335+093 & FSRQ & 1.84 &  1.51$\pm0.57$& &y&B\\ 
J2358.9-3034 & 359.734 & -30.567 & 1.09$\pm0.20$ & 5.4 & RBS 2070 & BLLAC & 0.17 &  3.00$\pm1.10$& & &F\\ 

\enddata
\tablenotetext{a}{Countepart in the third EGRET catalog \citep{hartman99} ?}
\tablenotetext{b}{Detected by {\it Fermi}-LAT \citep{lat_3m}?}
\tablenotetext{c}{References for the optical classification and redshift: B=\cite{massaro09}, 
F=\cite{falomo94},  C=\cite{healey08}, I=\cite{masetti06c}, K=\cite{stickel93},
 M=\cite{marcha96}, R=\cite{schwope00}, S=\cite{sbarufatti05}, 
T=\cite{stanghellini98}, V=\cite{veron06}.
}
\end{deluxetable}

%
%
\section{The evolution of Blazars}
\label{sec:vvm}
We first test the  evolution  of the BAT blazars by 
applying the $V/V_{\rm MAX}$ method proposed by \cite{schmidt68}.
For a non-evolving source population,  $V/V_{\rm MAX}$
is expected to be uniformly distributed between 0 and 1.
Thus, in case of no evolution the average $V/V_{\rm MAX}$ is 0.5.
The error on the average value can be computed as $\sigma=1/(12N)^{1/2}$.
A value of the $<$$V/V_{\rm MAX}$$>$ significantly different from 0.5
indicates positive (if $>0.5$) or negative evolution (otherwise).
Computing the $<$$V/V_{\rm MAX}$$>$ for the whole sample of blazars
we obtain 0.666$\pm0.045$. This indicates that the blazars population
detected by BAT evolves positively at $>$3\,$\sigma$. This means
that the luminosity or density of the blazars is increasing with
redshift. 
Moreover, we computed the $<$$V/V_{\rm MAX}$$>$ for FSRQs and BL Lacs
separately in order to test if the two sub-populations evolve differently.
  We find a value of $<$$V/V_{\rm MAX}$$>$ of 0.728$\pm0.056$ and 
0.576$\pm0.083$ for FSRQs and BL Lacs, respectively.
Thus, this preliminary analysis shows that while FSRQs are evolving strongly
BL Lacs show only mild evolution (compatible at $\sim2$\,$\sigma$ with no
evolution).
As a control sample we used the sample of the Seyfert-like AGNs detected
by BAT.  Since this population is truly local \citep{ajello08b,tueller08}
we expect the $<$$V/V_{\rm MAX}$$>$ to return a test value of 0.5.
As expected, we obtain 0.509$\pm0.021$. These results are summarized
in Table~\ref{tab:vvmax}.

\begin{deluxetable}{lcccc}
\tablewidth{0pt}
\tablecaption{V/V$_{MAX}$ and log $N$-- log $S$ tests. 
\label{tab:vvmax}}
\tablehead{
\colhead{Sample} & \colhead{$<$V/V$_{MAX}$$>$} & 
\colhead{$\beta$\tablenotemark{a}} & \colhead{A\tablenotemark{b}} & 
\colhead{\# Objects} }
\startdata
Seyferts & 0.509$\pm0.021$ & 1.496$\pm0.073$ & 6.70$\pm0.48$ & 199\\
BLAZARs  & 0.666$\pm0.045$ & 1.932$\pm0.206$ & 1.27$\pm0.20$ & 38 \\
FSRQs    & 0.728$\pm0.056$ & 2.077$\pm0.269$ & 0.83$\pm0.16$ & 26 \\
BL Lacs  & 0.576$\pm0.083$ & 1.694$\pm0.316$ & 0.38$\pm0.10$ & 12\\
\enddata
\tablenotetext{a}{Best-fit exponent of the log $N$--log $S$ distribution
(e.g. $N(>S)= A\ S^{-\beta}$).}

\tablenotetext{b}{Surface density of objects above 10$^{-11}$\,erg cm$^{-2}$
s$^{-1}$ in units of $10^{-3}$\,deg$^{-2}$.}

\end{deluxetable}

Another test of cosmological evolution, although less powerful
than the  $V/V_{\rm MAX}$ method, is the log $N$--log $S$ test
which is based on the distribution of source counts above a given
flux. This distribution is generally expressed as:
 $N(>S)$=A $ S^{-\beta}$ where $S$ is the source flux.
In this test, a non-evolving population shows a distribution which
is consistent with an Euclidean distribution with $\beta=3/2$.
If the source population is positively evolving  then
a $\beta>3/2$ is expected. The results of the log $N$--log $S$ test 
are in excellent agreement with those of the  $<$$V/V_{\rm MAX}$$>$
method (see  Table~\ref{tab:vvmax}). They confirm that
while the Seyfert population is not evolving, the blazar populations
(and in particular the FSRQs) are evolving positively.
Also the normalization of the log $N$--log $S$ distributions
(e.g. $A$ parameter in  Table~\ref{tab:vvmax}) highlights that,
at the current flux limit of BAT,
blazars are 5 times rarer than normal Seyfert galaxies. Thus,
it is only thanks to the uniformly deep all-sky exposure that
BAT gathered a sizable sample of them.

\subsection{The evolving Luminosity Function}

The differential luminosity function of a population of objects
is defined as the number of objects per unit comoving volume and
per unit luminosity interval:

\begin{equation}
\Phi(L_X,z) = \frac{d^2N}{dVdL_X} (L_X,V)\times  \frac{dV}{dz} = 
\rho(L_X,V) \times  \frac{dV}{dz}
\label{eq:phi}
\end{equation}

where $dV/dz$ is the comoving volume element per unit redshift and 
unit solid angle \citep[see e.g.][]{hogg99}.
The present day XLF can be obtained for z=0 as $\Phi(L_X,z=0)$.
We model the present day XLF with two commonly used functions.
A simple power of the form:

\begin{equation}
\Phi(L_X,z=0) = \frac{dN}{dL_X} = 
\frac{A}{L_*}\left( \frac{L_X}{L_*}\right) ^{-\gamma_2},
\label{eq:pow}
\end{equation}

and a double power-law of the form \citep[see e.g.][]{ueda03,hasinger05}:
\begin{equation}
\Phi(L_X,z=0) = \frac{dN}{dL_X}=
\frac{A}{ln(10)L_X}\left[ \left(\frac{L_X}{L_*} \right) ^{\gamma_1} 
+ \left(\frac{L_X}{L_*}\right)^{\gamma_2} \right]^{-1}
\label{eq:2pow}
\end{equation}

Where in Eq.~\ref{eq:pow} we set $L_*= 10^{44}$\,erg s$^{-1}$ while in
Eq.~\ref{eq:2pow} $L_*$ is allowed to vary. 

\subsubsection{The luminosity and density evolutions}
The simplest scenarios of evolution are pure luminosity (PLE) and pure
density evolution (PDE).
In the PLE case the XLF becomes:
\begin{equation}
\Phi(L_X(z),z) = \Phi(L_X/e(z),z=0) 
\label{eq:ple}
\end{equation}

while in the PDE case:
\begin{equation}
\Phi(L_X,z) = \Phi(L_X,z=0) \times e(z)
\label{eq:pde}
\end{equation}

and  the evolution is parametrized using the common 
power-law evolutionary  factor:
\begin{equation}
e(z)= (1+z)^{ k+\gamma z }
\label{eq:ez}
\end{equation}

It is thus clear that in the PLE case the typical blazar luminosity
is changing with redshift while in the PDE case only their 
densities are changing with redshift.
We note that this formulation of the evolutionary factor
(suggested first by \cite{wall08})
reduces for $\gamma=0$ to the standard power law $(1+z)^k$.
For clarity,  we call PDE (PLE) or modified-PDE/PLE (or MPDE, MPLE) the case
in which $\gamma$ has been fixed to zero or was allowed to vary.

\subsection{Maximum Likelihood Analysis}
\label{sec:ml}
A classical approach to derive the XLF is based on
the 1/V$_{\rm MAX}$ method of \cite{schmidt68}.
However, this method is known to introduce a bias 
if there is significant evolution within bins of redshift. 
Moreover, considering the small number of objects in our
sample, binning would result in a loss of information.
We thus decided to apply the Maximum Likelihood method using the
formalism  introduced by \cite{marshall83}.

In this method, the luminosity-redshift plane is parsed into extremely
small intervals of size $dL_X dz$. In each element we compute the 
expected number of blazars with luminosity $L_X$ and redshift $z$:

\begin{equation}
\lambda(L_X,z)dL_Xdz = \rho(L_X,V)\Omega(L_X,z)\frac{dV}{dz}\ dL_X dz
\label{eq:lambda}
\end{equation}

where $ \Omega(L_X,z)$ is the sky coverage of the survey 
(see $\S$~\ref{sec:sample} for details).
The sampling of the luminosity-redshift plane is sufficiently fine
that in each $dL_X dz$ element the number of observed blazars 
is either 1 or 0. In this sparse sampling limit we can define a
likelihood  function  based on joint Poisson probabilities where the
Poisson model is:
\begin{equation}
f(x:m) = \frac{e^{-\mu}\mu^x}{x!}
\end{equation}
where $\mu$ is the expected number of blazars. If $x=1$, then the function
is $\mu e^{-\mu}$ and if $x=0$, it is $e^{-\mu}$.
In this case the likelihood function can be written as:
\begin{equation}
L = \prod_i \lambda(L_{X,i},z_i) dL_X dz e^{-\lambda(L_{X,i},z_i) dL_X dz}
\times \prod_j e^{-\lambda(L_{X,j},z_j) dL_X dz}
\end{equation}

This is the combined probability of observing one blazar at each
element $(L_{X,i},z_i)$ populated by one BAT blazar and observing
zero blazars everywhere else $(L_{X,j},z_j)$.
Transforming to the standard expression $S=-2ln\ L$ and dropping
terms which are not model dependent, we obtain:
\begin{equation}
S = -2\sum_i w_i\ {\rm ln}[\rho(L_{X,i},z_i)] + 2\int^{L_{X,max}}_{L_{X,min}} 
\int^{z_{max}}_{z_{min}} \lambda(L_{X,i},z_i) dL_X dz
\label{eq:s}
\end{equation}

where following \cite{borgani01} we have introduce a weighting
term $w_i$ which takes into account the uncertainties in the luminosity
of each single blazar. In this way, each blazar instead of being a point
in the $L_X,z$-plane is smoothed in the $L_X$-direction according to
a Gaussian distribution with a width set by the 1\,$\sigma$ luminosity error.
Thus, a weight is assigned to each element in the luminosity-redshift plane
based on the fractional contributions of all blazars in the same redshift 
interval:
\begin{equation}
w_i = \sum_k \frac{1}{\sqrt{2\pi \epsilon^{2}_{L_{X,k}}}}\ {\rm exp}\left[ 
-\frac{(L_{X,k} - L_{X,i})^2 } {2\epsilon^{2}_L{_{X,k}}}
\right] dL_X
\end{equation}

where the summation k is over the blazars with a redshift
between $z_i-dz/2$ and  $z_i+dz/2$.
The limits of integration of Eq.~\ref{eq:s}, unless otherwise stated, are:
$L_{X,min}$=$10^{41}$\,erg s$^{-1}$, $L_{X,max}$=10$^{50}$\,erg s$^{-1}$, 
$z_{min}$=0 and $z_{max}$=6. While, the results are independent of the
upper limits of the integration, as far as they are chosen to be large,
this is not the case for the value of $L_{X,min}$ if the local
XLF is modeled as a single power law. In this case, the
lower limit of integration needs to be set to the minimum observed
blazar luminosity (6$\times 10^{43}$\,erg s$^{-1}$ and 
2$\times 10^{44}$\,erg s$^{-1}$ for BL Lacs and FSRQs respectively).

The best fit parameters are determined by minimizing\footnotemark{}
\footnotetext{The MINUIT minimization package, embedded in ROOT (root.cern.ch),
has been used for this purpose. }
 $S$ and their
associated 1\,$\sigma$ errors are computed by varying the parameter
of interest, while the others are allowed to float, until an increment
of  $\Delta S$=1 is achieved. This gives an estimate of the 68\,\%
confidence region for the parameter of interest \citep{avni76}.

\subsection{Alternative Maximum Likelihood Formulation}

Another way of posing the Maximum Likelihood problem
is \citep[following e.g.][]{chiang98,narumoto06}:
\begin{equation}
L = {\rm exp} (-N_{\rm exp}) \prod_{i=1}^{N_{obs}} \Phi(L_{X,i},z_i)
\end{equation}

where $N_{exp}$ is the expected number of blazar detections:
\begin{equation}
N_{exp} = \int dz \int dL_X  \Phi(L_X,z) .
\end{equation}

In this case the function $S$ ($=-2 ln L$) is defined as:
\begin{equation}
S = -2\sum_i^{N_{obs}} ln(\Phi(L_{X,i},z_i) ) - 
2N\ln(N_{exp}).
\end{equation}

We tested that we get exactly the same results if we use one or the other
formulation of the Maximum Likelihood problem. Thus, the results that
we present in the following sections are independent on the ML function
chosen.

\subsection{Consistency Checks}
The Maximum Likelihood approach does not provide a goodness of fit
test and this implies that other methods have to be used to understand
if the fitted function is a good representation of the data.
A common procedure is to use the Kolmogorov-Smirnov (KS) test which
is based on the maximum distance ($D_{KS}$) between the (cumulative) distributions
under comparison. This  test computes the
 probability of observing a KS test statistics as large or larger than 
the observed 
one and it can be used to reject a model when too low. We apply a KS test
to both the cumulative  
redshift and luminosity distributions of the BAT blazars and we reject
XLF models which produce a KS probability $<$20\,\%.

As a further test we check that the best-fit XLF reproduces well the observed
source count distribution (also known as log $N$ - log $S$).
The all-sky number of blazars with a flux stronger than $S$ can be computed as:
\begin{equation}
N(>S) = 4\pi \int^{z_{max}}_{0} dz \frac{dV}{dz}\int^{\infty}_{L_{X(z,S)}}
dL_{X} \rho(L_X,V(z))
\label{eq:logn}
\end{equation}

where ${L_{X(z,S)}}$ is the luminosity of a blazar at redshift $z$ whose
flux is $S$.

\subsection{The Cosmic X--ray Background constraint}
\label{sec:cxbcon}

It is almost certain that the bulk of the CXB emission (below
200\,keV), even if presently
unresolved above 10\,keV, is due to Seyfert-like AGNs 
\citep[e.g.][]{ueda03,lafranca05,treister05,gilli07,silverman08}.
Even though one of our goal is to estimate, in the most robust way,
the contribution of blazars to the CXB spectrum, as a first step
the CXB emission can be used to reject invalid XLF models.
Indeed the blazar contribution to the CXB emission at X--ray energies
is expected to be of the $\sim$10\,\% \citep[e.g.][]{giommi06}.
A much larger fraction would conflict with the present estimates produced
by population synthesis models \cite[e.g.][]{ueda03,treister05,gilli07}
and can be used to rule out a given evolutionary model.
Thus, for each best fit XLF model we compute the integrated background
flux arising from the blazar population. This flux can be derived as:
\begin{equation}
F_{CXB} = \int_{z_{min}}^{z_{max}} dz \frac{dV}{dz} \int^{L_{X,max}}_{L_{X,min}}
dL_{X} F_{X}(L_{X},z) \rho(L_x,V(z))
\label{eq:cxb}
\end{equation}

where the limits of integration are the same as in Eq.~\ref{eq:s}
and $F_{X}(L_{X},z)$ is the flux of a source with luminosity $L_X$
at redshift $z$.

As a final note, we remark that we are not interpreting the CXB as a 
'hard constraint' in the sense that 
the integrated blazar emission of Eq.~\ref{eq:cxb} 
is not constrained to be $\sim$10\,\,\%, but a model XLF which overproduces
the entire CXB (e.g. producing more than 100\,\% of the CXB), 
in the 15--55\,keV band, can be certainly ruled out.

\section{Results}
\label{sec:results}

\subsection{The control sample: Local Seyfert galaxies}

Recently, \cite{tueller08}  computed the (non-evolving)
XLF of local Seyfert galaxies using a sample of 88 objects
detected by BAT in the first 9 months of operations.
Given the small redshift range spanned (z$\leq$0.1),
they did not test  for the evolution of Seyferts
in the local Universe. However, it is well established
that the population of radio-quiet AGNs evolves in density and
luminosity \cite[e.g.][]{hasinger05,silverman08}.
Thus, the sample of 199 Seyferts detected in this analysis represents
a good test for the ML method introduced in $\S$~\ref{sec:ml}. Here,
we aim at deriving a parametric representation of the Seyfert XLF
testing at the same time for their  evolution in the local Universe.

\begin{figure*}[ht!]
  \begin{center}
  \begin{tabular}{cc}
    \includegraphics[scale=0.43]{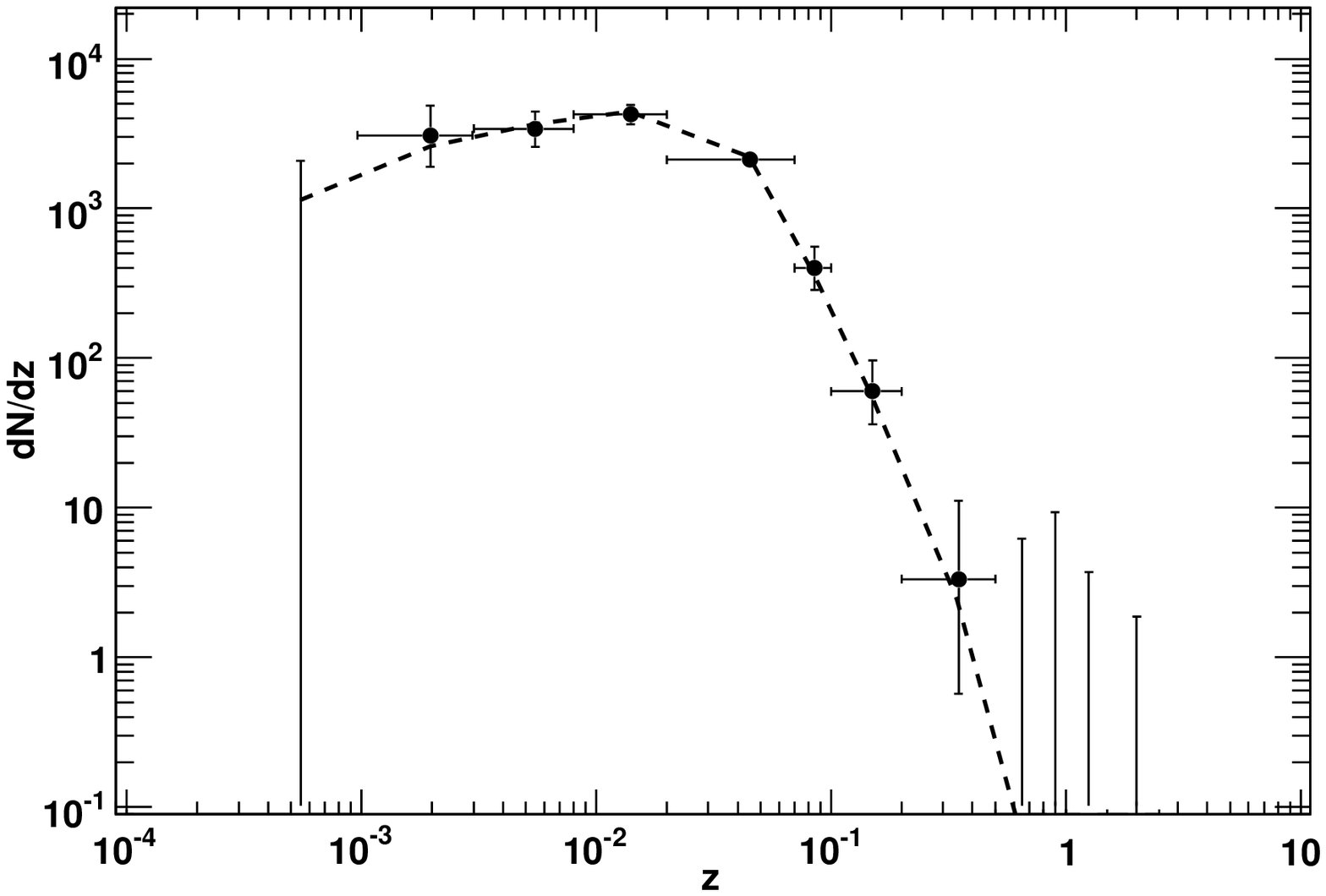} 
  	 \includegraphics[scale=0.43]{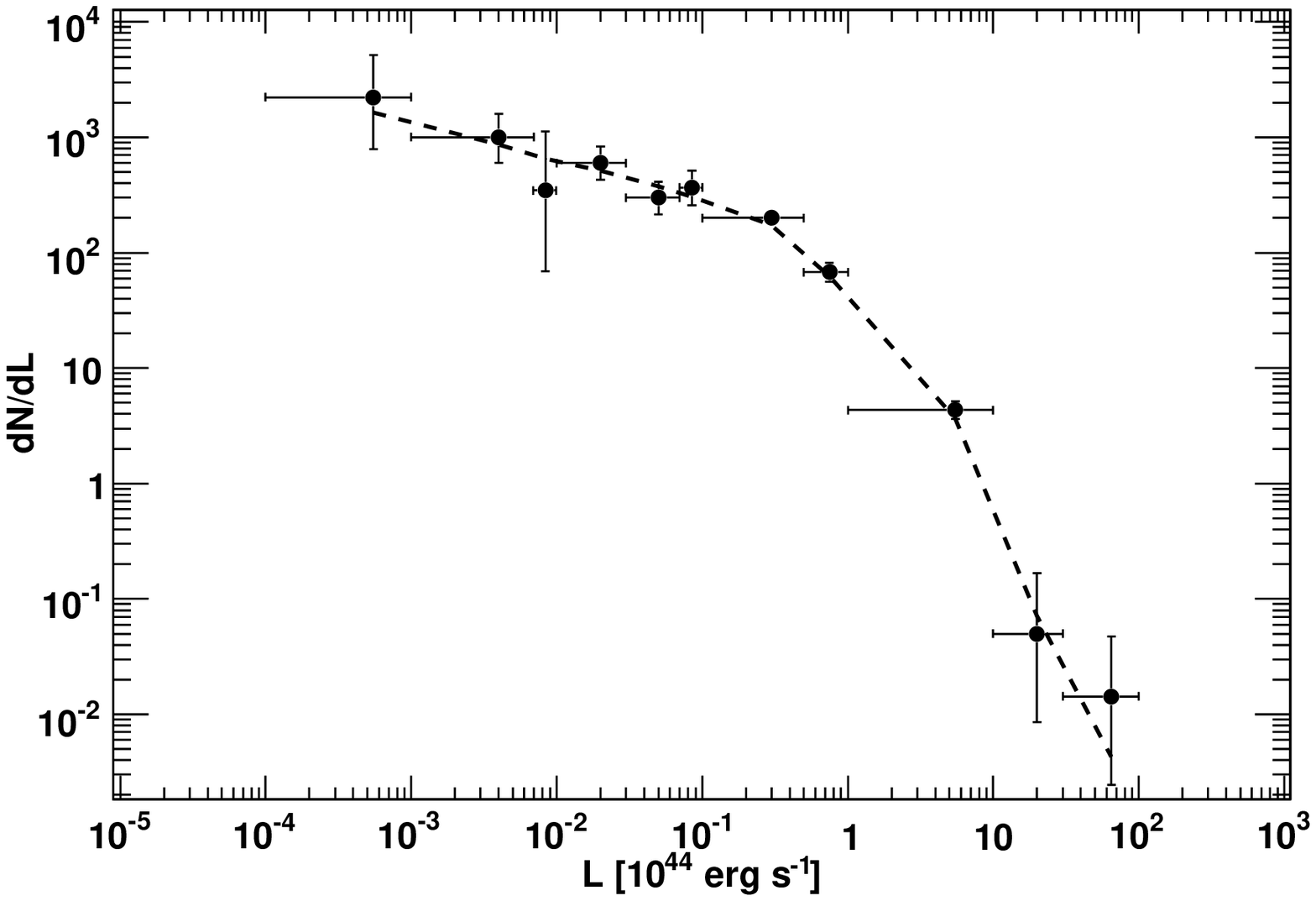}\\
\end{tabular}
  \end{center}
  \caption{Redshift (left) and luminosity (right) distribution of the 
BAT Seyferts. Long error bars consistent with zero are 1\,$\sigma$ upper 
limits 
in the case of observing zero events \citep{gehrels85}.
For both cases, the dashed line represents
a non-evolving XLF (model 1 in Tab~\ref{tab:pars}) 
convolved with the BAT sky coverage.
 \label{fig:agn_xlf}}
\end{figure*}

\begin{figure}[ht!]
  \begin{center}
  	 \includegraphics[scale=0.6]{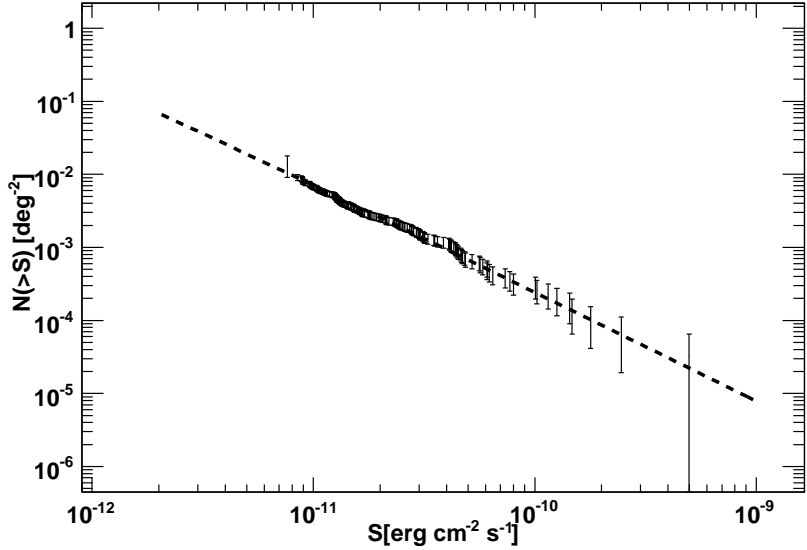}
  \end{center}
  \caption{Cumulative log $N$ - log $S$ of the BAT Seyferts. The dashed line
is  the prediction of the best-fit XLF (model 1 in Tab~\ref{tab:pars}).
\label{fig:agn_logn}}
\end{figure}

We model the local XLF (e.g. $\Phi(L_X,z=0)$) with a double power-law model
as in Eq.~\ref{eq:2pow}  and fix the evolutionary term $k$, of the PLE, to zero.
The best fit parameters, reported in Tab.~\ref{tab:pars} (model 1), are
$\gamma_1=0.80\pm0.08$, $\gamma_2=2.67\pm0.20$ and $L^*=6.1\pm1.4 \times
10^{43}$\,erg s$^{-1}$ and are in very good agreement with the
values reported by \cite{tueller08}\footnotemark{}.
\footnotetext{The different value of H$_0$ and energy band
that \cite{tueller08} adopt produce the net effect that 
the luminosities quoted here are directly comparable to those
quoted by  \cite{tueller08} without the need of a conversion factor
(i.e. the conversion factor is $\sim$1).}
 The error bars are generally smaller
because the sample we use is larger and because the fit is done
to the unbinned dataset. As shown by the redshift, luminosity
and log $N$--log $S$ distributions (reported in Fig.~\ref{fig:agn_xlf} 
and \ref{fig:agn_logn}), this non-evolving XLF model is an highly
acceptable description of the dataset (KS tests $\sim$1, 
see Table~\ref{tab:pars}).

\begin{deluxetable}{llcccccccccccc}
\tablewidth{0pt}
\tabletypesize{\scriptsize}
\rotate
\tablecaption{Parameters of fitted Luminosity Functions. Parameters without an
error estimate were kept fixed during the fitting stage. Kolmogorov-Smirnov 
(KS) values are the probabilities of the model and the data to be drawn
from the same parent population.
\label{tab:pars}}
\tablehead{\colhead{Sample} & \colhead{\# Objects} &
\colhead{Model \#} &
\colhead{Model} & \colhead{A\tablenotemark{a}} & \colhead{$\gamma1$} &
\colhead{$\gamma2$} & \colhead{$L_*$} & \colhead{k} & \colhead{z$_c^*$} &
\colhead{$\gamma$} &\colhead{KS$_{z}$} & \colhead{KS$_{L_X}$}
& \colhead{CXB \%}
}

\startdata

Seyferts & 199 & 1 &PLE+2pow  & 0.909$\pm0.064$\tablenotemark{b} & 0.80$\pm0.08$ & 2.67$\pm0.20$ & 0.61$\pm0.18$ & 0 &
\nodata & \nodata & 0.99 & 0.94 & 21.6\,\%\\

Seyferts & 199 & 2 & PLE+2pow  & 0.778$\pm0.055$\tablenotemark{b} & 0.84$\pm0.08$ & 3.01$\pm0.30$ & 0.61$\pm0.14$ &2.62$\pm1.18$ &
\nodata & \nodata & 0.99 & 0.93 & 55.2\,\%\\

BLAZAR\tablenotemark{c} & 38 & 3& PDE+1pow  & 0.757$\pm0.135$ & \nodata &2.67$\pm0.13$ & 1.0 & 4.00$\pm0.77$ & \nodata & \nodata & 0.43 & 0.22 & 22.0\,\%\\

BLAZAR\tablenotemark{c} & 38 & 4 & MPDE+1pow & 0.732$\pm0.117$ & \nodata & 3.08$\pm0.20$ & 1.0 & 8.95$\pm1.90$ & \nodata & -0.69$\pm0.27$ & 0.34 & 0.40 & 318.0\,\%\\

BLAZAR\tablenotemark{c} & 38 & 5 & MPLE+1pow & 0.804$\pm0.131$ & \nodata & 3.13$\pm0.21$ & 1.0 & 2.96$\pm0.47$ & \nodata & -0.23$\pm0.08$ & 0.43 & 0.41 & 18.4\,\%\\

BLAZAR & 38 & 6 & PDE+1pow  & 0.255$\pm0.041$ & \nodata &2.26$\pm0.07$ & 1.0 & 2.05$\pm0.57$ & \nodata & \nodata & 0.002 & 0.00 & 6.9\,\%\\

BLAZAR & 38 & 7 & MPLE+2pow  & 1.379$\pm0.224$ & -0.87$\pm1.31$ & 2.73$\pm0.38$ & 1.81$\pm0.77$ & 3.45$\pm0.44$ &  \nodata & -0.25$\pm0.07$ & 0.86 & 0.88 & $\sim$20.0\,\% \\

BLAZAR & 38 & 8 & MPDE+2pow  & 0.948$\pm0.152$ & -0.83$\pm1.43$ & 2.54$\pm0.21$ & 1.95$\pm0.93$ & 11.62$\pm1.40$ &  \nodata & -0.85$\pm0.18$ & 0.46 & 0.85 & 640.0\,\% \\

FSRQ\tablenotemark{d}  & 26 & 9 & MPLE+1pow & 0.533$\pm0.104$ & \nodata & 3.45$\pm0.20$ & 1.0 & 3.72$\pm0.50$ & \nodata & -0.32$\pm0.08$ & 0.86 & 0.87 & 9.0\,\% \\

FSRQ\tablenotemark{e} & 26 & 10 & MPLE+2pow  & 0.175$\pm0.034$ & $<-50.0$ & 2.49$\pm0.37$ & 2.42$\pm0.19$ & 3.67$\pm0.48$ &  \nodata & -0.30$\pm0.08$ & 0.85 & 0.89 & 8.3\,\% \\

BLLac\tablenotemark{c} & 12 & 11 & PLE+1pow &  0.830$\pm0.240$ & \nodata & 2.61$\pm0.37$ & 1.0 & -0.79$\pm2.43$ & \nodata & \nodata &  0.55 & 0.33 & 0.3\,\% \\

BLLac\tablenotemark{c} & 12 & 12 & PLE+1pow &  0.784$\pm0.226$ & \nodata & 2.73$\pm0.17$ & 1.0 &  0 & \nodata & \nodata &  0.55 & 0.33 & 0.3\,\% \\

BLLac & 12 & 13 & PLE+2pow &  1.506$\pm0.435$ & -0.89$\pm3.7$ & 2.51$\pm1.61$ & 1.78$\pm2.71$ & 1.54$\pm2.73$ & \nodata & \nodata &  0.71 & 0.93 & 2.2\,\% \\

\enddata
\tablenotetext{a}{In unit of $10^{-7}$\,Mpc$^{-3}$  erg$^{-1}$ s unless otherwise stated.}

\tablenotetext{b}{In unit of $10^{-5}$\,Mpc$^{-3}$.}

\tablenotetext{c}{\,L$_{Min}$ has been set to $6\times 10^{43}$\, erg$^{-1}$ s.}

\tablenotetext{d}{\,L$_{Min}$ has been set to 2$\times10^{44}$\, erg$^{-1}$ s.}

\tablenotetext{e}{Given the very small value of faint-end slope $\gamma_1$,
 this model is equivalent to a model where the local XLF is parametrized as a single power law
with a sharp cut-off at the lowest observed FSRQ luminosity of $\sim2\times10^{44}$\,erg s$^{-1}$.}

\end{deluxetable}

\begin{figure}[ht!]
  \begin{center}
  	 \includegraphics[scale=0.9]{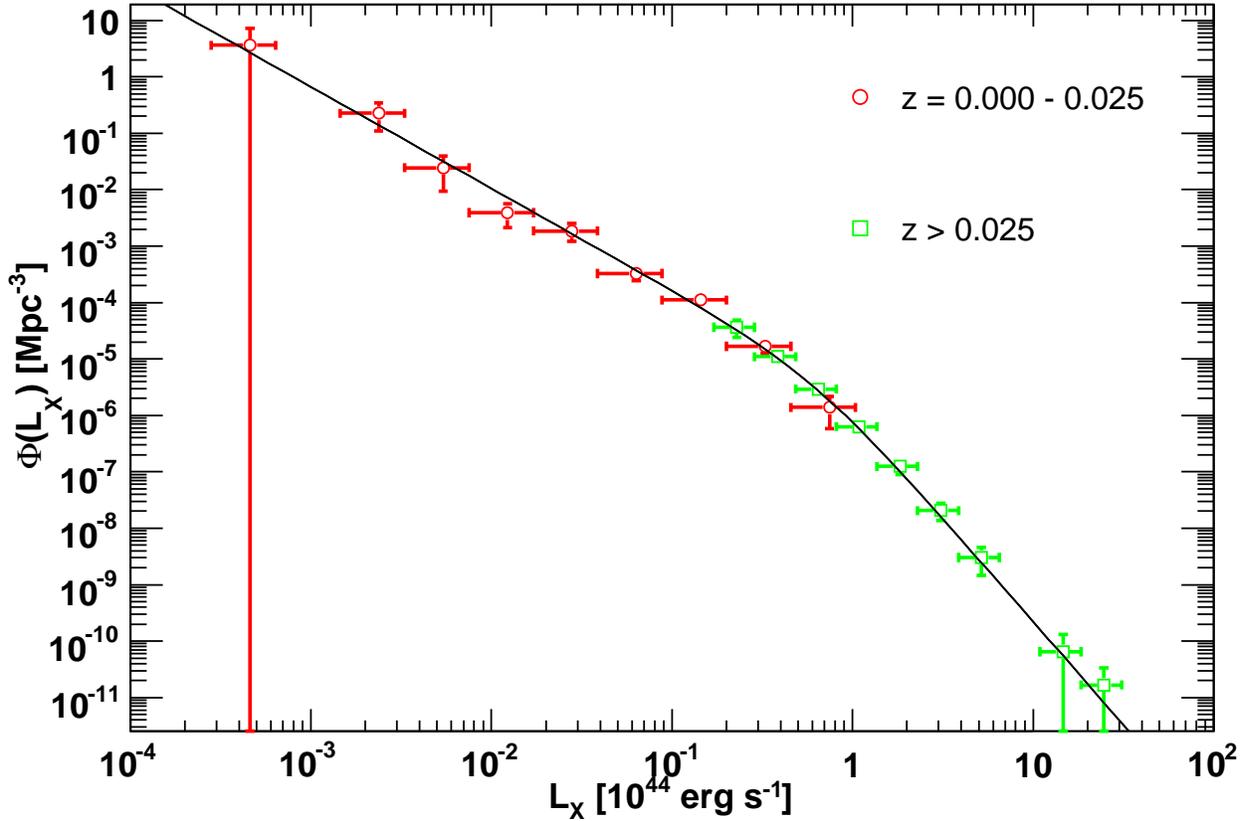}
  \end{center}
  \caption{Luminosity function of the Seyfert-like AGNs, derived using
the 1/V$_{MAX}$ method, in two redshift bins (data points). The solid
line is a non-evolving XLF double power-law model 
(model 1 in Tab.~\ref{tab:pars}) as derived from the ML algorigthm.
\label{fig:agnvmax}}
\end{figure}

Allowing the XLF to evolve in luminosity (see model 2 in Tab.~\ref{tab:pars})
produces an equally good fit with an evolution parameter ($k=2.62\pm1.18$)
which denotes positive evolution although constrained only at the 
$\sim$2\,$\sigma$ level.
It is interesting to note that the evolution parameter is in good 
agreement with the values of 2.29$\pm0.09$ and  2.7$\pm0.2$  found
for the PLE case by \cite{ueda03} and \cite{hasinger05} respectively.
However, since in our case the two models produce an equally good
fit (see KS test values), the non-evolving XLF has to be preferred
because of the lower number of free parameters.
Thus, we believe that the evidences of the evolution 
of radio-quiet AGN in the local Universe are, with the current dataset,
marginal. As a final proof, we built a non-parametric representation
of the luminosity function of the Seyferts using the 1/V$_{MAX}$ method
\citep{schmidt68}. In order to test for evolution we binned the dataset
in two redshift bins containing approximately the same number of sources.
This luminosity function is reported in Fig.~\ref{fig:agnvmax}.
An evolving XLF would show a shift (in luminosity or density) from
one redshift bin to the other. This is not the case for the BAT Seyferts
whose luminosity functions, derived in two different redshift bins,
are the continuation one of each other. In the same Figure, the 
best-fit non-evolving XLF (model 1) is also displayed and it is clear
that this represents the data well.
The fact that  the evolution, if detected, is only marginal
is not surprising, but 
consistent with  the results of the $V/V_{\rm MAX}$
test reported in Table~\ref{tab:vvmax} and the Euclidean behavior
of the log $N$ - log $S$ as reported by \cite{ajello08b} and \cite{tueller08}.

\subsection{A single Blazar population}
\label{sec:1pop}
As a first case, we start assuming that the local (present day) XLF can be
adequately approximated by a simple power law as in Eq.~\ref{eq:pow}.
It was already shown by \cite{marshall83} that in this case it is
impossible to discriminate between {\it density} and {\it luminosity}
evolution. So we refer to density evolution, but we note that the two
type of evolutions are formally equivalent in this case.

\begin{figure*}[ht!]
  \begin{center}
  \begin{tabular}{cc}
    \includegraphics[scale=0.43]{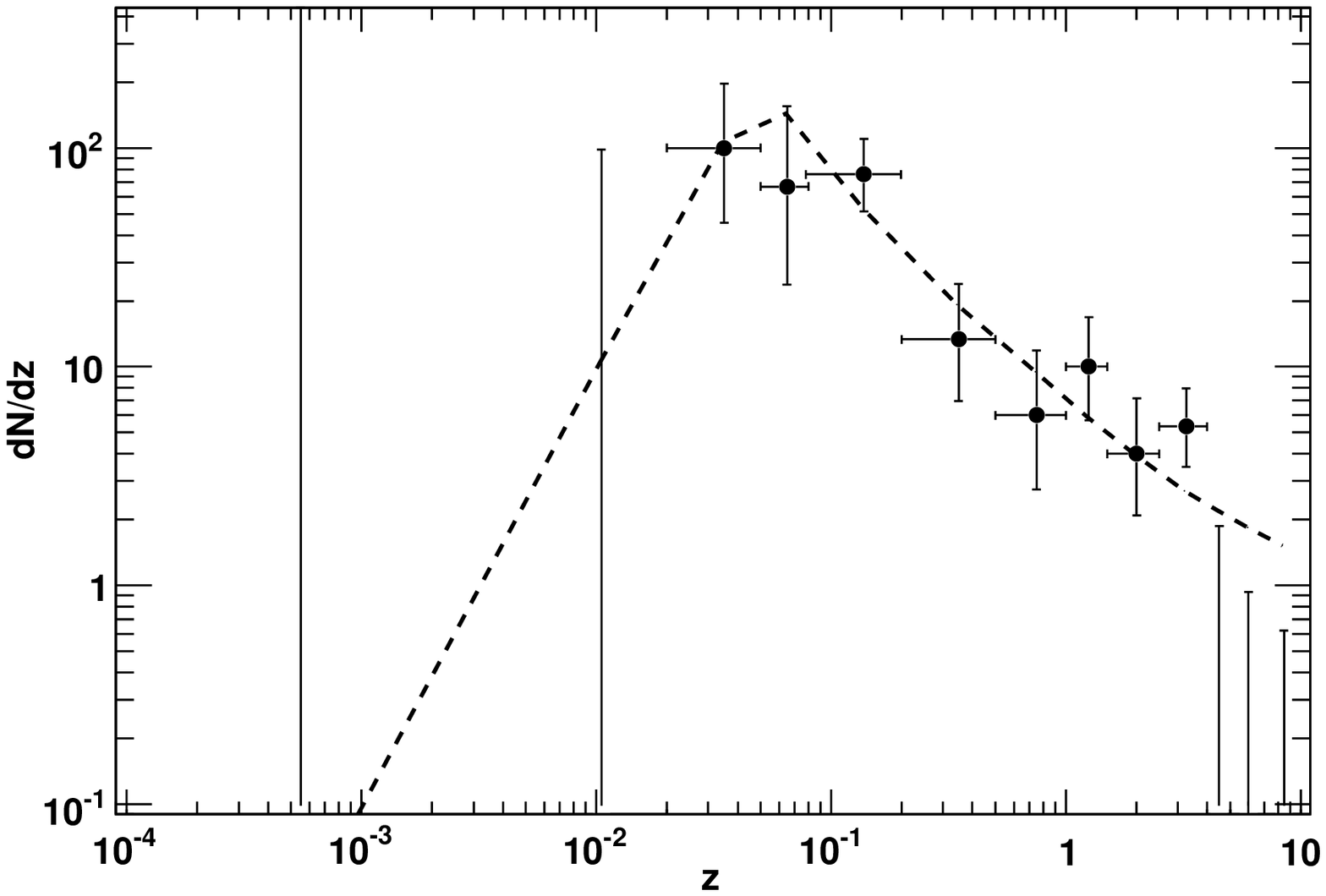} 
  	 \includegraphics[scale=0.43]{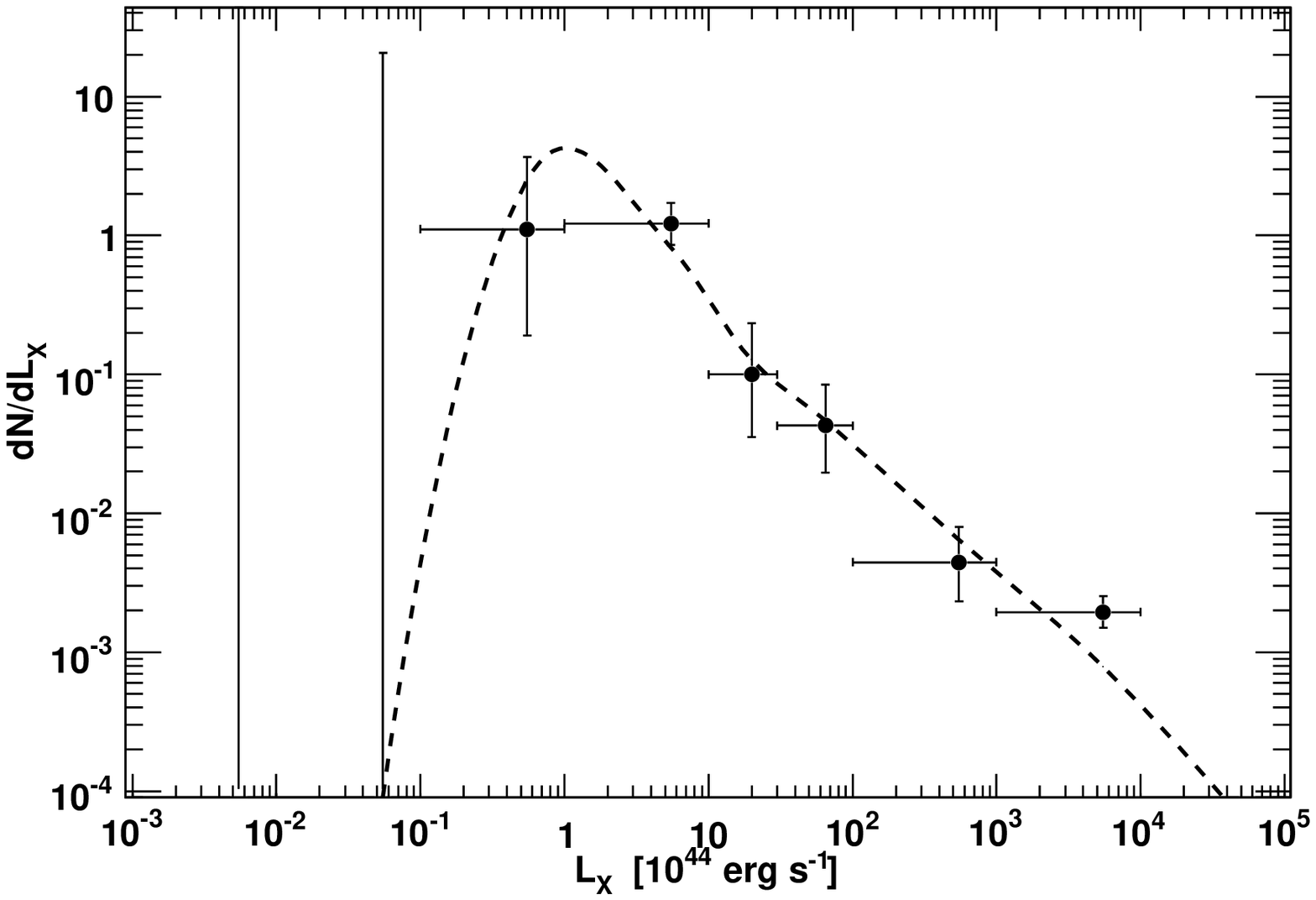}\\
\end{tabular}
  \end{center}
  \caption{Redshift (left) and luminosity (right) distribution of the 
BAT blazars with superimposed (dashed line)
 the prediction of the best-fit PDE XLF
(single power-law case, see model 3 in Tab.~\ref{tab:pars}) 
convolved with the BAT sky coverage.  
Error bars were computed taking into  account the Poisson
error \citep{gehrels85}. Long error bars consistent with zero are
1\,$\sigma$  upper limits for the case of observing zero events in a given
bin.
 \label{fig:pde_1pow}}
\end{figure*}

In the simple power-law case (and PDE) we obtain a slope
of the XLF of $\gamma_2=2.67\pm0.13$ and an evolution
parameter of  4.00$\pm0.77$. The error on the evolution parameter $k$
confirms that the evolution is significantly detected 
in the BAT sample (see model 3 in Tab.~\ref{tab:pars}). 
We note that the XLF slope is in very good agreement with the bright end
slope of the Seyfert-like AGNs detected by BAT (see Tab.~\ref{tab:pars}).
Figure~\ref{fig:pde_1pow} shows the redshift and luminosity distribution
of the BAT blazars with superimposed the best fit XLF. The KS test
shows that this is already an acceptable description of the data.

We note, from left panel of Fig.~\ref{fig:pde_1pow}, that the best-fit PDE XLF 
fails to describe the drop in blazar counts above z$\sim$4. This might be 
a sign of a possible cut-off in the evolution. Thus, we decided to model
the evolution factor as $e(z) = (1+z)^{k+\gamma z}$ as done by \cite{wall08}
(see Eq.~\ref{eq:ez}) calling this model a modified PDE (or MPDE). 
The best fit to the data shows
that the value of $\gamma$ is constrained to be negative at the 
$\sim$2.5\,$\sigma$ level ($\gamma$=-0.69$\pm0.27$). This shows that a cut-off
in the evolution is needed in our data. Although, this model reproduces
our data reasonably well, the integration of the XLF shows that it 
overpredicts the ``total'' CXB emission by a factor 3 (see Tab.~\ref{tab:pars}).


Thus the interpretation of {\it density} evolution might not be the
correct one. A similar effect was already noted by \cite{marshall83}
who concluded for their sample of optical QSOs that {\it luminosity}
evolution was likely occurring.

As already said, for a single power-law XLF there is no formal difference
between  density or luminosity evolution. The only difference is
that in the integral of Eq.~\ref{eq:s}, the integration limit $L_{X,min}$
is evolving and can be expressed as $L_{X,min} = L^0_{X,min}\times e(z)$,
where $ L^0_{X,min}$ is the present day luminosity cut-off 
(e.g. 6$\times$ 10$^{43}$\,erg s$^{-1}$) and $e(z) = (1+z)^{k+\gamma z}$.
The best fit confirms that indeed the luminosity evolution is a better
interpretation of the underlying evolution. Indeed,  integrating the
XLF we get that the blazar population, described by this MPLE function,
accounts for $~$20\,\% of the CXB emission in the 15-55\,keV band.
We also note that the best-fit value of the evolution parameter 
$k=2.96\pm0.46$ is in very good agreement with what found by 
\cite{ueda03} for X--ray selected AGNs and by \cite{wolter01}
for X--ray selected FSRQs.

All the  XLF models described so far (models 3, 4 and 5 in Tab.~\ref{tab:pars})
become unacceptable if the limit on the
minimum observed luminosity $L_{min}$ (see Eq.~\ref{eq:s}) is removed.
Indeed, in this case (see results of the KS tests for model 6) 
the luminosity and redshift
distributions are not reproduced correctly because the best-fit model
predicts many blazars at low luminosity and low redshift which are not
detected by BAT. Thus, a rather drastic change in the power-law behavior
of the local XLF is required in order to reproduce the lack of 
low-luminosity objects.

\begin{figure*}[ht!]
  \begin{center}
  \begin{tabular}{cc}
    \includegraphics[scale=0.43]{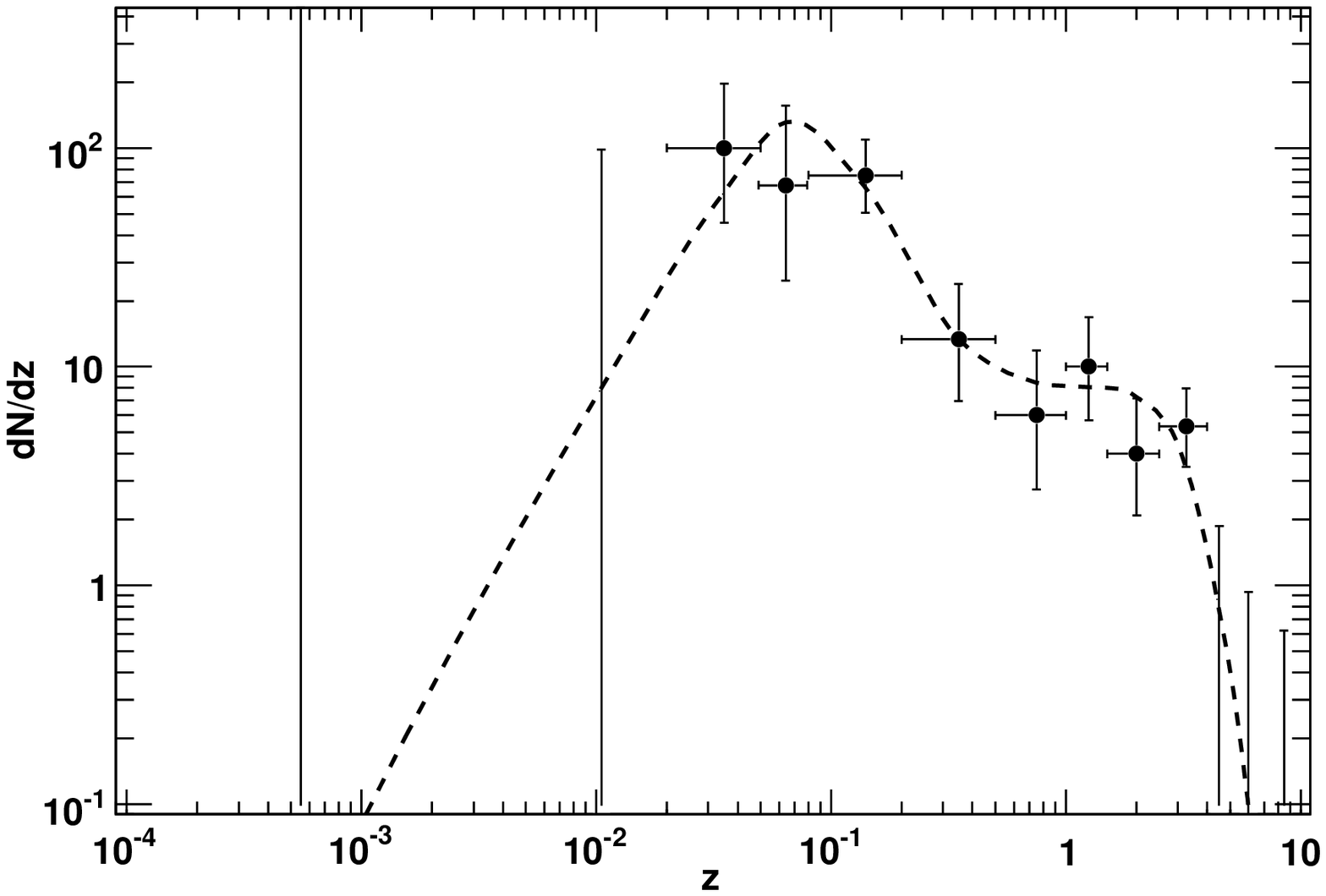} 
  	 \includegraphics[scale=0.43]{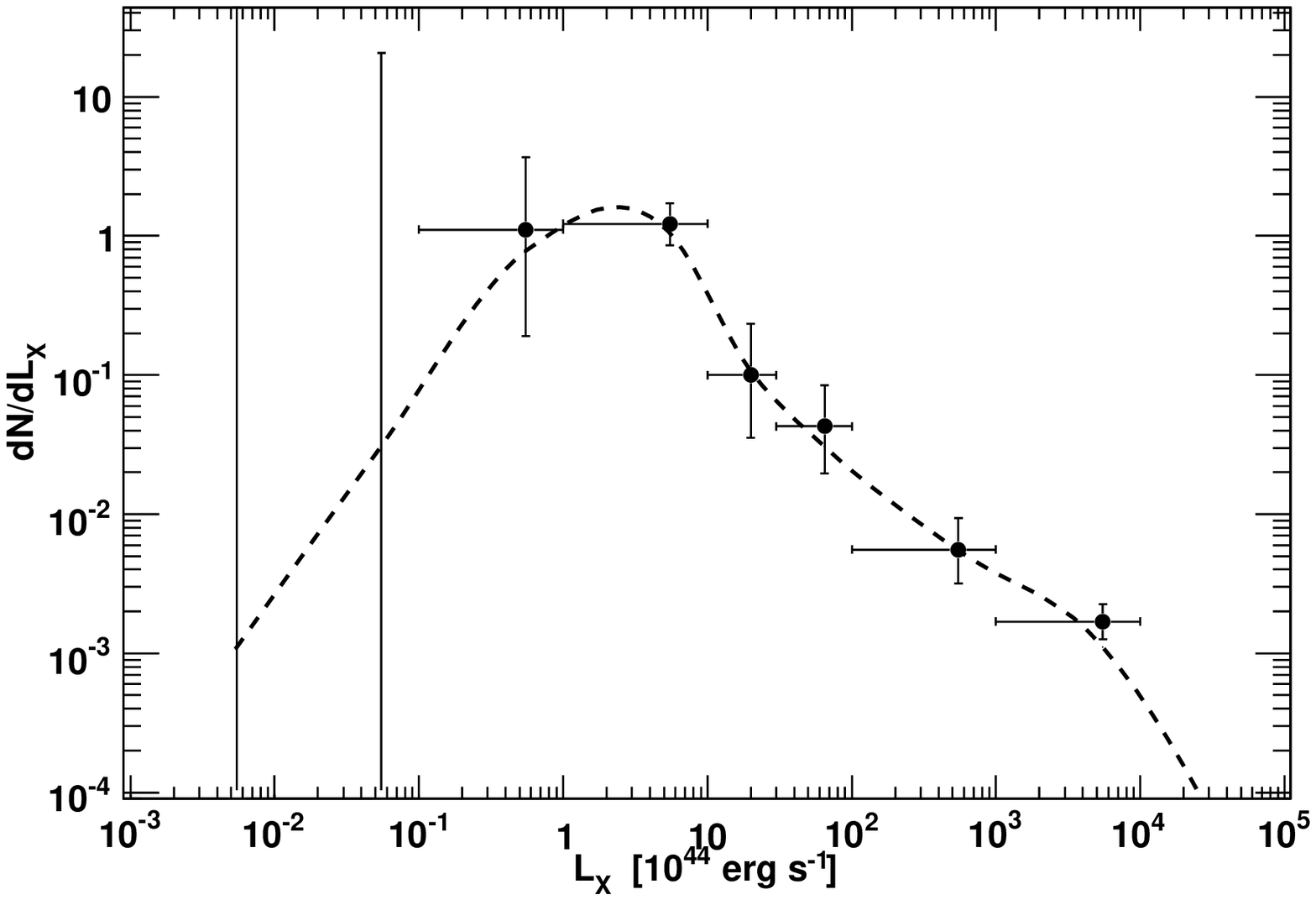}\\
\end{tabular}
  \end{center}
  \caption{Redshift (left) and luminosity (right) distribution of the 
BAT blazars. Error bars were computed taking into account the Poisson
error \citep{gehrels85}. For both cases, the dashed line represents
the MPLE XLF (double power-law model, see model 7 in Tab.~\ref{tab:pars}) 
convolved with the BAT sky coverage.
 \label{fig:2pow}}
\end{figure*}

To test this scenario, we model the local XLF as a double power law
(see Eq.~\ref{eq:2pow})  coupled to a MPLE model. 
In this model, we remove the constraint of a low luminosity cut-off
and the fit is performed to the whole luminosity-redshift plane.
This XLF model reproduces our data accurately 
(see model 7 in Tab.~\ref{tab:pars} and the distributions
reported in Fig.~\ref{fig:2pow}).
Given the lack of low-luminosity objects, the faint-end slope $\gamma_1$ is,
from our fit, required to be flat, but poorly constrained (-0.87$\pm1.31$).

\begin{figure*}[ht!]
  \begin{center}
  \begin{tabular}{cc}
    \includegraphics[scale=0.27]{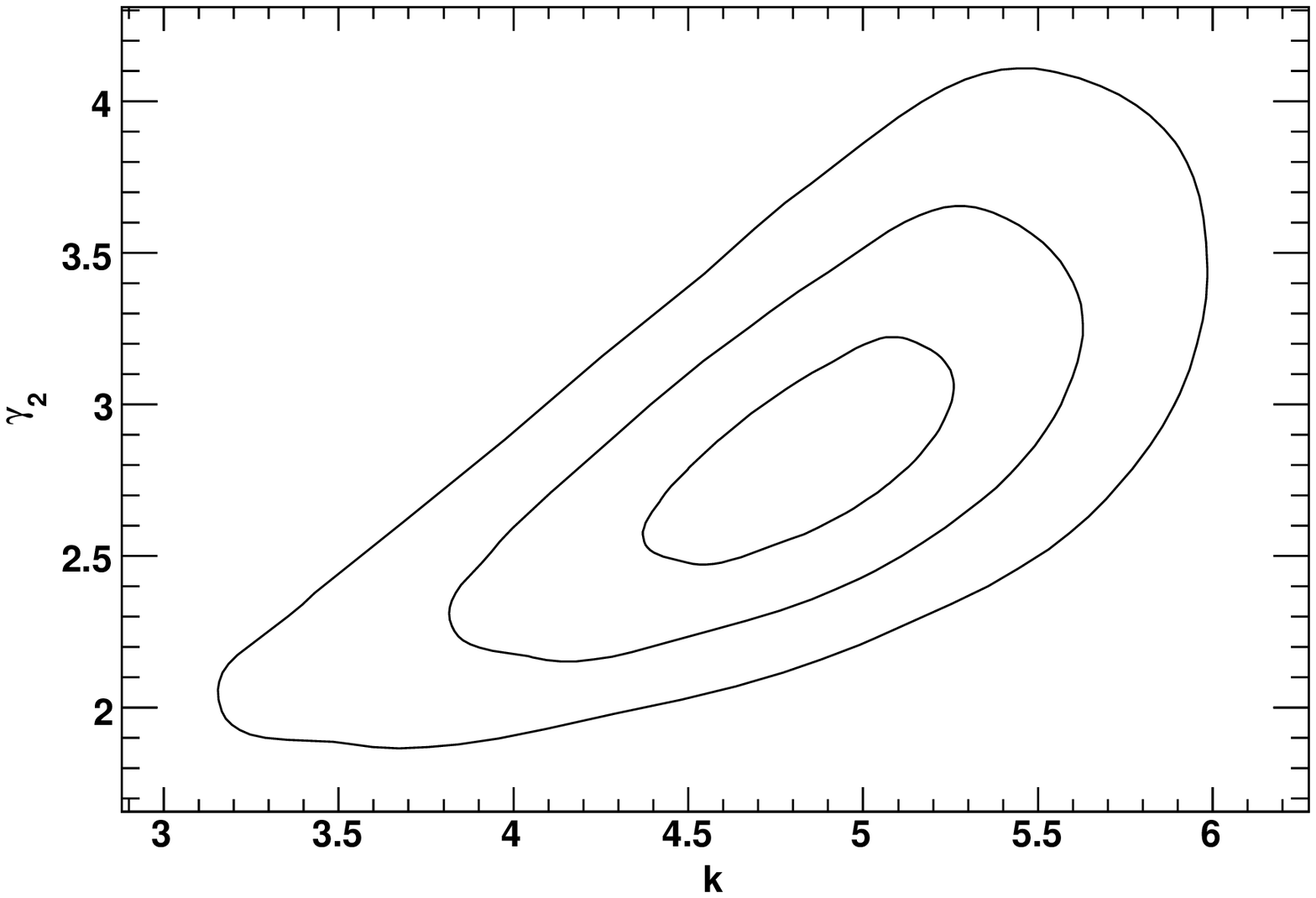} 
  	 \includegraphics[scale=0.27]{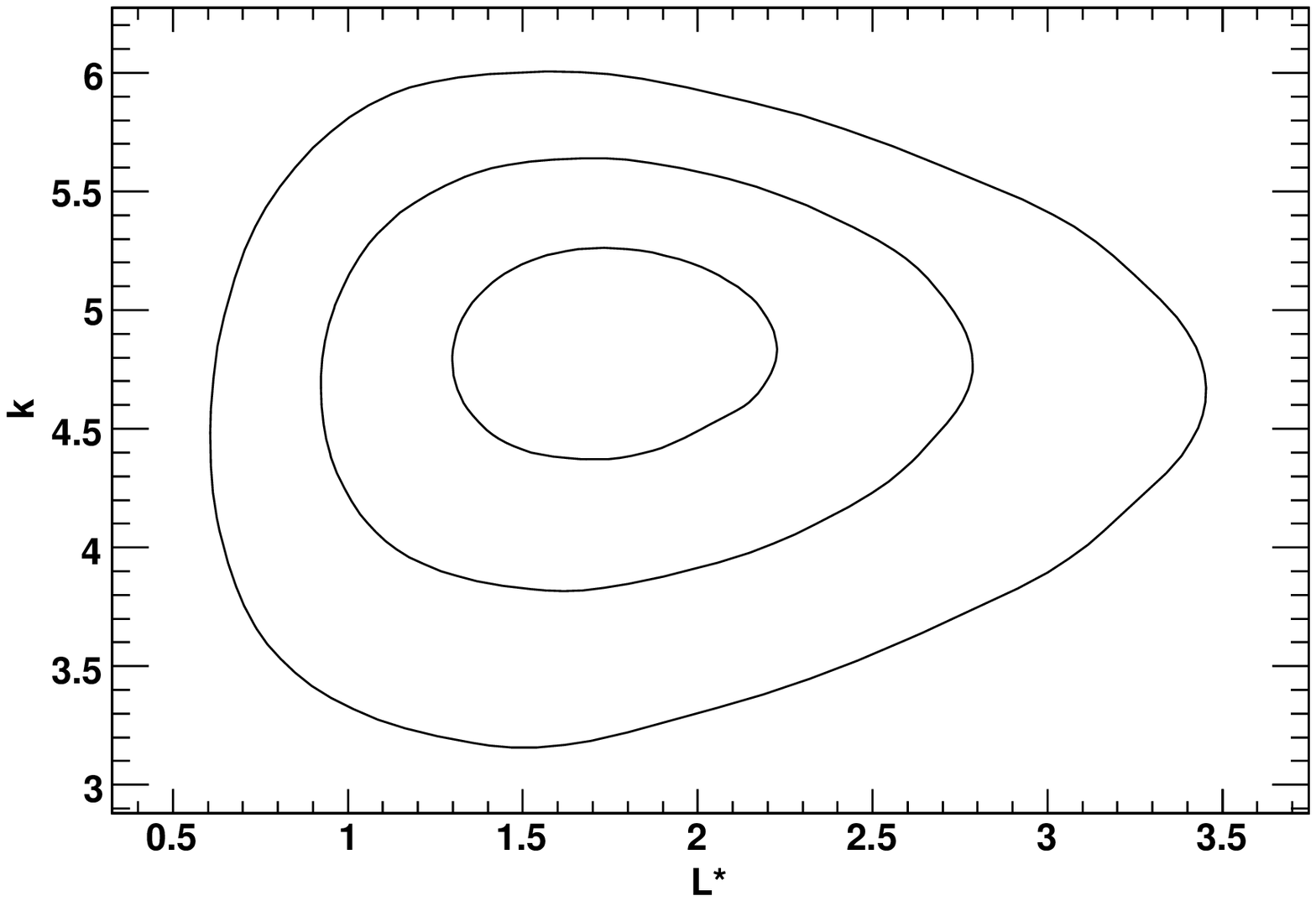}
	 \includegraphics[scale=0.27]{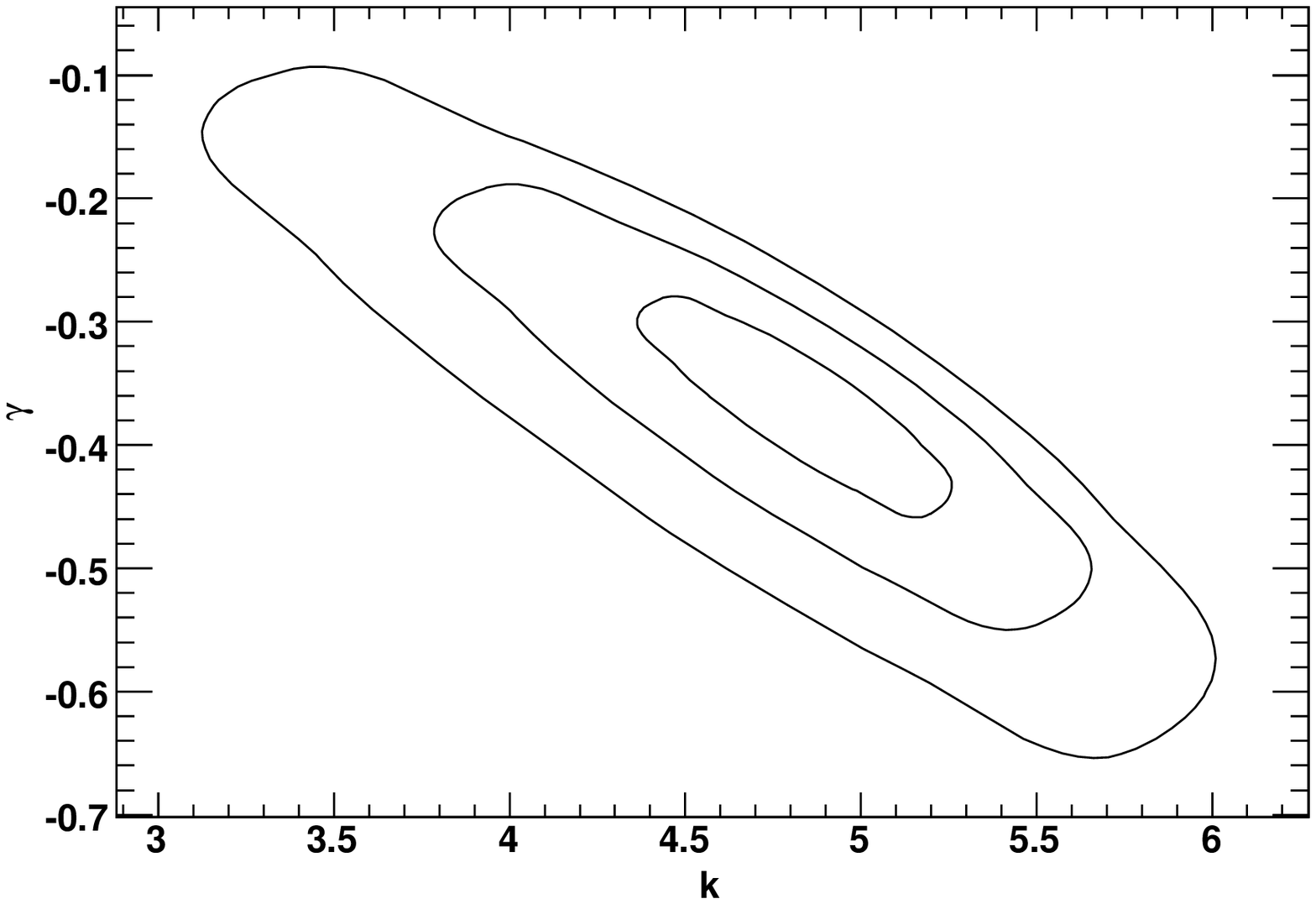}
\end{tabular}
  \end{center}
  \caption{
Confidence contours (1,2, and 3\,$\sigma$) for
the bright-end XLF slope ($\gamma_2$), the evolution parameters
($k$ and $\gamma$) and the break luminosity $L_*$ for the best-fit XLF model
(model 7 in Tab.~\ref{tab:pars}).
\label{fig:ldde_cont}}
\end{figure*}

The slope of the bright-end part of the XLF $\gamma_2=2.73\pm0.37$
is in good agreement with the same slope found for the Seyferts
(see model 1). The likelihood ratio test can be used
to assess whether a model produces a significant 
improvement over another one.
The  likelihood ratio test is the difference between the 
value of $S$ (see Eq.~\ref{eq:s}) produced by different models.
This value ($\Delta S$) is expected to be asymptotically distributed 
as the $\chi^2_n$ \citep{wilks38} where $n$ is the difference between the 
degrees of freedom of the two models. The $\Delta S$ for model 7 (double
power law plus MPLE)
with respect to  model 4 (single power-law plus MPDE) 
is $\sim$10.1 which translates in a probability
of 0.0015\footnotemark{} that the improvement was obtained by chance.
\footnotetext{The chance probability was computed using the $\chi^2$
distribution for one degree of freedom. Indeed, the difference between
model 7 and model 4 is given by the faint-end slope $\gamma_1$ which
is allowed to vary while the cut-off luminosity $L_{min}$ (imposed
in model 4) and the break luminosity L$_*$ (in model 7) represent
essentially  the same parameter.}
We also note that density evolution (with a double power law as a
local XLF) reproduces the data equivalently well (see model 8), but
it is ruled out since it overpredicts the CXB emission by a factor
$>5$. Thus, we consider model 7 as the best representation of our data.
Figure~\ref{fig:ldde_cont} shows the confidence 
contours for the best-fit parameters.

The extreme flattening of the XLF at low luminosities
can be the effect of beaming. As discussed by \cite{urry84} relativistic
beaming alters the observed luminosity function of blazars producing
a flattening at low luminosities. For common jet emission scenarios
\citep[see][for details]{urry84}, the faint-end slope of the XLF should
be $\sim1.0$.  Given the absolute lack of BAT blazars populating
the low-luminosity part of the XLF, it is not surprising that 
the best-fit value of $\gamma_1$ is $\sim$1.5\,$\sigma$ away from
the \cite{urry84} prediction. On the other hand, relativistic beaming
should not affect the bright end slope which should reflect the
slope of the intrinsic luminosity function. It thus becomes interesting
to compare the value of  $\gamma_2$ derived here with other surveys.
Recently \cite{cara08} derived the intrinsic radio luminosity function of 
the Fanaroff-Riley (FR) classes I and II which are thought to be the 
parent populations of respectively FSRQs and BL Lacs. For these
two classes they derive that the slope of the intrinsic luminosity
function is respectively 2.53$\pm0.06$ and 2.65$\pm0.06$ which
are in good agreement with the value of 2.73$\pm0.38$ derived here.

\begin{figure*}[ht!]
  \begin{center}
  \begin{tabular}{cc}
    \includegraphics[scale=0.43]{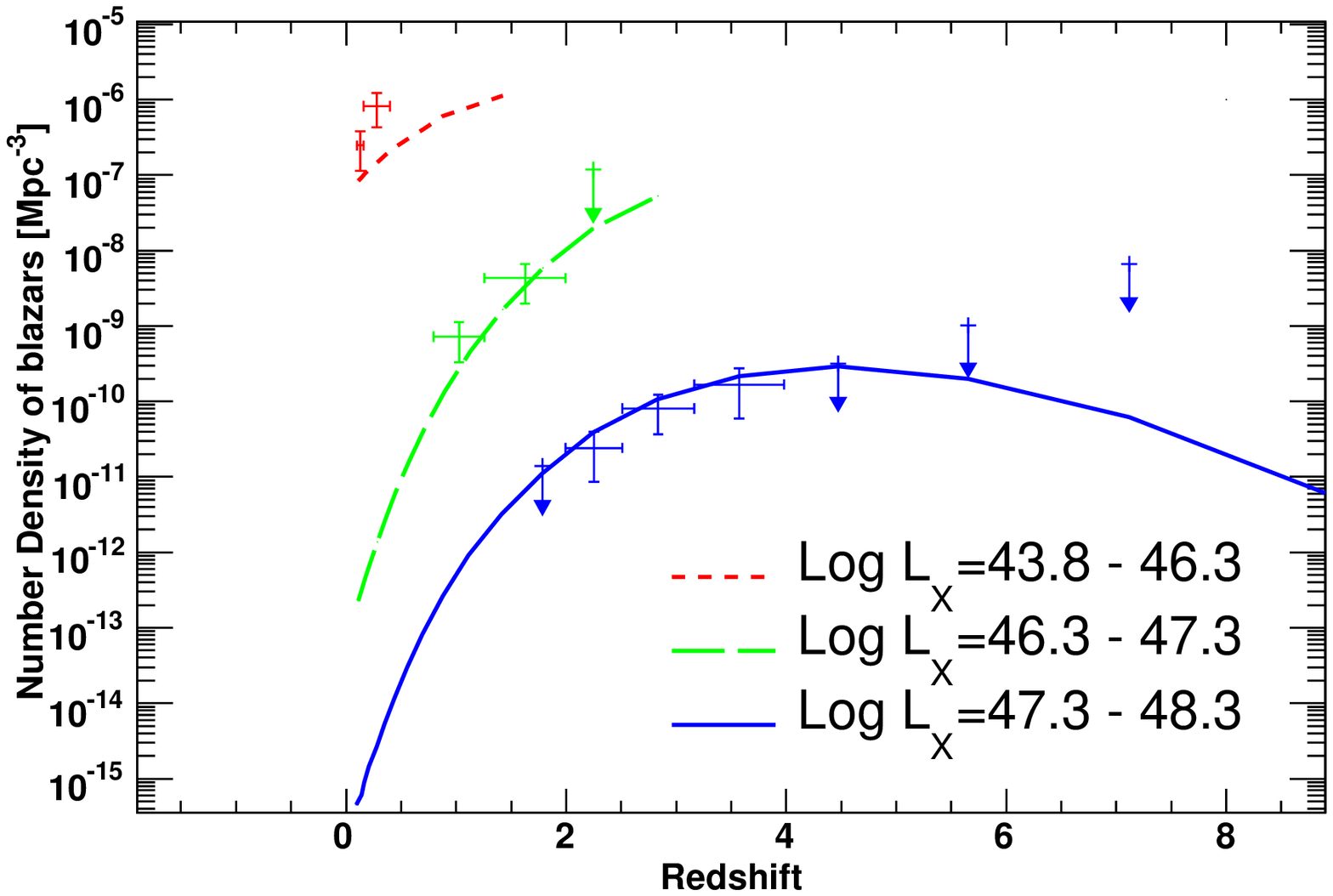} 
  	 \includegraphics[scale=0.43]{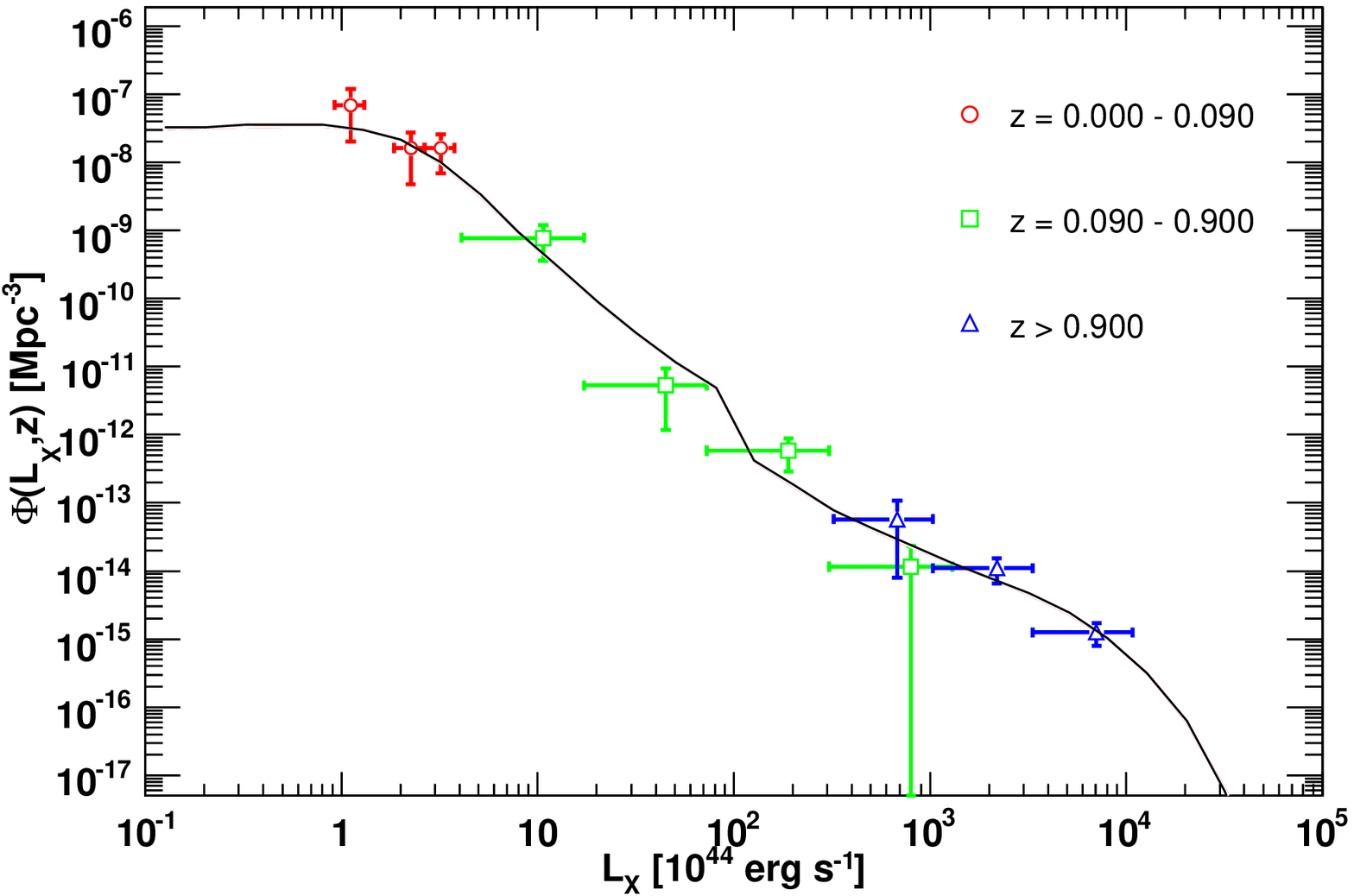}\\
\end{tabular}
  \end{center}
  \caption{Left panel:  Number density  of
 blazars (FSRQs and BL Lacs) as a function of redshift and 
luminosity class. The solid lines represent the best-fit XLF  model
(model 7 in Tab.~\ref{tab:pars}) . The BAT data (points with
errors) were ``deconvolved'' taking into account the BAT sensitivity
(see $\S$~\ref{sec:1pop} for details). 
Right panel:
Luminosity function of the BAT blazars built
using the 1/V$_{MAX}$ method (datapoints) with superimposed
the best fit XLF model (model 7 in Tab.~\ref{tab:pars}).
\label{fig:has_plot_blaz}}
\end{figure*}

A visual representation of the best fit XLF model (double power law 
plus MPLE model) is shown in the left panel
Fig.~\ref{fig:has_plot_blaz} which reports 
the volume density of blazar  as a function
luminosity class and redshift. The datapoints are the ``deconvolved''
BAT observed data, that is the number (or density) of blazars
which an instrument with optimum sensitivity would see.
In order to deconvolve the BAT data, we computed for each bin
of redshift and luminosity, the ratio between the integrals 
of $\Phi(L_X,z)$ and $\lambda(L_X,z)$ (see Eqs.~\ref{eq:phi} and \ref{eq:lambda} for a definition of both). This gives  a correction factor which allows
 to deconvolve the BAT data.
Also note, that given the sparseness of the BAT data, the correction factor
is sometimes averaged over large bins of redshift and luminosity where the
XLF is strongly varying, thus it might be somewhat uncertain. Nevertheless,
Fig.~\ref{fig:has_plot_blaz} highlights that BAT is sampling with good accuracy
the redshift peak of some of the most  luminous objects in the Universe.
From the same figure it is clear that the density of very luminous
blazars (Log L$_X>10^{47}$\,erg s$^{-1}$) peaks at large redshift
and precisely at z=4.3$\pm0.5$. This is much
 larger than the value of $\sim$1.9
derived (or assumed) for X--ray and Optical surveys
 \cite[see e.g.][]{ueda03,hasinger05,bongiorno07,silverman08}.
The likely reason of this difference will be addressed in details in
 $\S$~\ref{sec:highz}. The right panel of Fig.~\ref{fig:has_plot_blaz}
shows the non-parametric blazar XLF built using the 
1/V$_{MAX}$ method along with the best-fit analytical XLF model (model 7).
It is apparent the good agreement between the two representations.

\subsection{Two populations: FSRQs and BL Lacs}

Previous works \citep[e.g.][]{wolter91,rector00,wolter01,beckmann03,padovani07}
have reported evidence about the different evolutionary behaviors of 
FSRQs and BL Lacs. The  $V/V_{\rm MAX}$ test reported in $\S$~\ref{sec:vvm}
showed that also in our sample the two classes of objects might evolve
differently. In the next sections we test this hypothesis.

\subsubsection{FSRQs}
\label{sec:fsrq}

\begin{figure*}[ht!]
  \begin{center}
  \begin{tabular}{cc}
    \includegraphics[scale=0.43]{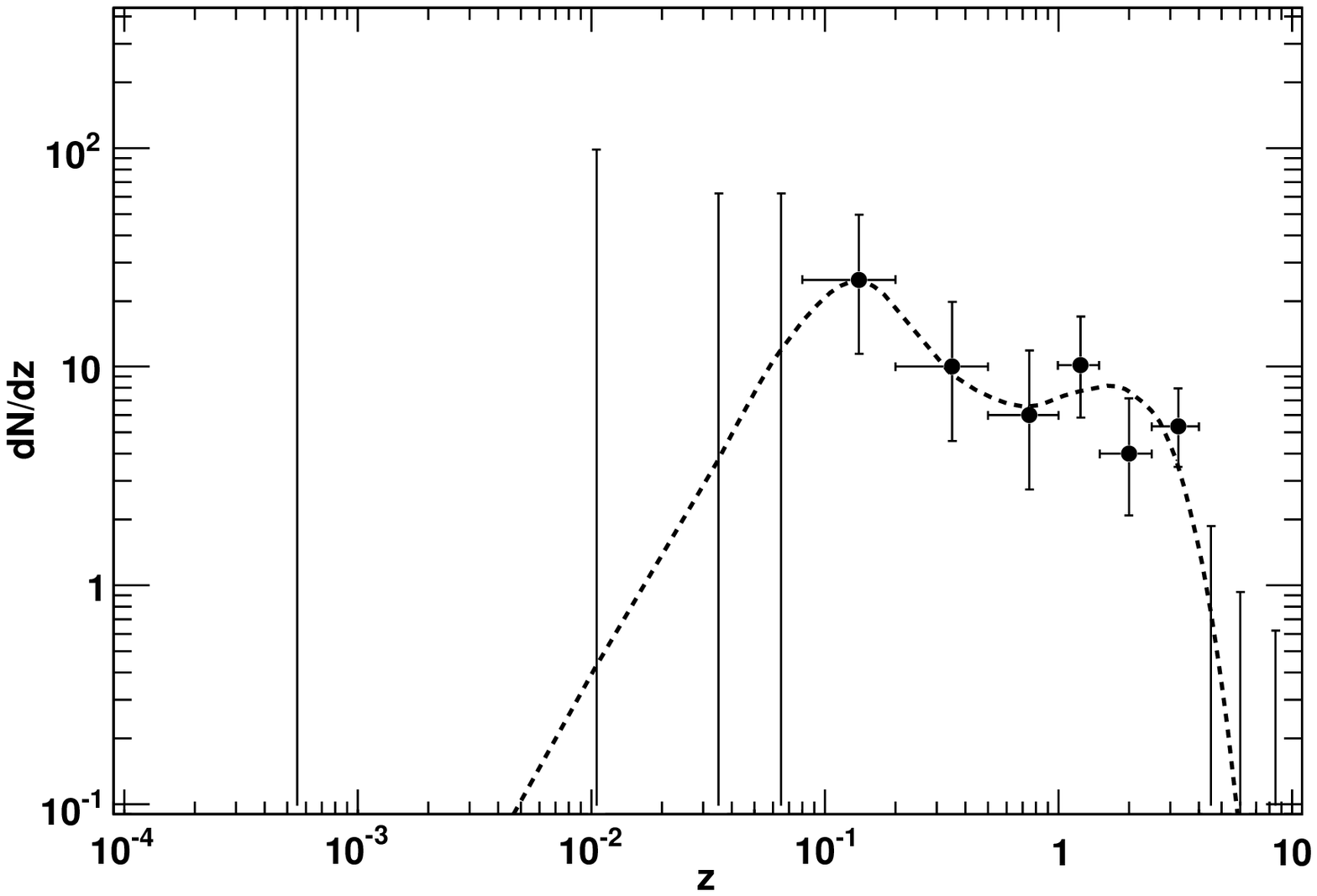} 
  	 \includegraphics[scale=0.43]{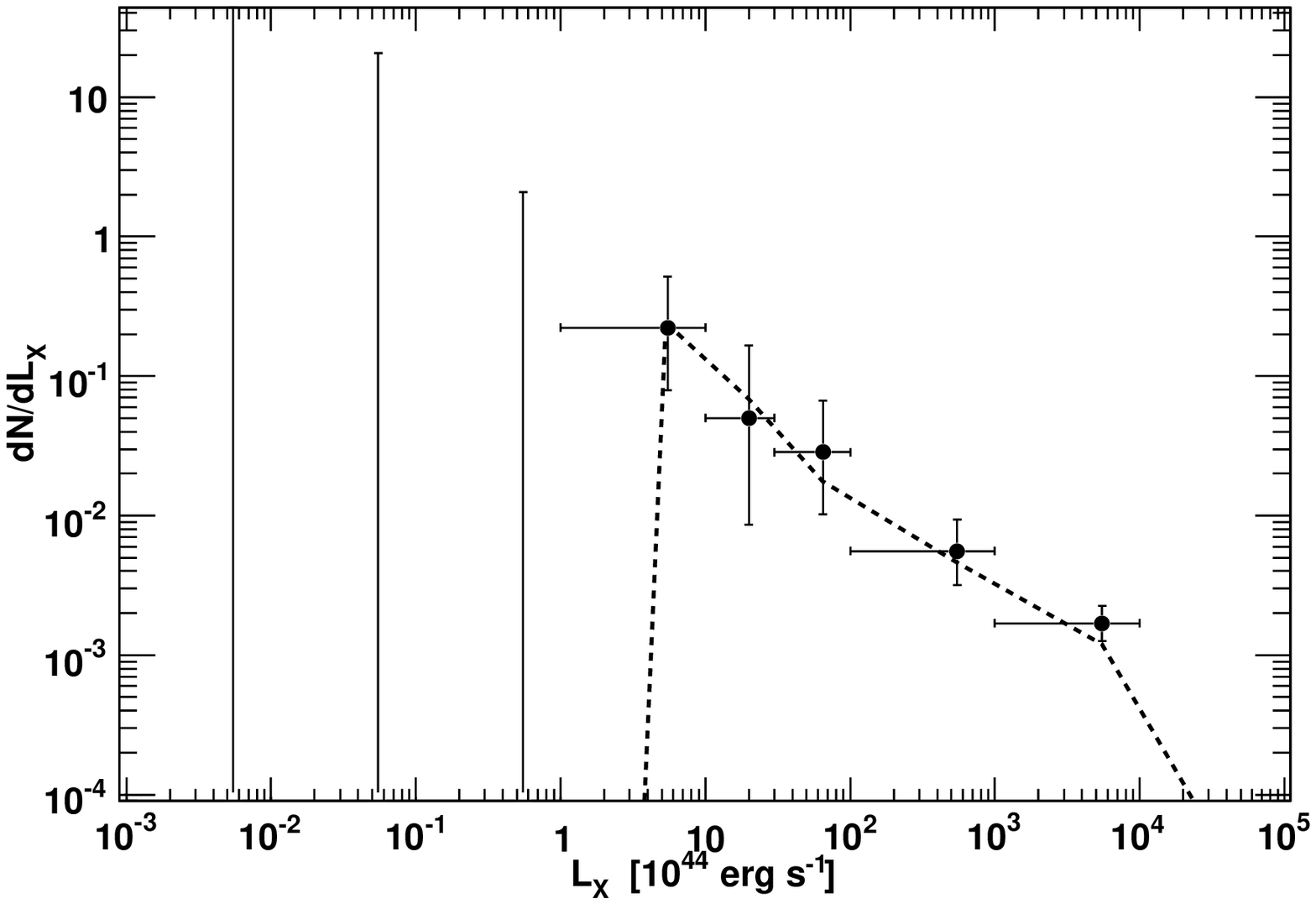}\\
\end{tabular}
  \end{center}
  \caption{Redshift (left) and luminosity (right) distribution of the 
BAT FSRQs. Error bars were computed taking into account the Poisson
error \citep{gehrels85}. For both cases, the dashed line represents
the best fit XLF model (model 10 in Tab.~\ref{tab:pars})
convolved with the BAT sky coverage.
 \label{fig:fsrq_dblpow}}
\end{figure*}

\begin{figure*}[ht!]
  \begin{center}
  \begin{tabular}{cc}
    \includegraphics[scale=0.43]{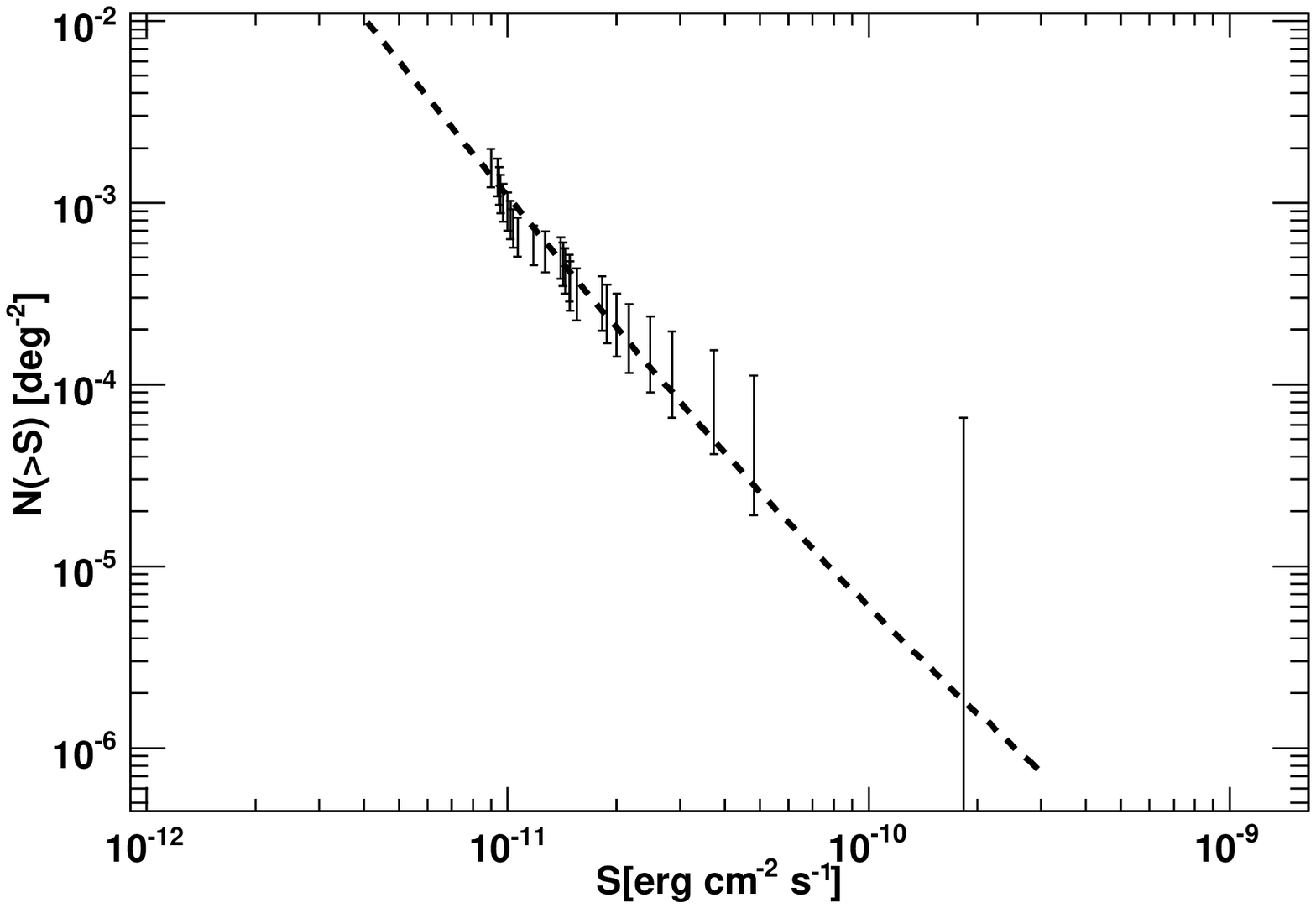}
  	 \includegraphics[scale=0.43]{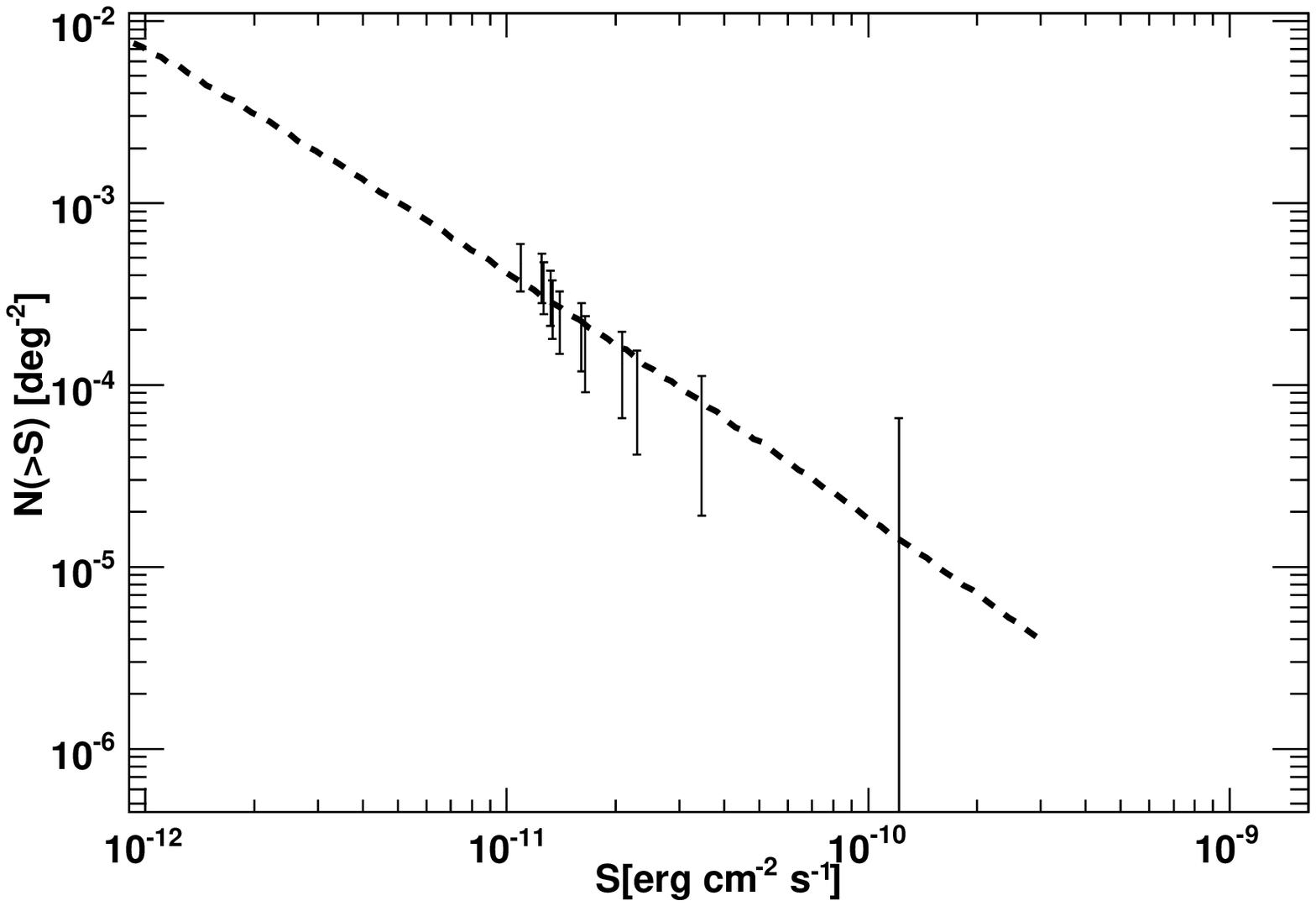}\\
\end{tabular}
  \end{center}
  \caption{Cumulative log $N$ - log $S$ distributions 
for the BAT FSRQs (left) and 
for the BAT BL Lacs (right). The dashed lines are the predictions from
the best fit XLF models (model 10 for FSRQs and model 11 for BL Lacs). 
The different evolution of the two source populations is apparent in the 
different slopes of the two log $N$ - log $S$ distributions (steeper
for FSRQs and flatter for BL Lacs).
 \label{fig:fsrq_logn}}
\end{figure*}

We applied the two best fit models of the previous section (MPLE coupled
to a single and double power-law local XLF respectively)
to the FSRQ class. The best-fit parameters are reported in Tab.~\ref{tab:pars}.
We note that both XLF models produce essentially the same result.
When the local XLF is modeled as a double power-law model, the
faint-end slope $\gamma_1$ is required to be largely negative ($<$-50) and 
the break luminosity $L_*$ coincides with the minimum observed luminosity
of FSRQs in the BAT sample. Under this conditions, the double power-law
model reduces to a single power-law distribution with a sharp cut-off
at L$_X<2\times10^{44}$\,erg s$^{-1}$.
Figures~\ref{fig:fsrq_dblpow} and \ref{fig:fsrq_logn} (right panel)
 show how well the best fit XLF models (model 9 and 10 in Tab.~\ref{tab:pars})  
reproduce the observed distributions (in redshift, luminosity 
and source counts).

\begin{figure*}[ht!]
  \begin{center}
  \begin{tabular}{cc}
    \includegraphics[scale=0.43]{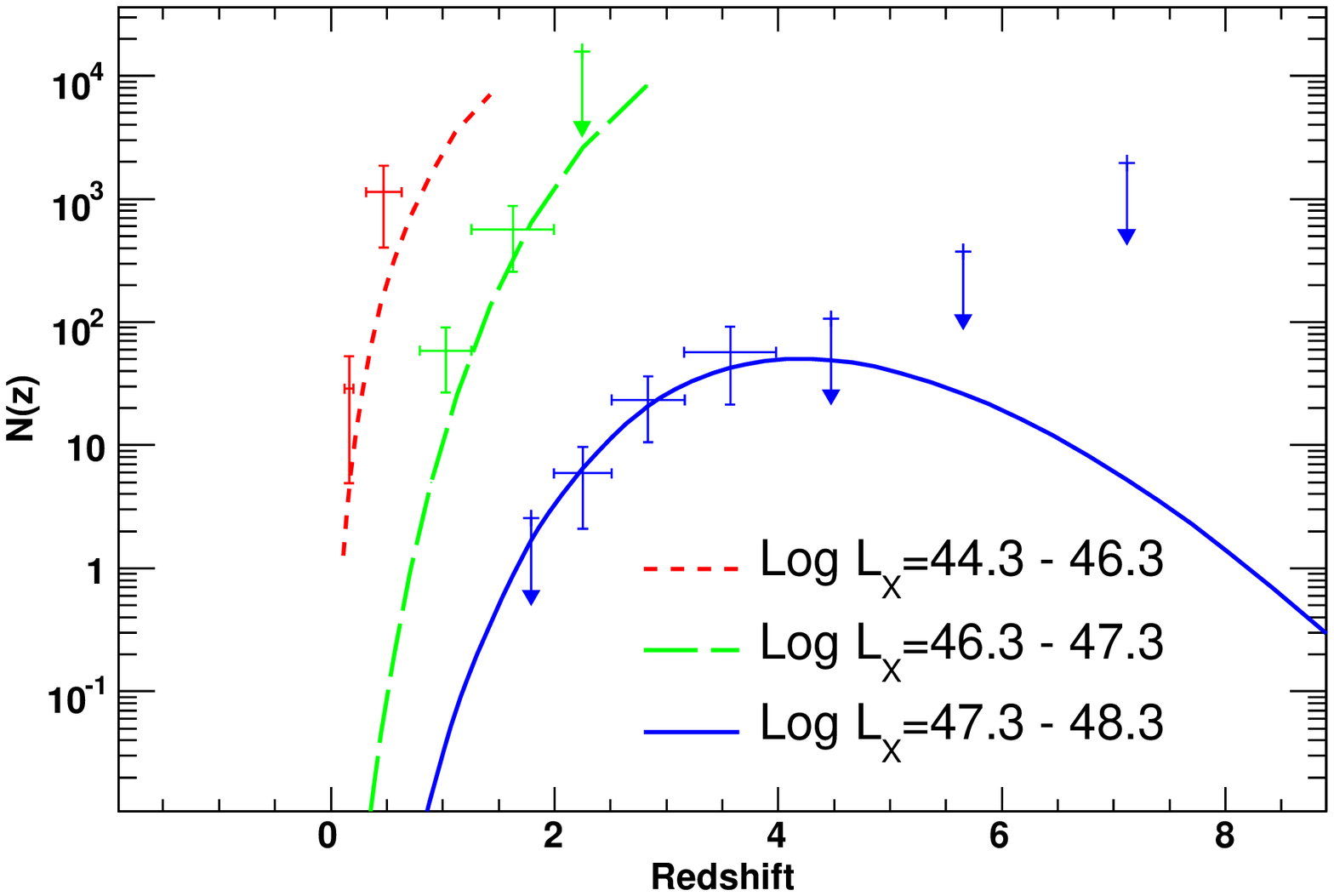} 
  	 \includegraphics[scale=0.43]{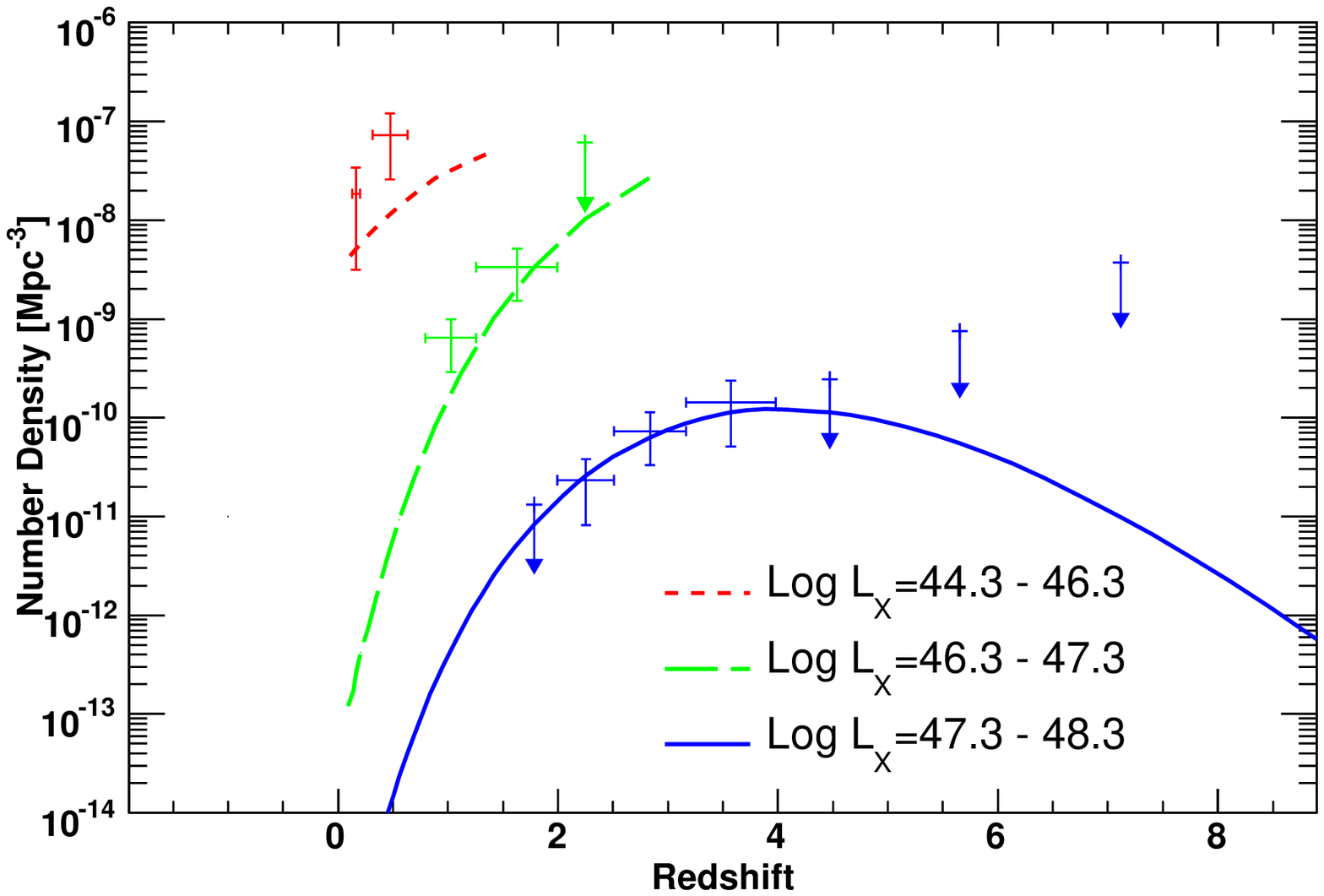}\\
\end{tabular}
  \end{center}
  \caption{All-sky Number (left) and number density (right) of
 FSRQs as a function of redshift and 
luminosity class. The solid line is the best-fit XLF model
(model 10 in Tab.~\ref{tab:pars}). The BAT data (points with
errors) were ``deconvolved'' taking into account the BAT sensitivity.
 \label{fig:has_plot}}
\end{figure*}

Fig.~\ref{fig:has_plot} shows the number, and its
volume density, of FSRQs
 in the Universe for different luminosity classes as derived from 
the best fit.  The BAT data were ``deconvolved'' with
the method outlined in $\S$~\ref{sec:1pop}.
It is clear that BAT is very effective in constraining the density
of FSRQs at high luminosity and large redshifts. The same figure shows
that the cut-off in the evolution is, at least for luminosities
larger than $10^{47}$\,erg s$^{-1}$, well constrained.

\subsubsection{BL Lacs}
\label{sec:bllac}

\begin{figure*}[ht!]
  \begin{center}
  \begin{tabular}{cc}
    \includegraphics[scale=0.43]{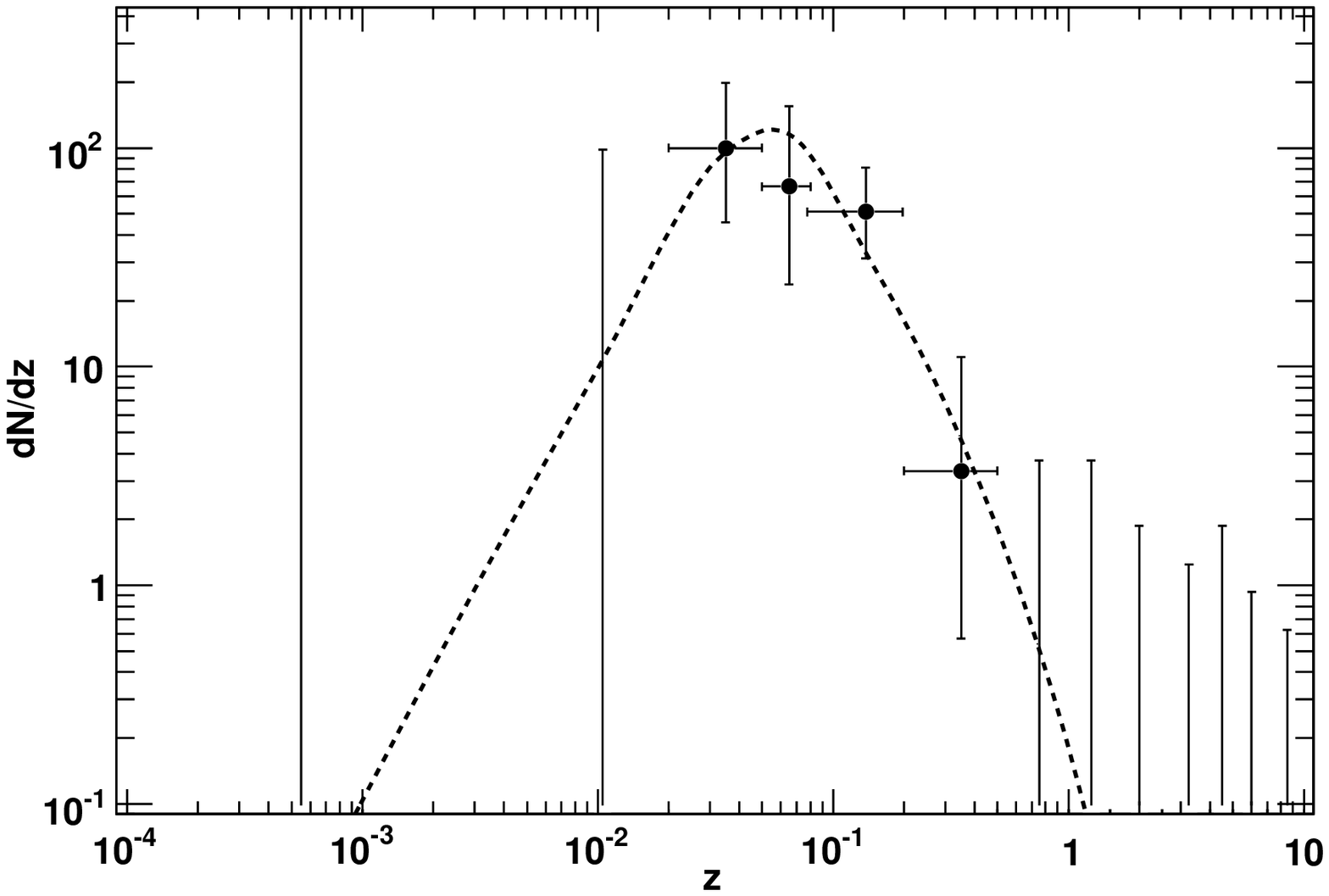} 
  	 \includegraphics[scale=0.43]{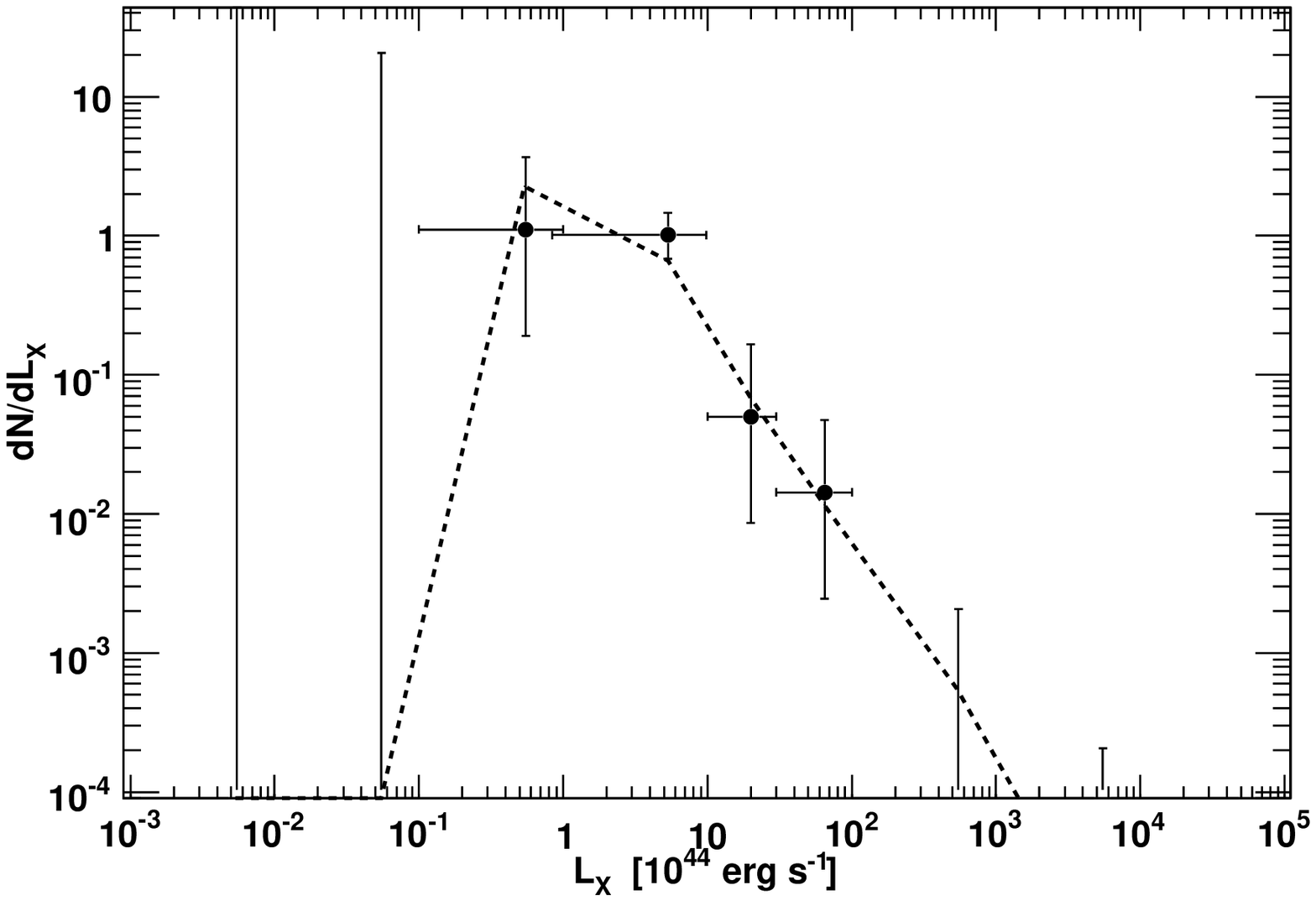}\\
\end{tabular}
  \end{center}
  \caption{Redshift (left) and luminosity (right) distribution of the 
BAT BL Lacs. Error bars were computed taking into account the Poisson
error \citep{gehrels85}. For both cases, the dashed line represents
the PLE XLF (single power law model, see model 11 in Tab.~\ref{tab:pars}) 
convolved with the BAT sky coverage.
 \label{fig:bllac_pde}}
\end{figure*}

Given the small number of BL Lac objects (12) and the relatively low
redshift range that they span ($0.01<$z$<1.0$) we cannot use complex
evolutionary models. We thus tried to fit a simple PLE
model to the data. We obtain an excellent fit which implies mild negative
evolution (albeit with large errors).
Indeed, the best-fit value of the evolutionary parameter is
-0.79$\pm2.43$.
As the results reported in
Tab.~\ref{tab:pars} show, fixing the evolution parameter at zero produces
an equally good fit (see model 11 and 12). 
Fig.~\ref{fig:bllac_pde} shows that the best-fit XLF models (model  11)
reproduces accurately the observed distributions in redshift and luminosity.
The best-fit XLF predicts the cumulative source count distribution always
within 1\,$\sigma$ (see right panel of Fig.~\ref{fig:fsrq_logn}).
We also tried to use a double power-law model for the local XLF (see model 13).
While this model reproduces the BAT data accurately, most of
its parameters are poorly constrained. However, we note that the best-fit
parameters are in good agreement with the values derived for the whole
blazars sample (see model 7) and that the evolution, although still consistent
with zero, became positive.
Thus, we believe that given the small number of BL Lac objects 
it is currently impossible to constrain the sign of the evolution 
(i.e. positive or negative evolution).

\section{Implication for the Cosmic X--ray Background}
\label{sec:cxb}
\begin{figure}[ht!]
  \begin{center}
  	 \includegraphics[scale=0.9]{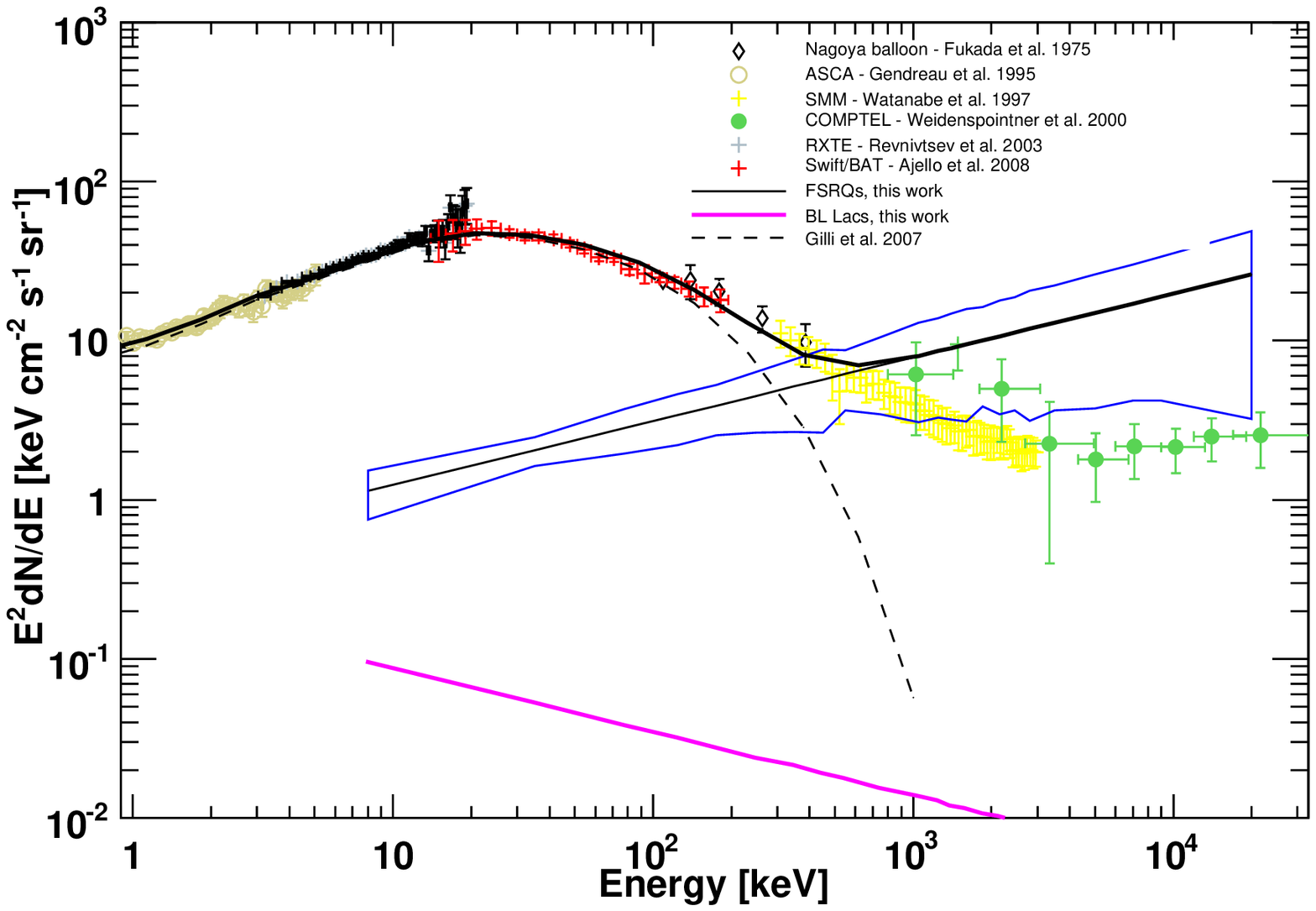}
  \end{center}
  \caption{Spectrum of the CXB and contribution of the FSRQs (blue region).
The data points are different measurements of the diffuse background as indicated
in the label \citep[][]{fukada75,gendreau95,watanabe97,
weidenspointner00,revnivtsev03,ajello08c}.
The dashed
line is the total contribution of Seyfert-like AGNs computed
with the model of \cite{gilli07} arbitrarily multiplied by 1.1 to fit
the CXB emission at 30\,keV. The solid
line is the sum of the Seyfert-like and FSRQs. 
The spectrum of FSRQs has been modeled as a power-law with a mean
photon index of 1.6.  The blue region
represents the range of values obtained from the Monte
Carlo realizations of best-fit parameter ranges.
The magenta solid line represents the contribution of BL Lac objects
whose uncertainty is not plotted for clarity, but is, due to the low
number of objects, $>30$\,\% at any energy.
\label{fig:cxb}}
\end{figure}

Eq.~\ref{eq:cxb} can be used to estimate the contribution of blazars
to the CXB outside the energy band of this survey. In this case,
$F_{X}(L_{X},z) =F_{X}(L_{X},z,E) $ becomes a function of energy
and represents the spectral energy distribution (SED) of the blazar source
class. To model the contribution of FSRQs and BL Lacs we used the
best-fit XLF models derived in the previous sections.
These are model 9 and model 11, in Tab.~\ref{tab:pars}, 
respectively for FSRQs and BL Lacs.
We also remark that the results presented here do not change
if other valid XLF models (e.g. model 10 for FSRQs and model 12 and 13
for BL Lacs) are used. 
In order to compute correctly the uncertainties we employed
a Monte Carlo simulation. We generated a large number ($>200$) of luminosity
functions starting from randomly sampled best-fit parameters 
drawn from the  covariance matrix derived during the fit stage. 
Moreover, for each randomly generated XLF, a random photon index has been
drawn from the index distributions of the given class (i.e. FSRQ or BL Lac).
We then computed the contribution to the CXB for each of these luminosity
functions and computed the 1\,$\sigma$ deviation, around the mean value,
at given fixed energies. As a first test, we model the SED using a simple
power-law model.
Fig.~\ref{fig:cxb} shows the contributions of
FSRQs and BL Lacs (evolving as different populations) to the CXB.
It is apparent that while the contribution of BL Lac objects appears
negligible in this hypothesis (i.e. no BL Lac evolution),
the contribution of FSRQs is substantial in hard X-rays. From our luminosity
function we derive that virtually 100\,\% of the CXB for energies $>500$\,keV
is produced by FSRQs. We also use the synthesis model
of \cite{gilli07}\footnotemark{}
\footnotetext{A web interface to the model of \cite{gilli07}
 is available at http://www.bo.astro.it/$\sim$gilli/xrb.html.}
 to take into account the contribution of Seyferts to the CXB. 
We arbitrarily renormalize the \cite{gilli07} model
by 1.1. This is justified by the fact that this synthesis 
model is tuned to reproduce the CXB as measured by HEAO-1 \citep{gruber99}
which is  known to underestimate the CXB emission of $\sim$10\,\%
at 30\,keV \citep[][and references therein]{ajello08c}. 
It is apparent from Fig.~\ref{fig:cxb} that
summing the contribution of blazars to the one of Seyferts achieves
a good estimate of the intensity of the CXB emission up to the MeV
range.

\begin{figure}[ht!]
  \begin{center}
  	 \includegraphics[scale=0.9]{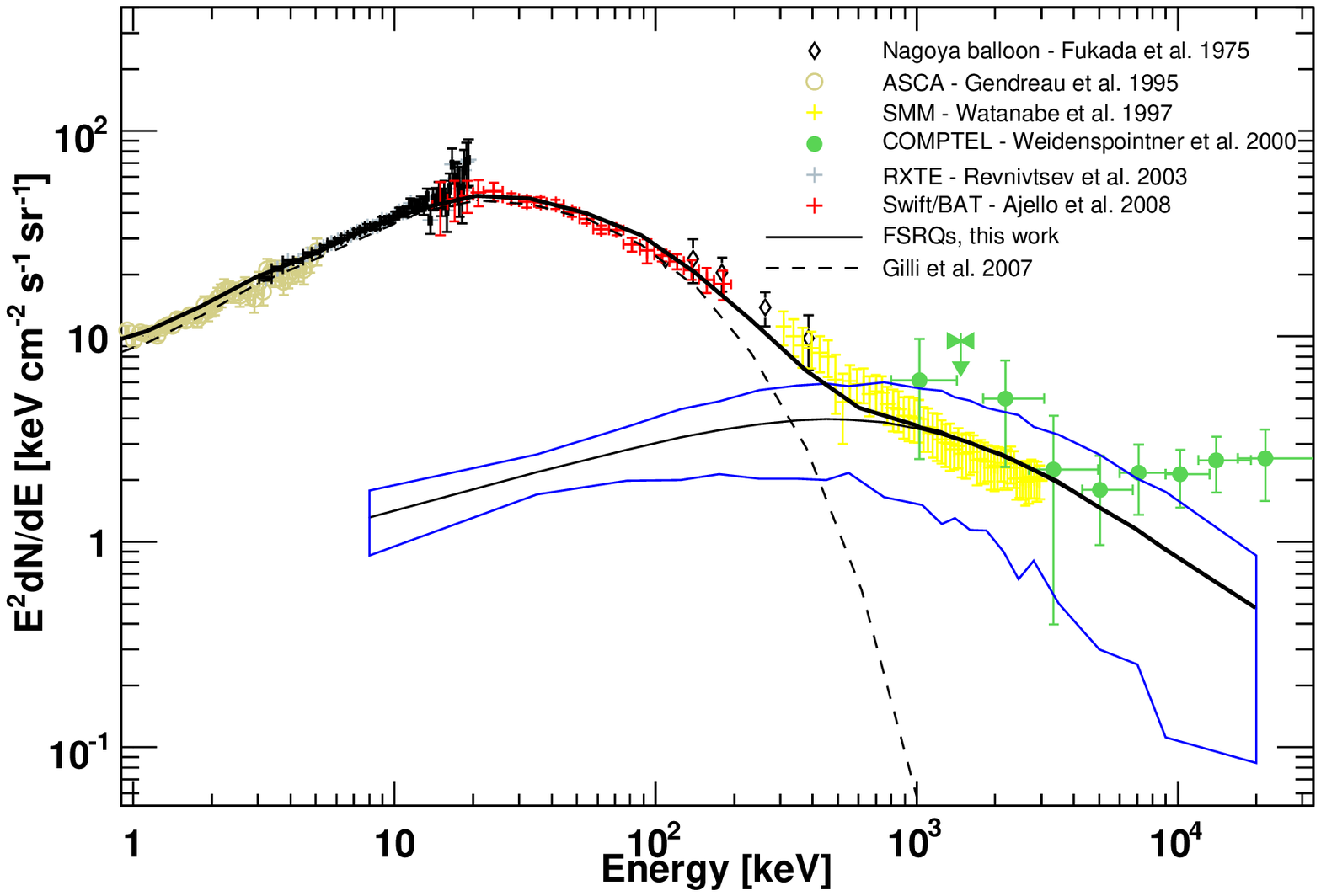}
  \end{center}
  \caption{Contribution of FSRQs (blue region) to the CXB. 
The data are the same as in Fig.~\ref{fig:cxb}, but in this case
the SED of  the  FSRQs has been modeled with
a double power-law function. The IC peak is located in the
$\sim$MeV region. The contribution of BL Lacs is the same as in 
Fig.~\ref{fig:cxb} and is not drawn here for clarity. The blue region
represents the range of values obtained from the Monte
Carlo realizations of best-fit parameter ranges.
\label{fig:cxb_double}}
\end{figure}

Modeling the SED with a simple power-law model is a straightforward
and robust hypothesis, but it remains accurate only for 
extrapolations close to the original 15-55\,keV band. 
Indeed, Fig.~\ref{fig:cxb}
shows that at 10\,MeV the contribution of FSRQs, computed in this way,
overestimates the 
diffuse background by an order of magnitude.
The $\nu F\nu$ spectrum of FSRQs 
exhibits an IC peak which is located somewhere in the MeV--GeV band.
While detailed modeling of the SEDs of each of the BAT blazars
is outside the scope of this paper, we note that some of the BAT FSRQs
were analyzed by several authors
\citep{zhang05,sambruna06,sambruna07,tavecchio07,watanabe08}.
In all cases, the authors find that the IC peak is located in the MeV
band. We thus represent the SED with an empirical double power-law
model of the type: 
$dN/dE \propto [ (E/E_b)^{-\Gamma_1} + (E/E_b)^{-\Gamma_2}]^{-1} $,
where $\Gamma_1$ and $\Gamma_2$ are the photon indices (1.6 and 2.5 
respectively) before and  after the energy break $E_b$.
For the energy break, $E_b$, we chose a value of 1\,MeV motivated
by the observations reported above. However, we note that the smooth
and large curvature of the model we employ, makes it virtually insensitive
to the exact value of $E_b$ if this is within 1 order of magnitude.
Fig.~\ref{fig:cxb_double} shows  the contribution of
FSRQs assuming that their IC peak is located in the MeV band.
We find that in this case  FSRQs
account for the entire CXB emission up to 10\,MeV.
While there is basically no difference with respect to the single power-law
case below 500\,keV, the curvature of the IC peak makes the contribution
of FSRQs to the CXB slightly smaller around 1\,MeV.
We also note that moving the IC peak beyond 10\,MeV produces
a negligible curvature in the FSRQ integral emission and thus
this case is well represented by the single power-law model.

Thus, the two analyses shown here cover well the case in which
the IC peak is either located at MeV or at GeV energies (double
and single power-law model respectively).
We must therefore conclude that the contribution of FSRQs
to the diffuse emission is relevant and likely accounts for a substantial
fraction (potentially $\sim$100\,\%)
of the CXB around 1\,MeV. Interpreting the CXB as a strong constraint,
we derive that  the population of FSRQ sampled by BAT
must have the IC peak located in the MeV band in order not to overproduce
the diffuse background at $\sim$10\,MeV.   
\cite{bhattacharya08} recently reported
for the FSRQs detected by EGRET a mean photon index of 2.30$\pm0.19$.
Since FSRQs have a mean photon index of 1.6 in BAT, this implies already
that the IC peak is located in between the BAT and EGRET energy bands.
However, as we noted already in $\S$~\ref{sec:sample} only 9 FSRQs
are in common between the EGRET and the BAT samples and this might
imply that the other FSRQs detected by BAT have an IC peak at even lower energies.
As it will discussed in $\S$~\ref{sec:disc_cxb} 
{\it Fermi}-LAT will certainly clarify   this scenario.
Indeed, very recently, \cite{lat_agn}, discussing the results
of the first three months of observations of Fermi-LAT, showed
that FSRQs are detected by Fermi with a mean photon index of 2.4.
Thus, FSRQs have soft spectra (photon index $>2.0$) in the GeV band
while they have hard spectra in the hard X--ray band. This confirms 
that their IC peak is located  between the two bands.
We note that the shape of the integrated emission of FSRQs 
is similar to the empirically-derived one
of \cite{comastri06}.

We also found that the contribution of BL Lacs is very small 
if they are a non-evolving (or mildly evolving) population.
 In agreement with \cite{georgakakis04}, \cite{galbiati05} and \cite{giommi06}
we find that the contribution of blazars (FSRQs and BL Lacs) to the 2-10\,keV 
CXB is $\sim$10\,\%.

\section{The high-redshift non-thermal Universe}
\label{sec:highz}

The fact that the shape of the blazar luminosity function and 
its evolution are in agreement with those of X--ray selected
AGNs suggests the presence of a link between accretion and jet activity
\citep[e.g.][]{merloni03}. In other words, it seems that the most luminous AGNs
(which in turn are the most luminous QSOs)
harbor the most powerful blazars. This scenario takes place mostly in the
very high-redshift Universe where, thanks to the abundance of dust and gas,
efficient accretion led to the build-up of massive QSOs.
However, their space density quickly decreases, with cosmic time, leaving the 
room for the bulk of low-luminosity QSOs. This ``anti-hierarchical''
scenario, where larger structures come first, was
also named ``cosmological downsizing'' \citep[e.g.][and 
references therein]{cowie99,hasinger05} and  constitutes a unique phenomenon
which is not predicted in most of the semi-analytic models based on
Cold Dark Matter structure formation \cite[][]{kauffmann00,wyithe03}.
The late evolution of low-luminosity AGNs coincides well with the
peak of the star formation in the Universe \citep[e.g.][]{hopkins06}
highlighting once more the interconnection between the host and its
nucleus. A mechanism of accretion with different efficiencies,
as a function of cosmic time,
has been invoked to explain the anti-hierarchical growth of AGNs 
\cite[e.g.][]{merloni04}.
 However, the fate of the very first and luminous quasars 
(i.e. the apparent disappearance of quasar activity in massive galaxies
at late times) remains still
unknown and there are doubts whether these objects can form at all
in a $\Lambda$CDM Universe \citep{springel05}.

\begin{figure}[ht!]
  \begin{center}
  	 \includegraphics[scale=0.9]{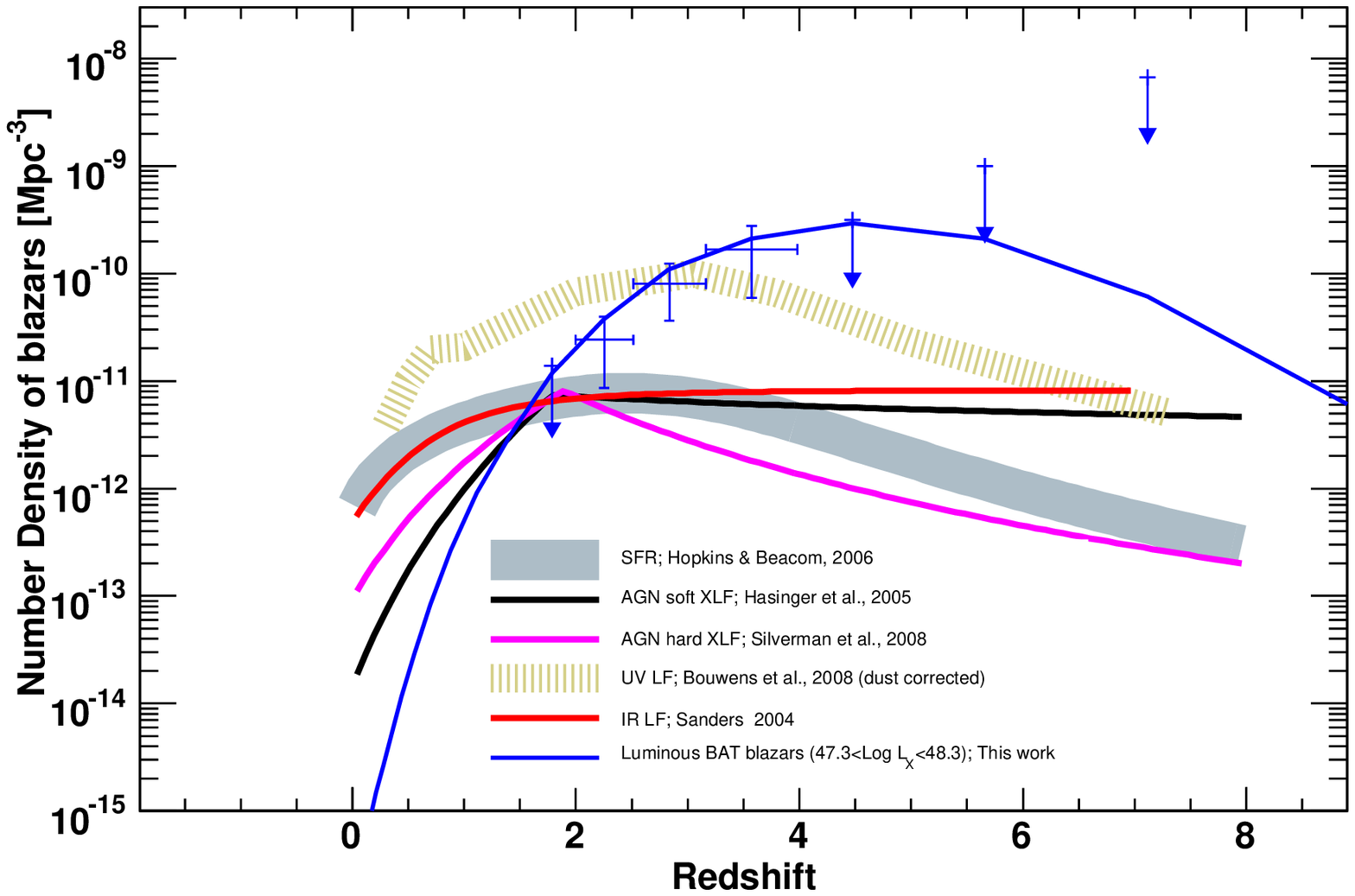}
  \end{center}
  \caption{Density of the most luminous BAT blazars 
(47.3$<$Log L$_X$ $<$48.3, blue datapoints and line)
 as a function of redshift compared to:  densities of AGNs from X--ray surveys 
\citep{hasinger05,silverman08}, star
formation rate \citep{hopkins06} and densities  of UV and IR galaxies
\citep[][respectively]{bouwens08,sanders04}. The different curves
were rescaled arbitrarily to match the evolution of the BAT blazars.
Thus, the comparison  involves only the shape.
\label{fig:highz}}
\end{figure}

In Fig.~\ref{fig:highz} we compare the shape of the evolution of the most
luminous BAT blazars with: 1) the evolution of luminous X--ray selected AGNs
\citep{hasinger05,silverman08}, 2) the star formation history (SFR)
of the Universe \citep{hopkins06}, 3) and the evolution of UV and IR
galaxies \citep[][respectively]{bouwens08,sanders04}.
The most direct comparison is clearly with AGNs. We note that the shape
of the evolution is very similar, but that the cut-off in the AGNs growth
is at lower redshifts (z$<$2) with respect to the cut-off of BAT blazars.
This means that the peak of the evolution of the BAT objects is at much
earlier times in the history of the Universe than the peak of the most
luminous AGNs detected in the deepest X--ray surveys.
The star formation history of the Universe shows a similar trend
and a peak around z=$\sim$2 as for normal AGNs and this has been interpreted
as the evidence of  the strong link between AGNs and the star formation
in its host galaxy.
Assuming that the activity of the BAT blazars is powered by accretion
onto super-massive black hole, this implies that the doppler-boosting
allows BAT to detect a class of objects (in luminosity) which
escaped even the deepest X-ray surveys.

The only objects which show a peak in the evolution at redshift $>$2 are
bright star-forming galaxies detected up to very high redshifts in the
GOODS fields  \citep[][and references therein]{wall08,bouwens08}.
The rapid brightening of galaxies within the first two billion years
is tightly connected to the assembly of large dark matter haloes 
\citep[e.g.][]{wang08}. An intriguing idea is that
black holes are formed and fueled, and AGN activity is triggered
during major galaxy mergers \citep{kauffmann00,wyithe03,croton06}.
Although this constitutes a plausible fueling mechanism, 
the intermediate-luminosity
AGNs are harbored by galaxies which display little or no merger events
\citep{hasan07}. However, to form a super-massive black hole, a more
violent process such as a major merger event may be required to 
funnel a large amount of gas into the central region of the galaxy.
It seems that the most massive galaxies undergo a major merger
event at earlier times, and within the first 2-4 billion years, 
than less massive galaxies \citep{wang08,stewart08}. If the bulk of the black
hole mass is formed in this way, then it would explain
the lack of growth of powerful AGNs at present times.
The abundance of luminous blazars at large redshift fits well in this
scenario since blazars are found in giant elliptical galaxies which
in turn are supposed to have undergone a major merger event
\citep[][]{toomre72,negroponte83,wang08}.
Thus, we argue that the luminous  blazars can be used as a
tracer of massive galaxies and merging activity in the very early Universe.

\begin{figure}[ht!]
  \begin{center}
  	 \includegraphics[scale=0.9]{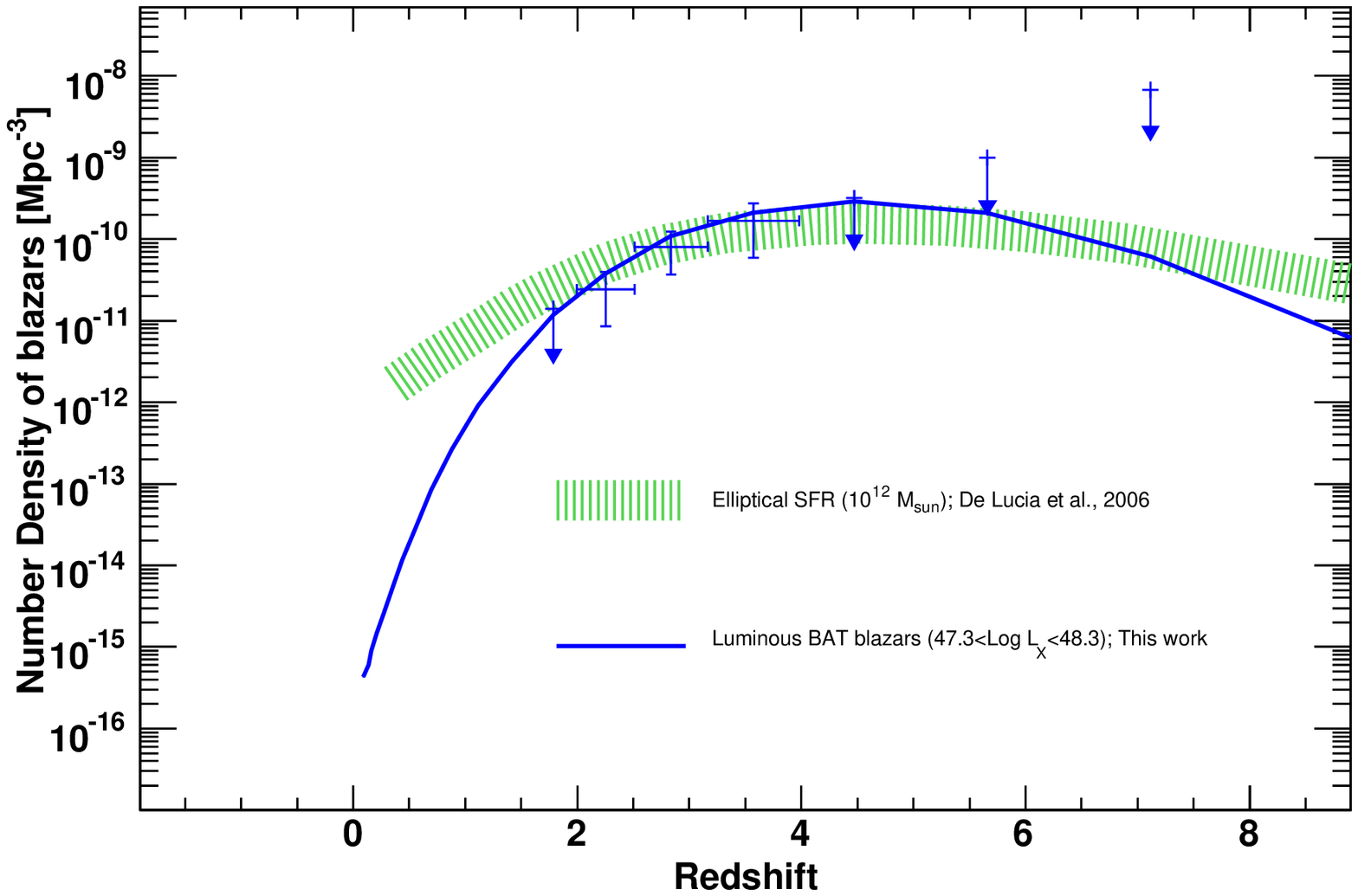}
  \end{center}
  \caption{Density of the most luminous BAT blazars 
(47.3$<$Log L$_X$ $<$48.3, blue datapoints and line) compared to the
prediction of the evolution of massive elliptical galaxies of 
\cite{delucia06}. The curve of \cite{delucia06} was rescaled to
match the BAT luminosity function. In order to consider also the contribution
of misaligned jets, the space density of blazars (y-axis) needs to
be multiplied by a factor in the 200-800 range (see $\S$~\ref{sec:highz}
for details).
\label{fig:ellipticals}}
\end{figure}

Understanding the formation of massive galaxies is an important 
astrophysical issue because as much as 50\,\% of the stellar mass
in the local Universe appears to be in early-type systems \citep{bell03}.
The most massive galaxies (i.e. M$_{*}>10^{11.5}$\,M$_{\odot}$) appear
to be already in place at z$\sim$2 suggesting that they formed in the very
early Universe \cite[see][for a review]{conselice08}.
Unfortunately present surveys, both in the optical and in X--rays,
are not sensitive enough to make a statistical census beyond redshift 2.
Here, BAT and more  in general blazar surveys, can play an important role.
In  Fig.~\ref{fig:ellipticals} we compare the redshift evolution of
the luminous BAT blazars with the prediction of the evolution of massive 
elliptical galaxies as determined, using simulations,
 by \cite{delucia06}. The similarity
between the two curves is apparent and reinforces our idea that
blazars can be used to study the formation of massive systems in the
early Universe.

The number density of massive ellipticals hosting active blazars is larger
than the one reported in Fig.~\ref{fig:ellipticals} since we 
consider only those blazars pointing at us. Since there should be
no difference between the population of 
blazars pointing at us and those pointing
in all the other directions we might try to estimate their total number.
The number of  misaligned blazars, can be estimated
as: $N_{TOT}\approx 2 \Gamma^2$, where $\Gamma$ is the bulk Lorentz factor
\citep[e.g.][]{scheuer79}.
Assuming a standard value of $\Gamma=10$ (20) 
\citep[see e.g.][]{urry95,sambruna07}
leads to a space density of ellipticals, hosting active blazars,
which is 200 (800) times larger than
that one shown in Fig.~\ref{fig:ellipticals} and thus of the order
of 2.0--8.0$\times10^{-8}$\,Mpc$^{-3}$ at a redshift of $\sim$4.

Moreover, the excellent agreement between the evolution
of massive galaxies and blazars  suggests that there should be 
no intrinsic difference, at least in the early Universe,
between ellipticals and ellipticals hosting active blazars.
This means that the fraction of ellipticals which host active blazars
is not changing dramatically as a function of cosmic time.

Recently, 
\cite{sikora07} showed that on a radio-loudness/Eddington-ratio diagram
ellipticals and spirals
form two distinct and well-separated parallel sequences 
\citep[see also][]{wilson95}.
They argue that this different behaviour might be given by the spin.
Indeed, within the hierarchical cosmological framework, the main difference 
in the evolution of giants ellipticals and spirals is that the first 
ones underwent at least one major merger event in the past.
These events can produce a maximally rotating black hole by
coalescence of the two black holes \citep{escala04,escala05,dotti07}
  and by triggering large-gas
accretion events that spin up the hole \citep{barnes96,escala07,bodganovic07}.
It is important to note that 
under the assumption that jets are powered by rotating black holes
via the  \cite{blandford77} mechanism, the efficiency
of the jet production is determined by the black hole spin.
We believe that gathering a large sample of blazars which spans
adequately luminosity and redshift will allow in detail to understand
the role of spin and accretion in triggering jet activity.
In this respect, BAT and, in particular, {\em Fermi} will play an important 
role.


\section{Summary and Conclusions}
\label{sec:discussion}

\subsection{The blazar XLF}
\label{sec:disc_xlf}
We have used a complete sample of blazars detected by BAT
to derive the first luminosity function of blazars in the 
15-55\,keV band.
We have shown using several methods that
BAT blazars are evolving strongly while the Seyfert-like AGNs
detected by BAT are not. The evolution of the blazars is implicit
in the fact that BAT detects
10 objects at redshifts larger than 2.0.
The local luminosity function (e.g. $\Phi(L_X,z=0)$) is compatible
with a double power-law model where the faint-end slope is required
to be flat 
mainly by the absence of low-luminosity BAT blazars.
According to \cite{urry84} the flattening of the local XLF at low luminosities
might be produced by beaming which boosts intrinsically low-luminosity sources
to high luminosities.
The best-fit XLF models imply an evolution in luminosity which is epoch
dependent (e.g. the evolution parameter changes with redshift)  
as was found for other samples of blazars or FSRQs
\citep[e.g.][]{wall08b,padovani07}. In all cases, it appears that 
blazars are evolving strongly up to a redshift cut-off which
is, at least for the most luminous objects, well constrained by
our data.

We find that the strong evolution of blazars is driven by the evolution
of FSRQs which are the only objects detected by BAT at large redshifts.
Our best-fit  XLF, to the sample of BL Lacs, shows that BL Lacs have
a negligible evolution. 
This is found to be in agreement with the results of \cite{rector00},
\cite{padovani07} and of \cite{bhattacharya08}. Given the small number
of objects in our sample we cannot rule out, nor confirm, the claims of 
  negative evolution of BL Lacs \citep[e.g.][]{wolter91,beckmann03}.
These last two samples reach a flux which is lower than the 
current BAT sensitivity, thus with a few more years of exposure
BAT might be able to test this negative-evolution scenario.
Finally, we also remark that 
given the small number of objects (12), the evolution of BL Lacs
is marginally consistent (at 1.5\,$\sigma$) with the evolution
of the FSRQ class.

The log $N$-- log$S$ distribution of 
blazars is steeper than the one of Seyfert-like AGNs. Its slope of
1.9 is larger than the Euclidean value of 1.5 which characterizes
the Seyfert galaxies (see Tab.~\ref{tab:vvmax} for details). We, thus, expect
that the fraction of blazars will steadily increase 
among the total AGN population detected by BAT. We expect
that in a relative short timescale (e.g. a couple of years) and,
depending also on the systematic errors of the BAT survey,
the blazar sample might contain more than 60 objects. This will be very
important as it will allow to improve the results and the prediction of the 
XLF models for both FSRQ and BL Lac objects. As we  have shown, blazars are
extremely rare objects which can be detected in hard X--rays only through
large-area sky surveys. Our best-fit XLF model predicts a flattening
in the log $N$--log $S$ at  fluxes lower than 5$\times 10^{-13}$\,erg cm$^{-2}$
s$^{-1}$.  Thus a mission like EXIST \citep{grindlay05} 
would detect FSRQs with a surface density of $\sim$0.5\,deg$^{-2}$
at fluxes of 5$\times 10^{-13}$\,erg cm$^{-2}$ s$^{-1}$. 
If confirmed that most of the BAT FRSQs are MeV blazars 
(see $\S$~\ref{sec:cxb} and \ref{sec:disc_cxb}), then an optimum band
to select and study them would be the MeV band. A mission like GRIPS
\citep{greiner08} would gather a fairly large  sample (500--1000)
of blazars.

The redshift distribution of the BAT blazars (e.g. Fig.~\ref{fig:2pow})
shows a peak at low redshift and a flat tail extending up to z$\approx4$.
This distribution differs from the redshift distributions
of radio-selected blazars which display a peak at z$\approx$1.0--1.5
\citep[e.g.][]{dunlop90,wall05}. The reasons for this difference
lie in the different selection effects and sensitivity
of these surveys. Since BAT is sensitive only to bright X-ray fluxes
most of the low-luminosity low-redshift sources are currently undetected.
Additionally, the discrepancy in the redshift distributions between
radio and hard X-rays might also be due to the different shapes
of the evolution and of the local luminosity functions in these bands.
This would not be surprising in view of the fact that radio and X-rays
probe different scales in these systems. A larger dataset of 
hard X-ray-selected blazars will allow us to test these hypothesis.

From our XLF we derive that the density of FSRQs at fluxes
of 10$^{-13}$\,erg cm$^{-2}$ s$^{-1}$ is 5.2$^{+12.1}_{-3.7}$\,deg$^{-2}$
which is compatible  with previous estimates in other X--ray bands 
\citep[][]{wolter01,giommi06b}. Within the large uncertainties
of our estimate, we derive that at faint fluxes the density
of FSRQs is not negligible when compared to the total AGN population.
As an example, the density of all AGNs in the XMM-COSMOS field
is  24.0$\pm3$\,deg$^{-2}$ for equivalent fluxes as above \citep{cappelluti07}.
This means that deep X--ray surveys necessarily contain a $\sim$10\,\% 
fraction of blazars.
This seems in agreement with the finding of \cite{dellaceca94}
who report a fraction of radio-loud objects (among X-ray selected AGNs)
of $\sim$10\,\% 
at fluxes of 10$^{-13}$\,erg cm$^{-2}$ s$^{-1}$. Similar fractions
of radio-loud AGNs were
also found by  \cite{hooper96} and \cite{zickgraf03}.

The main point of concern is however the selection of  very absorbed,
Compton-thick, AGNs in  deep X--ray fields. Generally, given the lack
of sufficient signal, the source intrinsic absorption is  
derived by an hardness-ratio analysis \citep[e.g.][]{fiore08,brusa08}.
The current 'paradigm' is that exceptionally hard X--ray spectra
(photon indices of 1.0-1.5)
are likely produced by strong absorption. Our analysis shows that 
FSRQs have intrinsically hard X--ray spectra with photon indices
sometimes lower than  1.6 \citep[see Fig.~\ref{fig:phidx}, 
but also ][]{tavecchio07,watanabe08}. We believe that, if radio
properties are not properly taken into account, selection of absorbed sources
based solely on hardness ratios  will produce a sample which can be
contaminated by a substantial fraction of FSRQs.

\subsection{The Cosmic X-ray Background}
\label{sec:disc_cxb}
The origin of the MeV background has been a long-standing issue in astrophysics.
Several astrophysical processes have been put forward to explain it.
Among them, dark matter annihilation \citep{ahn05b}, nuclear decays 
from Type Ia supernovae \citep{clayton75} and non-thermal emission from Seyfert
galaxies \citep{inoue08} were the most important ones.
We used our best-fit XLF model to make a prediction of the integrated
emission due to FSRQs and derived that FSRQs account for most
of the diffuse background emission for energies $>$500\,keV.
Moreover,
assuming that most of the FSRQs have an IC peak
in the MeV band, as some of the BAT blazars 
\citep[e.g.][]{zhang05,sambruna07,tavecchio07,watanabe08}, we showed
that the sum of the contribution of emission-line AGNs \citep{gilli07} 
and blazars reproduces well the CXB emission from 1\,keV to 10\,MeV.
Our prediction of the contribution of blazars to the CXB is well
in agreement, in the 2-10\,keV band, with the findings of several
authors
\cite[e.g.][]{galbiati05,comastri06,giommi06}.

Recently \cite{inoue08} proposed that a population of non-thermal 
electrons present in the hot AGN coronae can account for a substantial
part of the MeV background. Our finding shows that the non-thermal
contribution from AGN coronae should be small as most
of the diffuse background emission is accounted for by blazars.
In a more recent work, \cite{inoue08b} derived the luminosity
function of EGRET blazars taking into account the blazar sequence; this
is then used  to compute the contribution of blazars to the diffuse background.
From their best fit model, it arises that blazars contribute negligible 
emission around 10\,keV. This is in conflict with the  main
finding of this paper that  blazars contribute $\sim$10-20\,\%\footnotemark{}
\footnotetext{The BAT blazar XLF's main uncertainty is given by the small
number of objects. Moreover, the CXB fraction increases if 
FSRQs and BL Lacs are treated as a single population.}
of the 
CXB emission in the 15-55\,keV band. 
As a matter of fact, $\sim$17\,\% of all  
BAT AGNs are blazars and thus their contribution to the CXB must be of the
same order.

The scenario which we derive from our data can be easily tested
by the {\it Fermi}-Large Area Telescope (LAT). Indeed, we showed
that in order not to overproduce the MeV background, most FSRQs are required
to 'peak' at MeV energies for a large fraction of their time. 
Thus, the detection by LAT of soft FSRQs
(e.g. photon indices of 2.2-2.5) would constitute a final evidence
that the IC peak should be between the BAT and the LAT energy bands.
On the other hand,
the detection of hard FSRQs (indices of 1.4-1.8) would invalidate
our prediction. The first {\it Fermi}-LAT results convalidate
our results \citep{lat_agn}. Indeed, FSRQs are detected by {\it Fermi}
with a mean photon index of 2.4 (and a tail extending up to 3.0) confirming
that the IC peak must be located somewhere between the keV and the GeV band.

\subsection{Tracing the star formation history of 
massive ellipticals at high redshift}
\label{sec:disc_highz}

The similar evolution of radio-quiet AGNs and star formation history
of normal galaxies has been interpreted as the evidence of the co-evolution
of AGNs and their hosts \citep[e.g.][]{madau96,hasinger05}.

The main, serendipitous, finding of our analysis is that the evolution
of the BAT blazars shows a redshift cut-off which is larger than previously
found for other, mostly radio-quiet, AGN samples. 
This is found to be z$_c=$4.3$\pm0.5$
for blazars of typical luminosities exceeding  
10$^{47}$\,erg s$^{-1}$ .
The large redshift cut-off shows that the most luminous blazars formed 
very early in the Universe and then their number density quickly decreased.
To our knowledge, no other source class displays a similar extreme evolution.
X-ray surveys show that the redshift cut-off increases with luminosity
\citep{ueda03,hasinger05,lafranca05,silverman08} and thus we believe
that doppler-boosting, due to the relativistic beaming,
allows BAT to detect rare objects which escaped even the deepest surveys.

We compared the blazars luminosity function  and the prediction of the 
star formation history of  massive elliptical galaxies \citep{delucia06}
and found  good agreement. However, this agreement is not entirely 
surprising if one realizes that blazars are normally found in giant 
elliptical galaxies \citep{urry95,falomo00,odowd02}. This represents another evidence
that AGNs (jet activity in this case) and their hosts
 co-evolve through the history
of the Universe. However, tracing
the evolution of giant  galaxies is currently at the limit, or beyond,
of the present-generation instruments and thus the use of blazars might
represent the only approach to understand the formation of the most
massive galaxies in the early Universe.

Elliptical galaxies are thought to be the only objects which undergo one major
merger \citep[e.g.][]{wang08} and in particular
this seems to happen in the first billion years of the Universe.
As a natural consequence, it is believed that 
merging activity would produce a rapidly spinning black hole 
\citep[e.g.][]{volonteri07} which on  theoretical grounds, 
is required to explain the production of a collimated,
relativistic, outflow \citep{blandford77,blandford90}.
\cite{sikora07} found out that on a radio-loudness/Eddington-ratio
diagram elliptical and disk/spiral galaxies form different sequences
and  invoke the spin as the black hole parameter which 
might explain this different behavior.
Larger blazar samples, better understanding
of the evolution of massive systems, and direct black hole spin measurements
will help in clarifying the jet-spin-merger scenario.

\clearpage
\acknowledgments
It is a pleasure to thank the referee of his/her comments 
which improved the paper.
MA acknowledges very helpful suggestions 
from A.~Comastri, R.~Gilli and A.~Reimer and 
interesting discussions with A.~Tramacere about blazars
(and his guidance during the first  {\it Fermi} flare-advocate duty shift).
The help of D.~Burlon with XMM-Newton data is acknowledged.
This research has made use of the NASA/IPAC extragalactic Database (NED) which
is operated by the Jet Propulsion Laboratory, of data obtained from the 
High Energy Astrophysics Science Archive Research Center (HEASARC) provided 
by NASA's Goddard Space Flight Center, and of the SIMBAD Astronomical Database
which is operated by the Centre de Donn\'ees astronomiques de Strasbourg.

{\it Facilities:} \facility{Swift/BAT}.

\bibliographystyle{apj}
\bibliography{/Users/marcoajello/Work/Papers/BiblioLib/biblio}

\appendix
\section{The Seyfert Sample}
\label{sec:app}
In this section we report the 199 Seyfert objects which are used
as a control sample plus the 6 radio galaxies detected by BAT.
The sample, shown in Tab.~\ref{tab:seyferts},
is reported here as a reference for the reader to
demonstrate that the classifications reported in Tab.~\ref{tab:cat} are 
accurate. We note that 152 out of the 205 sources reported in
this table are also detected in the BAT 22 month survey of \cite{tueller09}.
The main differences among the two analyses are:
\begin{itemize}
\item the different energy band used (15-55\,keV band versus 
the 15--195\,keV band adopted by \cite{tueller09}),
\item the different exposure used (36 months versus the 22 months used
by \cite{tueller09}),
\item the slightly different data filtering and screening techniques
\citep[see][for details]{ajello08a,tueller09}.
\end{itemize}
Despite these differences, $\sim$75\,\% of the sources detected  in
this analysis are also contained in the sample of \cite{tueller09}.

\clearpage

\begin{deluxetable}{lcccclccc}
\tablewidth{0pt}
\tabletypesize{\scriptsize}
\tablecaption{Sample of Seyferts and Radio Galaxies\label{tab:seyferts}}
\tablehead{
\colhead{SWIFT NAME}       & \colhead{R.A.}                &
\colhead{Decl.}            & \colhead{Flux}                &
\colhead{S/N}              & \colhead{ID\tablenotemark{a}}                  &
\colhead{Type\tablenotemark{b}}             & \colhead{Redshift}          &
\colhead{In BAT 22 months\tablenotemark{c} ?}  \\
%
%
\colhead{}                        & \colhead{\scriptsize (J2000)}         &
\colhead{\scriptsize (J2000)}     & \colhead{\scriptsize(10$^{-11}$ cgs)} &
\colhead{}                        & \colhead{}                            &
\colhead{}                        & \colhead{}  & \colhead{} 
}
\startdata

J0006.4+2009 & 1.600 & 20.152 & 1.16$\pm0.20$ & 5.8 & Mrk 335 & Sy1 & 0.0254 & y\\ 
J0038.6+2336 & 9.650 & 23.600 & 1.10$\pm0.21$ & 5.3 & Mrk 344 & Sy & 0.0240 &  \\ 
J0042.7-2332 & 10.680 & -23.548 & 2.44$\pm0.21$ & 11.7 & NGC 235A & Sy2 & 0.0222 & y\\ 
J0048.7+3157 & 12.188 & 31.962 & 7.71$\pm0.20$ & 37.8 & Mrk 348 & Sy2 & 0.0150 & y\\ 
J0051.9+1726 & 12.998 & 17.447 & 1.81$\pm0.21$ & 8.6 & QSO B0049+171 & Sy1 & 0.0642 & y\\ 
J0059.9+3149 & 14.997 & 31.831 & 1.66$\pm0.21$ & 8.0 & SWIFT J0059.4+3150 & Sy1 & 0.0149 & y\\ 
J0101.0-4748 & 15.274 & -47.800 & 0.97$\pm0.18$ & 5.6 & 2MASX J01003469-478303 & GALAXY & 0.0753 & y\\ 
J0108.8+1321 & 17.201 & 13.351 & 1.78$\pm0.22$ & 8.2 & 4C 13.07 & Sy2 & 0.0596 & y\\ 
J0111.4-3805 & 17.867 & -38.086 & 1.52$\pm0.18$ & 8.3 & NGC 424 & Sy2 & 0.0116 &  \\ 
J0113.8-1450 & 18.453 & -14.850 & 1.24$\pm0.21$ & 5.8 & Mrk 1152 & Sy1 & 0.0522 & y\\ 
J0114.3-5524 & 18.600 & -55.400 & 0.92$\pm0.17$ & 5.3 & SWIFT J0114.4-5522 & Sy2 & 0.0121 &  \\ 
J0123.8-5847 & 20.952 & -58.785 & 2.65$\pm0.17$ & 15.3 & Fairall 9 & Sy1 & 0.0470 & y\\ 
J0123.8-3504 & 20.974 & -35.067 & 2.72$\pm0.18$ & 14.7 & NGC 526A & Sy1.5 & 0.0191 & y\\ 
J0127.9-1850 & 22.000 & -18.847 & 1.27$\pm0.20$ & 6.2 & MCG-03-04-072 & Sy1 & 0.0430 &  \\ 
J0134.0-3629 & 23.506 & -36.486 & 2.36$\pm0.18$ & 13.0 & NGC 612 & GALAXY & 0.0298 & y\\ 
J0138.6-4000 & 24.674 & -40.008 & 3.17$\pm0.18$ & 18.0 & ESO 297-018 & Sy2 & 0.0252 & y\\ 
J0142.6+0118 & 25.652 & 1.300 & 1.28$\pm0.22$ & 5.7 & [VV2003c] J014214.0+011615 & Sy1 & 0.0500 &  \\ 
J0152.9-0326 & 28.250 & -3.448 & 1.47$\pm0.22$ & 6.6 & IGR J01528-0326 & Sy2 & 0.0172 & y\\ 
J0201.2-0649 & 30.320 & -6.821 & 4.17$\pm0.22$ & 19.3 & NGC 788 & Sy2 & 0.0136 & y\\ 
J0206.5-0016 & 31.631 & -0.270 & 1.53$\pm0.22$ & 6.9 & MRK 1018 & Sy1.5 & 0.0424 & y\\ 
J0215.0-0044 & 33.751 & -0.749 & 1.30$\pm0.22$ & 5.9 & Mrk 590 & Sy1.2 & 0.0265 &  \\ 
J0226.0-6315 & 36.500 & -63.250 & 0.91$\pm0.18$ & 5.2 & FAIRALL 0926 & Sy1 & 0.0580 &  \\ 
J0226.8-2819 & 36.703 & -28.324 & 1.14$\pm0.18$ & 6.4 & 2MASX J02262568-2820588 & Sy1 & 0.0600 &  \\ 
J0228.4+3118 & 37.120 & 31.316 & 4.38$\pm0.23$ & 19.4 & NGC 931 & Sy1.5 & 0.0166 & y\\ 
J0232.0-3639 & 38.020 & -36.662 & 1.09$\pm0.17$ & 6.4 & IC 1816 & Sy2 & 0.0169 & y\\ 
J0234.4+3229 & 38.612 & 32.489 & 1.60$\pm0.23$ & 7.1 & NGC 973 & Sy2 & 0.0167 & y\\ 
J0234.8-0847 & 38.702 & -8.794 & 2.13$\pm0.21$ & 10.2 & NGC 985 & Sy1 & 0.0430 & y\\ 
J0235.6-2935 & 38.900 & -29.600 & 0.99$\pm0.18$ & 5.6 & ESO 0416-G0002 & Sy1.9 & 0.0592 & y\\ 
J0238.5-5213 & 39.647 & -52.220 & 1.31$\pm0.17$ & 7.6 & ESO 198-024 & Sy1 & 0.0452 & y\\ 
J0239.0-4043 & 39.767 & -40.732 & 0.97$\pm0.17$ & 5.8 & 2MASX J02384897-4038377 & Sy1 & 0.0610 &  \\ 
J0241.5-0813 & 40.381 & -8.220 & 1.34$\pm0.21$ & 6.4 & NGC 1052 & Sy2 & 0.0050 & y\\ 
J0242.9-0000 & 40.732 & -0.012 & 2.00$\pm0.22$ & 8.9 & NGC 1068 & Sy2 & 0.0038 & y\\ 
J0249.3+2627 & 42.349 & 26.451 & 1.25$\pm0.23$ & 5.5 & IRAS 02461+2618 & GALAXY & 0.0580 & y\\ 
J0252.8-0830 & 43.200 & -8.500 & 1.06$\pm0.21$ & 5.0 & MCG-02-08-014 & Sy2 & 0.0168 & y\\ 
J0255.4-0010 & 43.873 & -0.170 & 4.48$\pm0.22$ & 20.1 & NGC 1142 & Sy2 & 0.0288 & y\\ 
J0256.4-3212 & 44.117 & -32.208 & 1.31$\pm0.17$ & 7.7 & ESO 417-6 & Sy2 & 0.0164 & y\\ 
J0311.6-2045 & 47.919 & -20.760 & 1.27$\pm0.18$ & 6.9 & 2MASX J03111883-2046184 & Sy1 & 0.0660 & y\\ 
J0325.1+3409 & 51.296 & 34.152 & 1.61$\pm0.24$ & 6.8 & 2MASX J03244119+3410459 & Sy1 & 0.0629 & y\\ 
J0333.5+3716 & 53.397 & 37.278 & 1.63$\pm0.24$ & 6.8 & IGR J03334+3718 & Sy1 & 0.0574 &  \\ 
J0333.7-3608 & 53.433 & -36.141 & 3.12$\pm0.17$ & 18.9 & NGC 1365 & Sy1.8 & 0.0055 & y\\ 
J0342.2-2114 & 55.554 & -21.244 & 2.15$\pm0.18$ & 11.8 & SWIFT J0342.0-2115 & Sy1 & 0.0145 & y\\ 
J0347.3-3029 & 56.850 & -30.500 & 0.89$\pm0.17$ & 5.3 & RBS 0741 & Sy1 & 0.0950 &  \\ 
J0350.7-5022 & 57.679 & -50.377 & 1.29$\pm0.17$ & 7.5 & SWIFT J0350.1-5019 & GALAXY & 0.0365 & y\\ 
J0357.0-4039 & 59.268 & -40.666 & 0.89$\pm0.17$ & 5.4 & 2MASX J03565655-4041453 & GALAXY & 0.0747 &  \\ 
J0402.5-1804 & 60.639 & -18.077 & 1.35$\pm0.19$ & 7.0 &  ESO 549- G049 & Sy2 & 0.0262 & y\\ 
J0407.5+0342 & 61.883 & 3.717 & 1.90$\pm0.25$ & 7.6 & 3C 105 & Sy2 & 0.0890 & y\\ 
J0415.2-0753 & 63.800 & -7.900 & 1.31$\pm0.23$ & 5.6 & LEDA 14727 & Sy1 & 0.0379 & y\\ 
J0426.4-5712 & 66.603 & -57.201 & 1.40$\pm0.17$ & 8.2 & 1H 0419-577 & Sy1 & 0.1040 & y\\ 
J0433.4+0521 & 68.355 & 5.365 & 5.21$\pm0.26$ & 19.8 & 3C-120 & Sy1 & 0.0330 & y\\ 
J0438.5-1049 & 69.633 & -10.830 & 1.48$\pm0.23$ & 6.4 & MCG-02-12-050 & Sy1 & 0.0360 & y\\ 
J0444.7-2812 & 71.199 & -28.200 & 1.07$\pm0.18$ & 5.9 & 2MASX J04450628-2820284 & Sy2 & 0.1470 &  \\ 
J0451.8-5807 & 72.966 & -58.133 & 0.88$\pm0.17$ & 5.2 & RBS 0594 & Sy1 & 0.0900 & y\\ 
J0453.5+0403 & 73.380 & 4.060 & 2.11$\pm0.28$ & 7.6 & CGCG 420-015 & Sy2 & 0.0296 & y\\ 
J0455.3-7528 & 73.841 & -75.477 & 1.27$\pm0.18$ & 6.9 & ESO 33-2 & Sy2 & 0.0184 & y\\ 
J0505.9-2351 & 76.497 & -23.854 & 2.78$\pm0.20$ & 13.9 & XSS J05054-2348 & Sy2 & 0.0350 & y\\ 
J0516.2-0009 & 79.071 & -0.161 & 4.11$\pm0.28$ & 14.5 & QSO-B0513-002 & Sy1 & 0.0327 & y\\ 
J0519.7-3240 & 79.930 & -32.676 & 2.38$\pm0.19$ & 12.9 & SWIFT J0519.5-3140 & Sy2 & 0.0350 & y\\ 
J0519.8-4546 & 79.963 & -45.774 & 2.49$\pm0.17$ & 14.9 & Pictor-A & Sy1 & 0.0351 & y\\ 
J0524.2-1212 & 81.050 & -12.200 & 1.42$\pm0.25$ & 5.6 & LEDA 17233 & Sy1 & 0.0490 & y\\ 
J0552.3-0727 & 88.090 & -7.457 & 14.75$\pm0.29$ & 51.7 & NGC 2110 & Sy2 & 0.0078 & y\\ 
J0552.3+5929 & 88.100 & 59.500 & 1.14$\pm0.21$ & 5.3 & IRAS 05480+597 & Sy1 & 0.0585 &  \\ 
J0558.1-3820 & 89.549 & -38.347 & 2.12$\pm0.18$ & 11.6 & EXO 055620-3820.2 & Sy1 & 0.0340 & y\\ 
J0559.9-5026 & 89.980 & -50.441 & 0.94$\pm0.17$ & 5.7 & PKS 0558-504 & RG & 0.1370 &  \\ 
J0602.9-8633 & 90.749 & -86.555 & 1.82$\pm0.22$ & 8.4 & SWIFT J0601.9-8636 & Sy2 & 0.0064 & y\\ 
J0603.1+6523 & 90.799 & 65.399 & 1.38$\pm0.20$ & 6.8 & UGC 3386 & GALAXY & 0.0154 &  \\ 
J0615.8+7101 & 93.967 & 71.021 & 6.08$\pm0.20$ & 30.8 & Mrk 3 & Sy2 & 0.0135 & y\\ 
J0623.9-3214 & 95.994 & -32.248 & 1.53$\pm0.20$ & 7.5 & ESO 426-G 002 & GALAXY & 0.0224 & y\\ 
J0624.1-6059 & 96.028 & -60.998 & 1.25$\pm0.17$ & 7.4 & SWIFT J2141.0+1603 & Sy2 & 0.0410 &  \\ 
J0640.7-4324 & 100.200 & -43.400 & 0.92$\pm0.18$ & 5.2 & 2MASX J06400609-4327591 & GALAXY & 0.0570 & y\\ 
J0652.1+7425 & 103.044 & 74.425 & 3.29$\pm0.19$ & 17.1 & Mrk 6 & Sy1.5 & 0.0188 & y\\ 
J0656.1+3959 & 104.027 & 39.986 & 2.29$\pm0.26$ & 8.7 & UGC 3601 & Sy1 & 0.0172 & y\\ 
J0718.0+4405 & 109.517 & 44.084 & 1.67$\pm0.24$ & 7.1 & 2MASX  J07180060+4405271 & Sy1 & 0.0610 &  \\ 
J0742.5+4947 & 115.644 & 49.793 & 2.98$\pm0.21$ & 14.4 & Mrk 79 & Sy1.2 & 0.0222 & y\\ 
J0800.1+2322 & 120.032 & 23.370 & 1.62$\pm0.24$ & 6.6 & SDSS J0759.87+232448.3 & GALAXY & 0.0290 & y\\ 
J0800.3+2638 & 120.099 & 26.648 & 1.79$\pm0.24$ & 7.5 & IC 486 & Sy1 & 0.0272 & y\\ 
J0804.2+0506 & 121.050 & 5.101 & 3.18$\pm0.25$ & 12.6 & UGC 4203 & Sy2 & 0.0135 & y\\ 
J0811.1+7602 & 122.798 & 76.049 & 1.15$\pm0.19$ & 6.2 & PG 0804+761 & Sy1 & 0.1000 &  \\ 
J0814.4+0423 & 123.600 & 4.400 & 1.26$\pm0.24$ & 5.2 & CGCG 031-072 & Sy1 & 0.0331 &  \\ 
J0823.2-0456 & 125.800 & -4.947 & 1.35$\pm0.23$ & 6.0 & SWIFT J0823.4-0457 & Sy2 & 0.0218 & y\\ 
J0832.8+3706 & 128.200 & 37.100 & 1.03$\pm0.20$ & 5.2 & RB 0707 & Sy1.2 & 0.0919 &  \\ 
J0839.8-1214 & 129.950 & -12.248 & 1.28$\pm0.21$ & 6.1 & 3C 206 & Sy1 & 0.1978 & y\\ 
J0904.9+5537 & 136.250 & 55.632 & 1.04$\pm0.17$ & 6.0 & SWIFT J0904.3+5538 & Sy1 & 0.0370 & y\\ 
J0911.5+4528 & 137.898 & 45.471 & 1.23$\pm0.18$ & 7.0 & SWIFT J0911.2+4533 & GALAXY & 0.0268 & y\\ 
J0918.4+1618 & 139.615 & 16.316 & 1.65$\pm0.21$ & 8.0 & Mrk 104 & Sy1.5 & 0.0292 & y\\ 
J0921.0-0803 & 140.257 & -8.067 & 2.59$\pm0.20$ & 12.9 & SWIFT J0920.8-0805 & Sy2 & 0.0198 & y\\ 
J0923.8+2256 & 140.962 & 22.936 & 2.05$\pm0.20$ & 10.5 & MCG +04-22-042 & Sy1.2 & 0.0327 & y\\ 
J0925.2+5217 & 141.316 & 52.285 & 3.03$\pm0.17$ & 18.1 & Mrk 110 & Sy1 & 0.0353 & y\\ 
J0945.8-1419 & 146.468 & -14.332 & 1.28$\pm0.21$ & 6.1 & NGC 2992 & Sy2 & 0.0077 & y\\ 
J0947.7-3056 & 146.939 & -30.948 & 11.50$\pm0.22$ & 51.5 & ESO 434-40 & Sy2 & 0.0085 & y\\ 
J0947.9+0727 & 147.000 & 7.451 & 1.12$\pm0.21$ & 5.4 & 3C 227 & RG & 0.0860 & y\\ 
J0959.6-2250 & 149.916 & -22.834 & 4.44$\pm0.22$ & 19.8 & NGC 3081 & Sy2 & 0.0080 & y\\ 
J1001.8+5542 & 150.453 & 55.700 & 1.43$\pm0.16$ & 8.8 & NGC 3079 & Sy2 & 0.0037 & y\\ 
J1006.0-2306 & 151.500 & -23.100 & 1.25$\pm0.23$ & 5.5 & ESO 499-G 041 & Sy1 & 0.0127 &  \\ 
J1021.7-0327 & 155.450 & -3.450 & 1.35$\pm0.22$ & 6.3 & MCG+00-27-002 & Sy1 & 0.0409 &  \\ 
J1023.5+1951 & 155.888 & 19.864 & 7.35$\pm0.20$ & 36.8 & NGC 3227 & Sy1.5 & 0.0039 & y\\ 
J1031.8-3451 & 157.975 & -34.860 & 4.71$\pm0.26$ & 18.4 & NGC 3281 & Sy2 & 0.0107 & y\\ 
J1031.9-1417 & 157.996 & -14.300 & 2.13$\pm0.23$ & 9.3 & H 1029-140 & Sy1 & 0.0860 &  \\ 
J1044.0+7023 & 161.003 & 70.400 & 0.99$\pm0.17$ & 5.9 & MCG+12-10-067 & Sy2 & 0.0333 &  \\ 
J1046.5+2556 & 161.649 & 25.950 & 1.12$\pm0.19$ & 5.9 & UGC 05881 & GALAXY & 0.0200 &  \\ 
J1048.5-2512 & 162.149 & -25.200 & 1.42$\pm0.27$ & 5.3 & NGC 3393 & Sy2 & 0.0125 &  \\ 
J1049.3+2256 & 162.350 & 22.950 & 1.57$\pm0.20$ & 8.0 & SWIFT J1049.4+2258 & Sy2 & 0.0328 & y\\ 
J1106.6+7234 & 166.654 & 72.571 & 6.45$\pm0.17$ & 38.0 & NGC 3516 & Sy1.5 & 0.0088 & y\\ 
J1115.9+5426 & 168.999 & 54.450 & 0.88$\pm0.15$ & 5.7 & SDSS J111519.98+542316.6 & Sy2 & 0.0703 &  \\ 
J1125.4+5421 & 171.352 & 54.351 & 0.97$\pm0.15$ & 6.3 & ARP 151 & Sy1 & 0.0210 &  \\ 
J1127.5+1908 & 171.900 & 19.148 & 1.12$\pm0.20$ & 5.6 & 1RXS J112716.6+190914 & Sy1 & 0.1050 & y\\ 
J1132.7+5259 & 173.188 & 52.988 & 1.01$\pm0.15$ & 6.6 & UGC 6527 & Sy1 & 0.0277 & y\\ 
J1136.5+2132 & 174.150 & 21.548 & 1.13$\pm0.19$ & 5.9 & Mrk 739 & Sy1 & 0.0299 &  \\ 
J1139.0-3744 & 174.764 & -37.741 & 10.07$\pm0.27$ & 37.9 & NGC 3783 & Sy1 & 0.0097 & y\\ 
J1139.1+5912 & 174.783 & 59.212 & 1.25$\pm0.15$ & 8.1 & SBS 1136+594 & Sy1.5 & 0.0601 &  \\ 
J1139.4+3156 & 174.869 & 31.935 & 1.00$\pm0.17$ & 5.8 & NGC 3786 & Sy1.8 & 0.0089 &  \\ 
J1144.7+7939 & 176.190 & 79.662 & 2.13$\pm0.18$ & 11.9 & SWIFT J1143.7+7942 & Sy1.2 & 0.0153 & y\\ 
J1145.3+5859 & 176.349 & 59.000 & 0.81$\pm0.15$ & 5.3 & Ark 320 & GALAXY & 0.0099 &  \\ 
J1145.5-1825 & 176.393 & -18.428 & 2.84$\pm0.27$ & 10.5 & 2MASX J11454045-1827149 & Sy1 & 0.0329 & y\\ 
J1148.9+2938 & 177.230 & 29.634 & 1.06$\pm0.18$ & 6.1 & MCG+05-28-032 & LINER & 0.0230 &  \\ 
J1158.0+5526 & 179.502 & 55.449 & 1.04$\pm0.15$ & 6.9 & NGC 3998 & Seyfert & 0.0036 & y\\ 
J1201.0+0647 & 180.250 & 6.800 & 1.18$\pm0.21$ & 5.6 & SWIFT J1200.8+0650 & GALAXY & 0.0360 & y\\ 
J1203.0+4432 & 180.773 & 44.534 & 2.33$\pm0.15$ & 15.1 & NGC 4051 & Sy1.5 & 0.0023 & y\\ 
J1204.5+2018 & 181.149 & 20.301 & 1.32$\pm0.19$ & 7.1 & ARK 347 & Sy2 & 0.0225 & y\\ 
J1206.2+5242 & 181.565 & 52.710 & 1.24$\pm0.15$ & 8.3 & NGC 4102 & GALAXY & 0.0028 & y\\ 
J1209.1+4700 & 182.300 & 47.000 & 0.76$\pm0.15$ & 5.0 & Mrk 198 & Sy2 & 0.0246 & y\\ 
J1209.4+4341 & 182.370 & 43.686 & 1.56$\pm0.15$ & 10.1 & NGC 4138 & Sy1.9 & 0.0030 & y\\ 
J1210.5+3924 & 182.633 & 39.406 & 24.60$\pm0.16$ & 153.4 & NGC 4151 & Sy1.5 & 0.0033 & y\\ 
J1210.6+3819 & 182.667 & 38.333 & 0.96$\pm0.16$ & 6.0 & LEDA 38759 & Sy1 & 0.0230 &  \\ 
J1217.2+0711 & 184.300 & 7.200 & 1.21$\pm0.20$ & 5.9 & NGC 4235 & Sy1 & 0.0080 & y\\ 
J1218.3+2950 & 184.593 & 29.839 & 1.53$\pm0.17$ & 9.1 & Mrk 766 & Sy1.5 & 0.0127 & y\\ 
J1219.0+4715 & 184.750 & 47.252 & 0.96$\pm0.15$ & 6.3 & NGC 4258 & LINER & 0.0015 & y\\ 
J1222.0+7518 & 185.503 & 75.311 & 1.27$\pm0.17$ & 7.4 & Mrk 205 & Sy1 & 0.0700 & y\\ 
J1225.7+1239 & 186.447 & 12.665 & 12.58$\pm0.19$ & 65.6 & NGC 4388 & Sy2 & 0.0084 & y\\ 
J1225.8+3330 & 186.466 & 33.513 & 1.25$\pm0.16$ & 7.7 & NGC 4395 & Sy1 & 0.0010 & y\\ 
J1235.6-3955 & 188.902 & -39.919 & 10.21$\pm0.26$ & 39.4 & NGC 4507 & Sy2 & 0.0118 & y\\ 
J1238.8-2718 & 189.723 & -27.308 & 4.39$\pm0.28$ & 15.9 & ESO 506-027 & Sy2 & 0.0240 & y\\ 
J1239.0-1611 & 189.769 & -16.196 & 2.02$\pm0.26$ & 7.7 & XSS J12389-1614 & Sy2 & 0.0360 & y\\ 
J1239.5-0520 & 189.898 & -5.341 & 4.52$\pm0.23$ & 19.8 & NGC 4593 & Sy1 & 0.0090 & y\\ 
J1246.6+5434 & 191.661 & 54.575 & 1.34$\pm0.15$ & 9.0 & NGC 4686 & GALAXY & 0.0168 & y\\ 
J1302.8+1624 & 195.700 & 16.400 & 0.90$\pm0.17$ & 5.1 & Mrk 0783 & Sy1.2 & 0.0672 &  \\ 
J1306.7-4024 & 196.698 & -40.415 & 2.37$\pm0.27$ & 8.9 & ESO 323-077 & Sy1.2 & 0.0150 & y\\ 
J1309.1+1137 & 197.279 & 11.632 & 2.19$\pm0.18$ & 12.0 & SWIFT J1309.2+1139 & GALAXY & 0.0251 & y\\ 
J1315.4+4424 & 198.852 & 44.404 & 1.28$\pm0.15$ & 8.4 & IGR J13149+4422 & Seyfert & 0.0367 &  \\ 
J1322.3-1642 & 200.591 & -16.716 & 2.57$\pm0.27$ & 9.5 & MCG -03-34-064 & Sy1.8 & 0.0165 & y\\ 
J1325.4-4301 & 201.366 & -43.017 & 49.77$\pm0.27$ & 187.6 & Cen A & Sy2 & 0.0018 & y\\ 
J1334.8-2323 & 203.700 & -23.400 & 1.49$\pm0.29$ & 5.1 & ESO 509-38 & Sy2 & 0.0265 &  \\ 
J1335.7-3418 & 203.944 & -34.302 & 4.86$\pm0.29$ & 16.6 & MCG -06-30-015 & Sy1.2 & 0.0077 & y\\ 
J1338.1+0433 & 204.547 & 4.552 & 3.63$\pm0.20$ & 17.9 & NGC 5252 & Sy2 & 0.0230 & y\\ 
J1341.4+3022 & 205.356 & 30.369 & 1.15$\pm0.16$ & 7.2 & Mrk 268 & Sy2 & 0.0404 &  \\ 
J1349.5-3018 & 207.390 & -30.304 & 17.87$\pm0.31$ & 58.2 & IC 4329A & Sy1 & 0.0161 & y\\ 
J1353.2+6919 & 208.305 & 69.327 & 2.78$\pm0.17$ & 16.7 & Mrk 279 & Sy1.5 & 0.0305 & y\\ 
J1356.1+3835 & 209.033 & 38.583 & 1.22$\pm0.16$ & 7.7 & Mrk 464 & Sy1 & 0.0507 & y\\ 
J1408.4-3024 & 212.100 & -30.400 & 1.65$\pm0.32$ & 5.1 & PGC 050427 & Sy1 & 0.0235 &  \\ 
J1413.5-0312 & 213.375 & -3.201 & 14.38$\pm0.24$ & 59.0 & NGC 5506 & Sy1.9 & 0.0062 & y\\ 
J1418.2+2507 & 214.568 & 25.133 & 3.12$\pm0.17$ & 18.2 & NGC 5548 & Sy1.5 & 0.0172 & y\\ 
J1419.5-2639 & 214.893 & -26.663 & 3.49$\pm0.34$ & 10.4 & ESO 511-G030 & Sy1 & 0.0224 & y\\ 
J1421.6+4750 & 215.420 & 47.838 & 1.03$\pm0.16$ & 6.4 & QSO B1419+480 & Sy1 & 0.0720 & y\\ 
J1424.3+2435 & 216.100 & 24.600 & 0.89$\pm0.17$ & 5.1 & NGC 5610 & GALAXY & 0.0169 &  \\ 
J1429.6+0117 & 217.400 & 1.300 & 1.26$\pm0.23$ & 5.4 & QSO B1426+015 & Sy1 & 0.0860 &  \\ 
J1436.5+5847 & 219.149 & 58.798 & 1.37$\pm0.16$ & 8.3 & QSO J1436+5847 & Sy1 & 0.0312 & y\\ 
J1441.2+5330 & 220.300 & 53.500 & 0.85$\pm0.16$ & 5.1 & Mrk 477 & Sy2 & 0.0380 &  \\ 
J1442.6-1713 & 220.664 & -17.223 & 4.83$\pm0.34$ & 14.4 & NGC 5728 & Sy2 & 0.0095 & y\\ 
J1453.1+2556 & 223.282 & 25.936 & 1.29$\pm0.18$ & 7.0 & RX J1453.1+2554 & Sy1 & 0.0465 &  \\ 
J1504.2+1025 & 226.073 & 10.417 & 1.51$\pm0.22$ & 6.8 & Mrk 841 & Sy1 & 0.0364 & y\\ 
J1515.4+4201 & 228.868 & 42.033 & 1.05$\pm0.18$ & 5.9 & NGC 5899 & Sy2 & 0.0085 & y\\ 
J1536.2+5753 & 234.061 & 57.890 & 1.43$\pm0.18$ & 7.9 & Mrk 290 & Sy1 & 0.0296 & y\\ 
J1548.4-1344 & 237.106 & -13.749 & 2.91$\pm0.39$ & 7.4 & NGC 5995 & Sy2 & 0.0251 & y\\ 
J1554.8+3242 & 238.700 & 32.700 & 1.04$\pm0.20$ & 5.2 & 2MASX J15541741+3238381 & Sy1 & 0.0483 &  \\ 
J1618.4+3224 & 244.600 & 32.400 & 1.07$\pm0.21$ & 5.1 & 3C 332 & RG & 0.1500 &  \\ 
J1628.3+5147 & 247.082 & 51.793 & 2.45$\pm0.20$ & 12.4 & SWIFT J1628.1+5145 & Sy1.9 & 0.0547 & y\\ 
J1653.2+0224 & 253.319 & 2.404 & 4.20$\pm0.35$ & 12.1 & NGC 6240 & Sy2 & 0.0245 & y\\ 
J1822.1+6421 & 275.541 & 64.361 & 1.10$\pm0.21$ & 5.3 & QSO B1821+643 & Sy1 & 0.2970 & y\\ 
J1824.2-5620 & 276.057 & -56.348 & 1.98$\pm0.29$ & 6.9 & IC 4709 & Sy2 & 0.0169 & y\\ 
J1835.1+3240 & 278.791 & 32.683 & 4.67$\pm0.21$ & 22.0 & 3C 382 & Sy1 & 0.0579 & y\\ 
J1837.1-5922 & 279.284 & -59.368 & 1.79$\pm0.28$ & 6.3 & FAIRALL 49 & Sy2 & 0.0200 & y\\ 
J1838.6-6523 & 279.658 & -65.394 & 6.23$\pm0.28$ & 22.4 & ESO 103-035 & Sy2 & 0.0133 & y\\ 
J1842.4+7946 & 280.616 & 79.771 & 5.82$\pm0.19$ & 29.9 & 3C 390.3 & Sy1 & 0.0561 & y\\ 
J1845.1-6223 & 281.297 & -62.399 & 2.52$\pm0.28$ & 9.0 & ESO 140-43 & Sy1 & 0.0141 & y\\ 
J1857.3-7827 & 284.341 & -78.464 & 1.86$\pm0.26$ & 7.1 & LEDA 140831 & Sy1 & 0.0420 & y\\ 
J1921.2-5840 & 290.323 & -58.677 & 3.54$\pm0.28$ & 12.6 & ESO 141-55 & Sy1 & 0.0366 & y\\ 
J1942.7-1018 & 295.680 & -10.316 & 4.30$\pm0.30$ & 14.3 & NGC 6814 & Sy1 & 0.0052 & y\\ 
J2009.1-6103 & 302.289 & -61.064 & 3.03$\pm0.26$ & 11.5 & SWIFT J2009.0-6103 & Sy1 & 0.0149 & y\\ 
J2018.3-5538 & 304.598 & -55.649 & 1.65$\pm0.27$ & 6.1 & PKS 2014-55 & RG & 0.0600 &  \\ 
J2042.7+7508 & 310.685 & 75.136 & 3.23$\pm0.19$ & 16.9 & 4C +74.26 & RG & 0.1040 & y\\ 
J2044.1-1043 & 311.039 & -10.731 & 5.62$\pm0.29$ & 19.7 & Mrk 509 & Sy1.2 & 0.0344 & y\\ 
J2052.0-5703 & 313.017 & -57.063 & 4.63$\pm0.25$ & 18.3 & IC 5063 & Sy2 & 0.0113 & y\\ 
J2109.1-0939 & 317.300 & -9.652 & 1.58$\pm0.27$ & 5.9 & 1H 2107-097 & LINER & 0.0265 &  \\ 
J2132.1-3343 & 323.028 & -33.727 & 2.92$\pm0.27$ & 10.7 & CTS 109 & Sy1 & 0.0297 & y\\ 
J2136.0-6223 & 324.006 & -62.400 & 2.42$\pm0.23$ & 10.6 & QSO J2136-6224 & Sy1 & 0.0589 & y\\ 
J2138.8+3206 & 324.713 & 32.115 & 1.29$\pm0.20$ & 6.3 & LEDA 67084 & Sy1 & 0.0250 &  \\ 
J2200.7+1033 & 330.199 & 10.565 & 1.76$\pm0.21$ & 8.6 & SWIFT J2200.9+1032 & Sy1.9 & 0.0266 & y\\ 
J2202.1-3152 & 330.526 & -31.878 & 8.01$\pm0.25$ & 31.8 & NGC 7172 & Sy2 & 0.0087 & y\\ 
J2204.5+0335 & 331.149 & 3.600 & 1.33$\pm0.21$ & 6.3 & IRAS 22017+0319 & Sy2 & 0.0610 &  \\ 
J2209.5-4709 & 332.387 & -47.166 & 3.02$\pm0.23$ & 13.4 & NGC 7213 & Sy1.5 & 0.0277 & y\\ 
J2223.8-0207 & 335.962 & -2.121 & 1.93$\pm0.22$ & 8.9 & 3C 445 & Sy1 & 0.0564 & y\\ 
J2235.8-2603 & 338.966 & -26.054 & 2.76$\pm0.24$ & 11.6 & NGC 7314 & Sy1.9 & 0.0048 & y\\ 
J2236.1+3357 & 339.040 & 33.952 & 1.66$\pm0.19$ & 8.8 & Arp 319 & Sy2 & 0.0225 & y\\ 
J2236.8-1235 & 339.223 & -12.599 & 1.43$\pm0.23$ & 6.2 & Mrk 915 & Sy1 & 0.0240 & y\\ 
J2245.7+3941 & 341.449 & 39.695 & 1.71$\pm0.18$ & 9.2 & 3C 452 & Sy2 & 0.0811 & y\\ 
J2254.1-1734 & 343.535 & -17.578 & 5.67$\pm0.23$ & 24.7 & MR 2251-178 & Sy1 & 0.0640 & y\\ 
J2258.9+4053 & 344.749 & 40.899 & 1.31$\pm0.18$ & 7.2 & UGC 12282 & Sy1 & 0.0171 & y\\ 
J2259.5+2455 & 344.899 & 24.929 & 1.51$\pm0.19$ & 8.0 & LEDA 70195 & Sy1 & 0.0338 & y\\ 
J2303.2+0853 & 345.809 & 8.885 & 3.87$\pm0.20$ & 19.4 & NGC 7469 & Sy1.2 & 0.0163 & y\\ 
J2304.7-0841 & 346.194 & -8.686 & 6.09$\pm0.22$ & 27.9 & Mrk 926 & Sy1.5 & 0.0469 & y\\ 
J2304.7+1217 & 346.200 & 12.300 & 1.11$\pm0.20$ & 5.6 & NGC 7479 & Sy2 & 0.0079 &  \\ 
J2318.4-4221 & 349.614 & -42.360 & 4.09$\pm0.20$ & 20.7 & NGC 7582 & Sy2 & 0.0053 & y\\ 
J2319.0+0014 & 349.762 & 0.241 & 2.82$\pm0.21$ & 13.5 & NGC 7603 & Sy1 & 0.0293 & y\\ 
J2326.3+2154 & 351.600 & 21.900 & 0.97$\pm0.19$ & 5.1 & RBS 20005 & Sy1 & 0.1200 &  \\ 
J2342.0+3035 & 355.500 & 30.600 & 1.08$\pm0.19$ & 5.7 & UGC 12741 & GALAXY & 0.0174 & y\\ 
J2358.7-6052 & 359.699 & -60.876 & 1.10$\pm0.19$ & 5.9 & PKS 2356-61 & RG & 0.0963 &  \\ 

\enddata
\tablenotetext{a}{Sources with a SWIFT name were identified in the works
of \cite{tueller08,tueller09}.}
\tablenotetext{b}{RG are radio galaxies, while sources identified as Galaxies
are candidate radio-quiet AGNs which are of the XBONG type \citep[see][]{tueller09}
or for which an optical spectrum is not yet available.}
\tablenotetext{c}{Is the source detected in the 22 months BAT survey of \cite{tueller09} ?}
\end{deluxetable}

\end{document}